\documentclass[numbook,epjST]{svjour}

\usepackage{a4wide}   

\usepackage[nottoc,numbib]{tocbibind}

\usepackage{epsfig}
\usepackage{bbm}               
\usepackage{amsmath,amssymb,amsfonts}

\usepackage{cclicenses}

\usepackage{dsfont}            
\usepackage{exscale}           
\usepackage{rotating,color} 
\usepackage{dcolumn}         
\usepackage{bm}              

\usepackage{wrapfig}          
\usepackage[rightcaption]{sidecap}  
\usepackage{caption}
\usepackage{subcaption}      

\usepackage[latin2]{inputenc}    
\usepackage[T1]{fontenc}

\usepackage{graphicx}
\usepackage[pdftex,
 colorlinks,          
 bookmarksnumbered,   
 hypertexnames=false, 
 pdfstartview=FitH,   
 linkcolor=blue,      
 citecolor=green,     
 urlcolor=magenta,    
 filecolor=blue  
]{hyperref}

\usepackage[square,comma,numbers,sort&compress]{natbib} 

\newcommand{\mytitle}{
  Atomic and molecular dynamics 
  triggered by ultrashort light pulses 
  on the atto- to picosecond time scale
}

\newcommand{\bra}[1]{\left< #1 \right|}
\newcommand{\ket}[1]{\left| #1 \right>}
\newcommand{\braket}[2]{\left<\left. #1 \right| #2 \right>}
\newcommand{\ev}[1]{\left< #1 \right>}

\newcommand{\half}[0]{\frac{1}{2}}

\newcommand{\cre}[0]{\hat c^\dagger}
\newcommand{\ann}[0]{\hat c}

\newcommand{\rotmats}[2]{D^{*\left[#1\right]}_{#2}}

\newcommand{\IDM}[0]{\text{IDM}}

\definecolor{MyDarkGreen}{rgb}{0,0.6,0}
\definecolor{MyDarkBlue}{rgb}{0,0,0.8}
\definecolor{MyDarkRed}{rgb}{0.6,0,0.3}

\hypersetup{breaklinks=true,
colorlinks =true,
plainpages =true,
linktocpage=true,
linkcolor=MyDarkBlue,
citecolor=MyDarkGreen,
urlcolor =MyDarkRed,
pdfborder={0 0 0},
pdfauthor={Stefan Pabst},
pdftitle ={\mytitle},
pdfsubject  = {tutorial review},
pdfkeywords = {Ultrafast Science, Atomic and Molecular Motion, Alignment, Diffraction, Ionization, Correlation, Entanglement
               Multichannel Effects, Correlation Dynamics, Many-Body Physics,
               Attosecond, Femtosecond, Picosecond, 
               Ultrashort Pulse, Strong-Field, NIR Pulse, UV Pulse, X-ray Pulse, Pump-Probe Experiments},
pdfcreator  = {LaTeX with hyperref package},
pdfproducer = {pdflatex}
}

\begin{document}

\title{\mytitle}
\dedication{Dedicated to my father}

\author{Stefan Pabst \inst{1,2} \fnmsep\thanks{\email{stefan.pabst@cfel.de}}}
\institute{Center for Free Electron Laser Science, Deutsches Elektronen-Synchrotron, D-22607 Hamburg, Germany \and Department of Physics, University of Hamburg, D-20355 Hamburg, Germany
}

\abstract{
Time-resolved investigations of ultrafast electronic and molecular dynamics were not possible until recently.
The typical time scale of these processes is in the picosecond to attosecond realm.
The tremendous technological progress in recent years made it possible to generate ultrashort pulses, which can be used to trigger, to watch, and to control atomic and molecular motion.
This tutorial focuses on experimental and theoretical advances which are used to study the dynamics of electrons and molecules in the presence of ultrashort pulses.
In the first part, the rotational dynamics of molecules, which happens on picosecond and femtosecond time scales, is reviewed.
Well-aligned molecules are particularly suitable for angle-dependent investigations like x-ray diffraction or strong-field ionization experiments.
In the second part, the ionization dynamics of atoms is studied.
The characteristic time scale lies, here, in the attosecond to few-femtosecond regime.
Although a one-particle picture has been successfully applied to many processes, many-body effects do constantly occur.
After a broad overview of the main mechanisms and the most common tools in attosecond physics, examples of many-body dynamics in the attosecond world (e.g., in high-harmonic generation and attosecond transient absorption spectroscopy) are discussed.
}
\maketitle

\section{Overview}
\label{p1:intro}

Rapid technological progress in generating shorter and shorter pulses has driven the interest in studying rotational, vibrational, and electronic excitations on their characteristic time scales.~\cite{Bl-RMP-1999,Ze-JPCA-2000,KrIv-RMP-2009} 
This review starts by an overview of two areas in atomic, molecular, and optical physics: (1) laser alignment of molecules and (2) ultrafast ionization dynamics.
Both areas have in common that the dynamics of the system is induced by very short laser pulses.
Due to the different characteristic time scales of rotational excitations (alignment) and electronic excitations (ionization), the duration of these pulses ranges from hundreds of picosecond (1~ps$=10^{-12}$~s) down to tens of attosecond (1~as$=10^{-18}$~s), respectively. 

In Sec.~\ref{p1c1}, the alignment dynamics of molecules will be studied.
First, a review is given which discusses what have been done in aligning and orienting molecules with optical laser fields.
The underlying principle of laser alignment and orientation is presented in Sec.~\ref{p1c1:intro_method}.
In Sec.~\ref{p1c1:intro_align}, the details of one-dimensional (1D) and three-dimensional (3D) alignment is explained and what alignment and orientation means and how it can be measured.
The ratio between the pulse duration and the rotational dynamics determines whether the alignment happens adiabatically or non-adiabatically.
These dynamical aspects are discussed in Sec.~\ref{p1c1:intro_dynamics}.
Schemes of achieving orientation rather than alignment are reviewed in Sec.~\ref{p1c1:intro_orient}.

In Sec~\ref{p1c1:apps}, the some specific applications of laser alignment of molecules will be discussed.
It starts with a discussion in Sec.~\ref{p1c1:apps_ff3da} about a new schemes to improve field-free alignment in three dimension by using a multiple pulse scheme.
Laser alignment of molecules can be also used as a mean to prepare the molecular state for further scientific investigations.
Particularly interesting is the application of x-ray scattering from 3D laser-aligned molecules which is discussed in Sec.~\ref{p1c1:apps_xscatter}.
In Sec.~\ref{p1c1:apps_ionization} examples are discussed where the alignment and even the orientation of molecules is crucial for studying angle-dependent strong-field phenomena like tunnel ionization and high-harmonic generation.

In Sec.~\ref{p1c2}, the focus switches to ultrafast ionization dynamics of noble gas atoms on the sub-femtosecond (1~fs$=10^{-15}$~s) time scale.
Even though many processes and mechanisms can be well understood by focusing on a single electron, multi-electron effects can become important particularly for large atoms and molecules.
The advantage of atomic systems is their small size and their high degree of symmetry.
For ultrafast light-induced processes, this greatly helps to systematically study the importance of details in the electronic structure originating from electron-electron correlations.

In Sec.~\ref{p1c2:photo} and Sec.~\ref{p1c2:tunnel} the two most prominent mechanisms in ultrafast science: one-photon photoionization and tunnel ionization are discussed, respectively.
In Sec.~\ref{p1c2:hhg}, the fundamental aspects of high harmonic generation (HHG), one of the most essential process in ultrafast sciences which converts near-infrared (NIR) light with wavelengths of $\sim1~\mu$m into UV and x-ray light with wavelengths in the 1--100~nm regime.
Because of the coherence properties of this UV light, HHG is the most common method to generate pulses with durations in the sub-femtosecond regime and photon energies up to the extreme-ultraviolet (XUV). 

The combination of NIR and XUV pulses offers a wide range of new possibilities.
In the form of a pump-probe setup, where the pump pulse triggers the dynamics and the probe pulse detects the changes in the system, either pulse could be used as pump or probe.
In Sec.~\ref{p1c2:streaking}, attosecond streaking experiments are discussed, where the XUV and the NIR pulses are used as pump and probe pulses, respectively.
Streaking makes it possible to imprint the temporal dynamics of ionized electrons into the energy spectrum where it can be measured by a detector.
For attosecond transient absorption experiments discussed in Sec.~\ref{p1c2:tas} the function of the two pulses are reversed and the NIR pulse is used as a pump and the XUV pulse is used as a probe.
Here, the electron dynamics gets imprinted in the spectrum of the probe pulse, which can be subsequently measured by a detector.
Transient absorption, to be more precise, is more a statement about the probing step rather than the pumping step. 
Generally, any other mechanism can be used as a pump as long as it is well-defined in time and preferably well-synchronized with the probe pulse.
The latter is normally only possible when the pump process is triggered by a second pulse in the terahertz, IR, UV, or x-ray regimes.

In Sec.~\ref{p1c2:theory}, numerical schemes are presented that makes it possible to understand the underlying mechanisms.
In Sec.~\ref{p1c2:sae}, the widely used single-active electron (SAE) model is discussed, where only one specific electron is allowed to move and all others are frozen.
Hence, electron-electron correlations cannot appear.
The time-dependent configuration-interaction singles (TDCIS) approach generalizes the idea of the SAE model.
Here, only one electron can be fully active as well. 
However, it can come from any occupied orbital.
Furthermore, Coulomb interactions between the excited/ionized electron and the produced hole state in the ion can trigger correlation effects.
The TDCIS approach is discussed in detail in Sec.~\ref{p1c2:tdcis}.

In Sec.~\ref{p1c2:apps}, research on many-body effects in the strong-field and attosecond regime are presented. 
In Sec.~\ref{p1c2:apps_decoh} it is shown that correlation effects can already happen in simple processes like one-photon ionization.
Also the aspect of multi-orbital contributions in the HHG spectrum, which is commonly neglected, is discussed in Sec.~\ref{p1c2:apps_hhg}.
The dynamics of many-body effects is best studies by a pump-probe scheme. 
In Sec.~\ref{p1c2:apps_tas} attosecond transient absorption spectroscopy studies are presented, which reveals novel field-driven dressing mechanisms on a sub-cycle time scale.

Atomic units ($\hbar\!=\!|e|\!=\!m_e\!=\!1/(4\pi\epsilon_0)\!=\!1$)~\cite{NIST_website,MoTa-RMP-2012} are used throughout if not explicitly mentioned otherwise.

\section[Laser Alignment of Molecules]
{Laser-Induced Molecular Alignment: About Picosecond and Femtosecond Dynamics}
\label{p1c1}
\subsection{Principles of Molecular Orientation and Alignment}
\label{p1c1:intro_method}
A wide range of processes in nature depend on the relative orientation of the objects involved.
This is true in the macroscopic world (e.g., collisions of classical particles) and it is also true in the microscopic world of atoms and molecules (e.g., chemical reactions~\cite{Be-JCP-1985}).
Therefore, the interest is high in controlling the directionality of single molecules~\cite{Lo-AnnRevPhysChem-1995}.
A very elegant way to control the molecular motion is by using electric fields ${\cal E}$~\cite{KrBe-JCP-1965,FrHe-PRL-1995,FrHe-JPC-1995} and magnetic fields ${\cal H}$~\cite{FrHe-ZPhysD-1992,SlFr-PRL-1994,AuSt-JCP-1996,PeGa-JPCL-2010}.
In the following the focus lies on electric fields, since they have become the common technique to orient and align molecules~\cite{StSe-RMP-2003}.
The strengths of magnetic interactions are generally three orders of magnitudes weaker than electronic interactions~\cite{Siegman-book}.
Magnetic moments of molecules are normally around a Bohr magneton ($1\,\mu_B=0.5$~a.u.) and large magnetic fields of around 1~Tesla are needed to orient and to align molecules~\cite{FrHe-ZPhysD-1992}.
Static magnetic fields of this magnitude are quite difficult to generate than corresponding electric fields.

The underlying mechanism for orienting and aligning molecules with electric fields can be well understood by looking at the Taylor expansion of the vibronic (electronic + vibrational) energy levels $E_i$ in terms of the applied electric field ${\cal E}(\omega)$~\cite{Bi-RMP-1990}
\begin{subequations}
  \label{eq1:dressed-energy}
\begin{align}
  \label{eq1:dressed-energy-level}
  E_i
  &=
  E_i^{(0)}
  +
  \sum_{n>0} E_i^{(n)}(\omega)
  ,
\\
  \label{eq1:dressed-energy-correction}
  E_i^{(n)} (\omega)
  &=
  -\frac{1}{n!}\, \gamma_i^{(n)}(\omega) \cdot {\cal E}^n(\omega),
\end{align}
\end{subequations}
where $E_i^{(0)}$ are the field-free energy levels and $\gamma_i^{(n)}(\omega)$ are the $n$-th order dipole response functions depending on the laser frequency $\omega$.
The symbol $\cdot$ stands for the inner product between the two tensors
$\gamma_i^{(n)}(\omega)$ and ${\cal E}^n(\omega)$.
In Cartesian coordinates, the inner product of two tensors $A$ and $B$ of rank $n$ reads $A \cdot B = \sum_{i_1,\cdots,i_n \in\{X,Y,Z\}} A_{i_1,\cdots,i_n} B_{i_1,\cdots,i_n}$.
Note that the electric field ${\cal E}(\omega)$ is a vector (i.e., tensor of rank 1).

The homogeneous character of the electric field over the size of a molecule allows one to neglect contributions from quadrupole or even higher moments~\cite{MeFo-MolPhys-2000}.
The first three dipole response functions appearing in Eq.~\eqref{eq1:dressed-energy-correction} are well known under the name permanent dipole moment $\mu:=\gamma^{(1)}$, dipole polarizability tensor $\alpha(\omega):=\gamma^{(2)}(\omega)$, and hyperpolarizability tensor $\beta(\omega):=\gamma^{(3)}(\omega)$.
Note, the permanent dipole moment is not $\omega$-dependent. 
The field frequency $\omega$ is commonly much lower than any vibronic transition.
Therefore, the $\omega$-dependence of the response functions can be dropped and the static limit ($\omega \rightarrow 0$) can be used~\cite{BuSa-JCP-2008}.
The Taylor expansions in Eq.~\eqref{eq1:dressed-energy-level} is particularly powerful when relatively weak electric fields are applied such that the infinite sum converges quickly for small $n$.

\subsubsection{Rotational Potential}
\begin{figure}[ht]
  \centering
  \includegraphics[width=.5\linewidth]{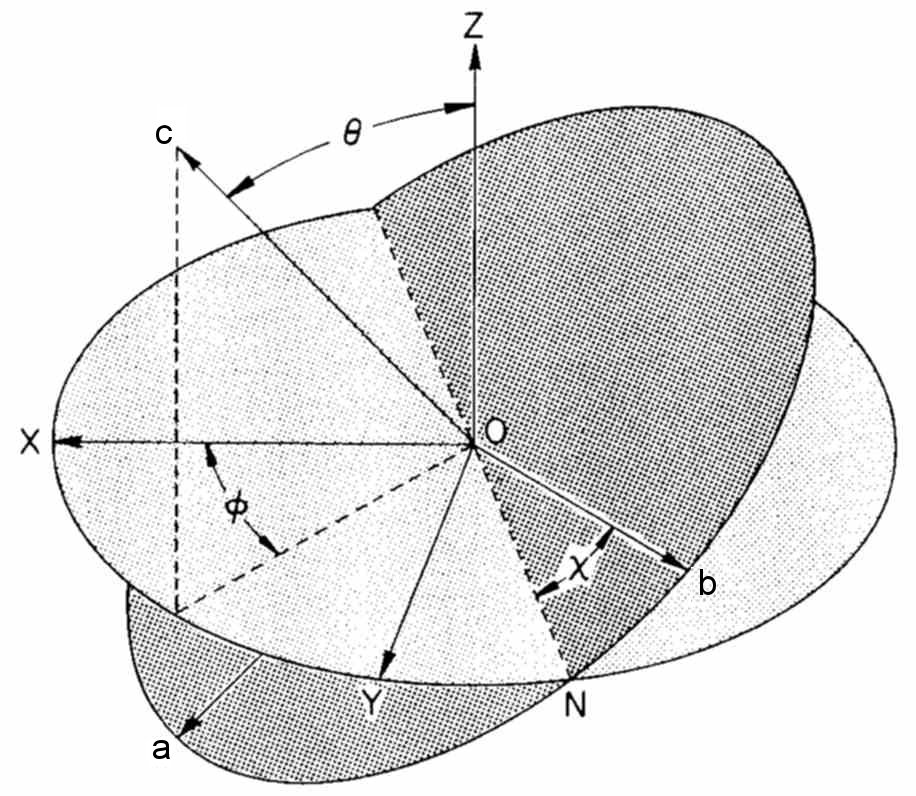}
  \caption{
    An illustration of a coordinate transformation from the space-fixed frame defined by the axes $X,Y,Z$ into the body-fixed frame defined by the axes $a,b,c$.
    The rotations defined by the Euler angles $\varphi,\vartheta,\chi$ are explicitly shown.
    The illustration is taken from Ref.~\cite{Zare-book}.
    Copyright~\copyright 1988 John Wiley \& Sons, Inc.
  }
  \label{fig1:euler}
\end{figure}

For common alignment scenarios, the field intensity is in the range 10$^{12}$-10$^{13}$~W/cm$^2$~\cite{StSe-RMP-2003} and lies in the perturbative regime such that it is sufficient to focus on the first three energy corrections
in the infinite sum of Eq.~\eqref{eq1:dressed-energy-level}.
The energy corrections $E_i^{(n)}$ of the vibronic states can be viewed as a potential,
\vspace{-.5ex}
\begin{align}
  \label{eq1:potential_rotational}
  \hat U_i
  &=
  \sum_{n=1}^3
    E_i^{(n)}(\omega)
  ,
\end{align}
for the remaining rotational degrees of freedom.
The corrections $E_i^{(n)}$ depend on the relative orientation of the molecule with respect to the polarization of the electric field, since $\gamma^{(n)}$ are defined in the body-fixed molecular frame.
The orientation of any molecule can be characterized by the Euler angles $\varphi,\vartheta$, and $\chi$~\cite{Zare-book}.
Therefore, all potentials $\hat U_i$ can be expressed in terms of the Euler angles.
Geometrically, the Euler angles define three rotations, which transform the space-fixed laser frame defined by the axes $X,Y$, and $Z$ into the body-fixed molecular frame defined by the axes $a,b$, and $c$.
Figure~\ref{fig1:euler} illustrates this 3D coordinate transformation.
Commonly, the angle $\varphi$ refers to the rotation around the space-fixed $Z$ axis, the angle $\chi$ refers to a rotation around the body-fixed $c$ axis, and $\vartheta$ defines the angle between the space-fixed $Z$ and body-fixed $c$ axes.
In general, the angle between a body-fixed axis $g\in\{a,b,c\}$ and a space-fixed axis $F\in\{X,Y,Z\}$ is labeled $\vartheta_{Fg}$.
In Sec.~\ref{p1c1:apps_xscatter}, these angles are used to analyze the 3D alignment of naphthalene molecules.

In Sec.~\ref{p1c1:intro_align}, the explicit form of the angle-dependence of $\hat U_i$ is discussed in detail for different types of molecules and laser fields.
If $U_i$ is energetically larger than the kinetic energy, the molecule becomes trapped in the potential and the orientation or alignment of the molecule becomes well-defined in the space-fixed coordinate system.
If an entire ensemble of molecules is exposed to such an electric field, a collective (ensemble-averaged) orientation or alignment is observed, which is not random anymore.
This is desirable, since the signal of a single molecule is generally too small to be statistically useful.
A collective orientation or alignment increases the signal strength by the number of molecules in the ensemble, which can be easily of the order of $10^7$ molecules~\cite{PeBu-AppPhysLett-2008}.

\subsubsection{Orientation and Alignment}
Orientation and alignment refer both to the directionality of the molecule.
However, there is a subtle difference between orientation and alignment.
For orientation in 3D, all three Euler angles are uniquely defined. 
In the case of alignment, there exists a ``head-tail'' ($C_2$) symmetry around each symmetry axis--meaning an 180$^\circ$-rotation of any symmetry axis does not matter.
For perfect alignment, there exist 4 unique orientations that correspond to the same alignment~\cite{SpSc-ActaCrystA-2005}.
The degree of orientation is measured by the quantity $\cos\vartheta$, where the sign indicates whether the molecules is oriented along or opposite to the desired orientation direction.
The measure for alignment is $\cos^2\vartheta$, which is invariant under a $180^\circ$-rotation of the alignment axis (i.e., $\vartheta \rightarrow 180^\circ + \vartheta$).
For a randomly oriented (aligned) ensemble of molecules, the corresponding ensemble-averaged degree of orientation (alignment) is $\ev{\cos\vartheta}=0$ $\left(\ev{\cos^2\vartheta}=\frac{1}{3}\right)$\footnote{The angle $\vartheta_{Zc}$ is identical to the Euler angle $\vartheta$.}.

\begin{figure}[ht]
  \centering
  \includegraphics[width=.6\linewidth]{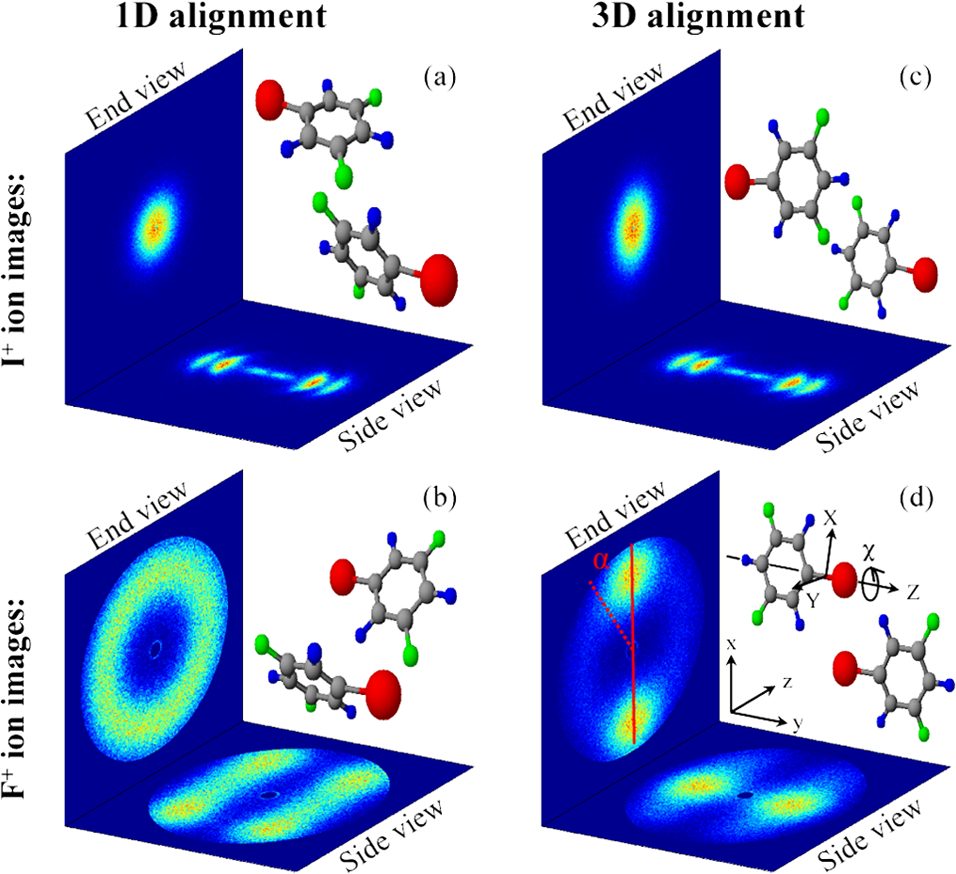}
  \caption{
    Illustration of measuring experimentally 3D alignment and/or orientation via the side view and end view detectors.
    The molecule shown is 3,5 difluoroiodobenzene, where the Iodine atom is depicted as red ball and the Fluorine atoms are depicted as green balls.
    Note, the space-fixed frame is labeled $x,y,z$ and the body-fixed molecular frame is labeled $X,Y,Z$.    
    The illustration is taken from Ref.~\cite{ViKu-PRL-2007}.
    Copyright~\copyright~2007 American Physical Society (APS).
  }
  \label{fig1:align_exp}
\end{figure}

The measure for 3D orientation and 3D alignment requires at least 3 angles, one for each axis~\cite{KaSa-JCP-2001}.
One specific choice are the three directional cosines $\cos^2\vartheta_{Xa},\cos^2\vartheta_{Yb}$, and $\cos^2\vartheta_{Zc}$~\cite{PaHo-PRA-2010}.
An average of these three angles can be also taken as a measure such that the 3D orientation/alignment is characterized by one quantity~\cite{MaRe-PRA-2012}.
Orientation and alignment can be measured experimentally via the angular distribution of ion fragments of the aligned molecule~\cite{DoLi-PRA-2003,StSe-RMP-2003}.
The ion fragments are produced by an intense second laser pulse, which forces the molecule to Coulomb explode.
Fig.~\ref{fig1:align_exp} illustrates how the degree of alignment is experimentally measured via Coulomb explosion.
Recently, it has been shown that HHG can also be used to measure alignment dynamics~\cite{LoRa-PRL-2012}. 

The potential resulting from $E_i^{(2)}$ leads to aligned molecules, since inverting the direction of the electric field (${\cal E \rightarrow -E}$) does not affect the potential due to the quadratic field dependence [cf. Eq.~\eqref{eq1:dressed-energy-correction}].
The terms $E_i^{(1)}$ and $E_i^{(3)}$, which depend on odd powers of ${\cal E}$, change sign when the electric field is pointing in the opposite direction.
Hence, the potential resulting from $E_i^{(1)}$ and $E_i^{(3)}$ leads to oriented molecules~\cite{FrHe-JCP-1999,DeZn-PRL-2009,HoNi-PRL-2009,SaMi-PRL-2003}.

\subsubsection{Structural Deformations}
In most experiments the goal of applying an external field is to influence and to control only the rotational state while at the same time leaving the molecule in the vibronic ground state~\cite{StSe-RMP-2003}.
Common way to achieve orientation and alignment is with vibronically non-resonant optical fields alone~\cite{DaGu-PRL-2005,RoGu-PRA-2006} and in combination with static electric fields~\cite{FrHe-JCP-1999,CaMa-PRL-2001,SaMi-PRL-2003}.
However, alignment can be also achieved with resonant frequencies coupling two or several vibronic states with each other~\cite{VrSt-ChemPhysLett-1997,GrPe-JCP-2004,YaGu-JCP-2007,HuSh-ChemPhysLett-2009}.
It has been shown that less intense electric fields are needed when resonant frequencies are used~\cite{Se-JCP-1995}.

Structural deformations may also occur when static fields or non-resonant optical fields are used~\cite{CoSe-JPCA-2003,MaMa-PRL-2009}.
Particularly for short laser pulses, where higher electric field strengths are needed to achieve the same degree of alignment~\cite{BuSa-JCP-2008}, structural deformations start to appear~\cite{DiKe-PRA-1999,CoSe-JPCA-2003}.
Nevertheless, the rigid-rotor approximation, where molecular deformations are ignored, is the most common model used~\cite{FrHe-PRL-1995,StSe-RMP-2003,UnSu-PRL-2005,RoGu-PRA-2006,ArSe-JCP-2008,MaRe-PRA-2012}.
This model has been successful due to its great simplicity and explanatory power for many experimental results.
Some works have been done to extend the rigid-rotor model in order to include centrifugal deformations in the rotational energies~\cite{Kroto-book,ReRe-PRL-2003,NaHo-JRS-2007,HoMi-JCP-2009}.
These deformations influence the rotational dynamics of the molecules, which becomes particularly prominent for linear molecules~\cite{CrBu-PRA-2009}.
A complete description of rotational and vibronic dynamics, which may be even coupled to each other, is quite challenging.

In the following, the molecule is assumed to be in the vibronic ground state at all times\footnote{
To be more precise, only the electronic state is considered to be in the quantum mechanical ground state.
The vibrational degrees of freedom are described classically and not quantum-mechanically--meaning the nuclei are classical objects with fixed, well-defined distances between each other.
}.
Hence, the vibronic index $i$ is dropped.
All energy corrections $E^{(n)}$, from now on, refer to the vibronic ground state of the molecule if not explicitly mentioned otherwise.

\subsection{1D and 3D Alignment}
\label{p1c1:intro_align}
Depending on the symmetry of the molecule and on the polarization of the 
electric field, molecules can be either aligned around one axis
(1D alignment) or around all three axes (3D alignment)~\cite{StSe-RMP-2003}.
To understand under which conditions 1D or 3D alignment can be achieved, it is
useful to consider the symmetries of the terms entering in the alignment potential.

For molecular alignment, only the term $E^{(2)}$ needs to be considered.
Aligning molecules has several practical advantages in comparison to orienting
them.
First, not all molecules have permanent dipole moments ($\mu\!=\!0\, \Rightarrow\, E^{(1)}\!=\!0$) and the third-order term $E^{(3)}$ has to be exploited, which requires large electric field strengths.
Since all molecules have a non-zero polarizability~\cite{FrHe-PRL-1995}, there always exists an aligning potential $E^{(2)}$.
Second, quasi-monochromatic optical pulses can only couple to even order response functions (i.e., $\gamma^{(2n)}$) and, therefore, generate aligning potentials.

The tensors ${\cal E}^{2n+1}(t)$ vanish for optical monochromatic fields, since the cycle-averaged quantities of ${\cal E}^{n}(t)$ only survive for even $n$.
Cycle-averaging of ${\cal E}^{n}(t)$ is justified when the field period is much smaller than the rotation dynamics of the molecule~\cite{DiKe-PRA-1999}.
This is the case for typical alignment pulses (e.g., wavelength of 800~nm) with cycle periods in the few femtosecond regime.
The typical time scale of rotational excitations is picoseconds (see Sec.~\ref{p1c1:intro_dynamics}).

As already mentioned, the energy correction of the vibronic ground state is an effective potential for the rotational degrees of freedom [cf. Eq.~\eqref{eq1:potential_rotational}].
The rotational Hamiltonian, where an optical pulse couples only to the polarizability $\alpha$, is given by~\cite{UnSu-PRL-2005,PaHo-PRA-2010}
\begin{align}
\label{eq1:ham}
  \hat H(t)
  &=
  \underbrace{A\,\hat J^2_a + B\,\hat J^2_b + C\,\hat J^2_c}_{\displaystyle \hat H_\text{rot}}
  -
  \underbrace{
  \half
  \sum_{LM} 
    (-1)^{L+M} \hat \alpha^{[L]}_M \
    F^{[L]}_{-M,\text{avg}}(t)
  }_{\displaystyle -\hat U(t)}
  ,
\end{align}
where $\hat H_\text{rot}$ is the kinetic operator for molecular rotations, and $\hat U(t)$ is the effective aligning potential induced by the electric field, ${\cal E}(t)$.
The rotational constants $A,B,C$ of the molecular axes $a,b$, and $c$ diagonalize the moment of inertia tensor.
The angular momentum operators are given by $\hat J_{g}$ with $g\in\{a,b,c\}$.
The potential $\hat U(t)$ is expressed via spherical tensor products~\cite{Zare-book}, where $F^{[L]}_M(t) = \left[{\cal E}(t) \otimes {\cal E}(t)\right]^{[L]}_{M,\text{avg}}$ is the spherical tensor of rank 2 with angular momentum $L$.
$M$ is the angular momentum projection onto the space-fixed $Z$ axis.
The subscript ``avg'' stands for the cycle-averaged quantity.
Spherical tensors with $L=1$ do not appear for $\alpha$ and $F(t)$, since both tensors are symmetric.

The electric field tensor $F(t)$ is ``well-defined'' in the space-fixed laser frame whereas the polarizability tensor $\alpha$ is ``well-defined'' in the molecular frame. 
Since the space-fixed $Z$ axis is the quantization axis in the tensor product of Eq.~\eqref{eq1:ham}, it is necessary to transform the known body-fixed spherical tensor components $\alpha^{[L]}_K$ (with the quantization axis $c$) into the space-fixed laser frame, which read~\cite{UnSu-PRL-2005,PaHo-PRA-2010}.
\begin{align}
  \label{eq1:polarizability_trafo}
  \alpha^{[L]}_M(\Omega)
  =
  \sum_K \rotmats{L}{M,K}(\Omega)\, \alpha^{[L]}_K
  ,
\end{align}
where $\rotmats{L}{M,K}(\Omega)$ are the Wigner-D matrices connecting both frames with each other~\cite{Zare-book}.
The three Euler angles are combined into $\Omega=(\varphi,\vartheta,\chi)$.
It is exactly through this coordinate transformation that the alignment potential becomes angle-dependent $\hat U(t) \rightarrow \hat U(\Omega,t)$.

The quantum number $K$ refers to the angular momentum projection onto the molecular $c$ axis.
The polarizability tensor $\alpha$ in the Cartesian representation is diagonal in the molecular frame with the values $\alpha_{a,a},\alpha_{b,b}$, and $\alpha_{c,c}$.
Note, the body-fixed frame that diagonalizes $\alpha$ is generally not the same frame that diagonalizes the moment of inertia tensor~\cite{Bernath-book}.
For small molecules (with a high symmetry), however, these two frames fall together.
In the following, only molecules are considered where these two molecular frames coincide.

Only terms in Eq~\eqref{eq1:ham} contribute to the aligning potential where both spherical tensor components, i.e., $\alpha^{[L]}_M$ and $F^{[L]}_{-M}(t)$, are non-zero.
For example, $\alpha^{[0]}_0$ is the only non-zero component for a spherically symmetric molecule.
Regardless of the complexity of $F(t)$ the resulting potential $\hat U(t)$ has no angle dependence, since only the Wigner-D matrix $\rotmats{0}{0,0}(\Omega)=1$ contributes.
This angle-independent potential results in a global energy shift for all rotational states and, therefore, has no influence on the alignment process and can be neglected~\cite{BuSa-PRA-2008}.

\subsubsection{1D Alignment}
The simplest alignment scenario is the 1D alignment of a linear ($A=B,C^{-1}=0$) or 
symmetric-top ($A=B\neq C$) molecules with linearly polarized light.
For linearly polarized fields, only the terms $F^{[0]}_0(t)$ and $F^{[2]}_0(t)$ are non-zero.
The polarizability tensor in the Cartesian basis has the diagonal entries $\alpha_{aa}=\alpha_{bb}\neq \alpha_{cc}$ and, therefore, only the terms $\alpha^{[0]}_{K=0}$ and $\alpha^{[2]}_{K=0}$ are non-zero.
By rotating from the molecular frame to the laser frame the Wigner-D matrix $\rotmats{2}{0,0}(\Omega)=\half(3\cos^2\vartheta-1)$ enters. 
The resulting angle-dependent potential reads~\cite{FrHe-PRL-1995}
\begin{align}
  \label{eq1:potential_1d}
  U_\text{1D}(\vartheta,t)
  &=
  -\frac{\alpha_{cc}-\alpha_{aa}}{2} \,
  \cos^2\vartheta \,
  [{\cal E}^2(t)]_\text{avg}
  .
\end{align}
The cycle-average of a quasi-monochromatic pulse ${\cal E}(t)={\cal E}_0(t)\,\sin(\omega\,t)$, where ${\cal E}_0(t)$ defines the pulse envelope, leads to the effective intensity $[{\cal E}^2(t)]_\text{avg}={\cal E}_0^2(t)/2$.
For linear molecules, $\alpha_{cc}>\alpha_{aa}$ and the potential $U_\text{1D}(\vartheta,t)$ has a minimum when the molecular $c$ axis is aligned with the space-fixed $Z$ axis ($\cos^2\vartheta=1$).
Depending on whether the symmetric-top molecule is prolate ($\alpha_{cc}>\alpha_{aa}$) or oblate ($\alpha_{cc}<\alpha_{aa}$), the molecular $c$ axis is preferably aligned along ($\cos^2\vartheta=1$) or perpendicular ($\cos^2\vartheta=0$) to the field polarization axis $Z$.

In 1D alignment, only one molecular axis is fixed in space. 
Rotations around the axes $c$ and $Z$ are unaffected by $U_\text{1D}(\vartheta,t)$, since no $\varphi$ and $\chi$ dependencies exist in $U_\text{1D}(\vartheta,t)$.
As a consequence, the molecular axes $a$ and $b$ are aligned within a plane but it is not possible to hold either axis in a particular direction within this plane.

\begin{figure}[h]
  \centering
  \includegraphics[width=.45\linewidth]{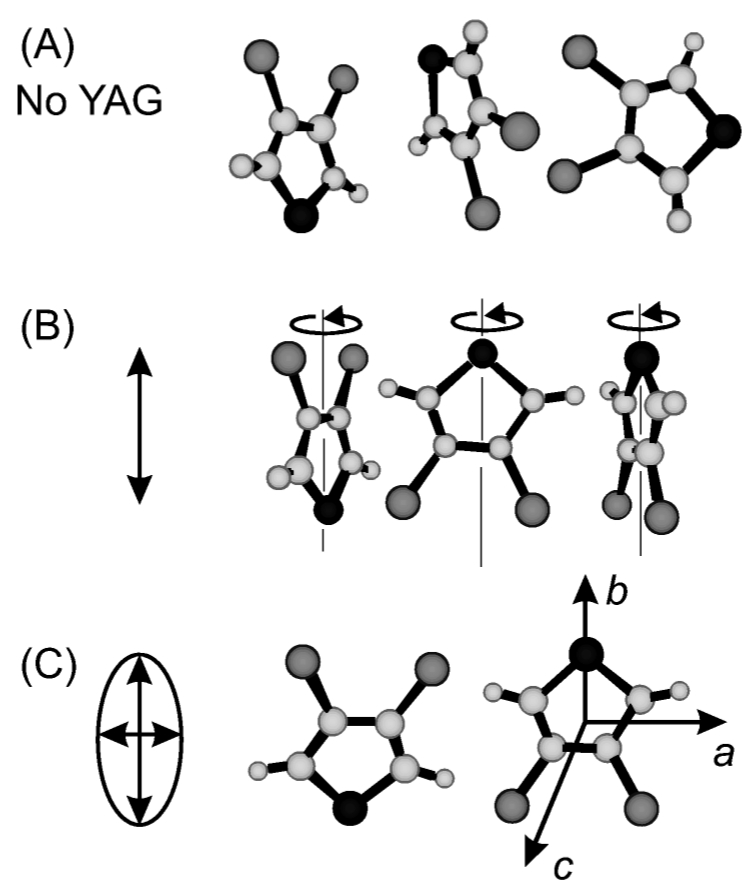}
  \caption{Illustration of non-aligned (A), 1D-aligned (B), and 3D-aligned (C) 3,4-dibromothiophene (black circles: sulfur, dark grey circles: bromine). 
    For non-aligned molecules, all orientations are equally likely. 
    1D alignment is achieved with a linearly polarized pulse, and 3D alignment is accomplished with an elliptically polarized pulse.
    The illustration is taken from Ref.~\cite{LaHa-PRL-2000}.
    Copyright~\copyright~2000 American Physical Society (APS).
  }
  \label{fig1:3dalign_potential}
\end{figure}

\subsubsection{3D Alignment}
3D alignment requires that all three molecular axes are fixed in space~\cite{MaRe-PRA-2012} (see Fig.~\ref{fig1:3dalign_potential}).
The potential $\hat U$ in Eq.~\eqref{eq1:ham} can only lead to 3D alignment, when $\hat U$ depends on all three Euler angels $\varphi,\vartheta,\chi$~\cite{ArSe-JCP-2008}.
The $\vartheta$ dependence enters, since $F^{[L]}$ and $\alpha^{[L]}$ have non-zero components for $L>0$, which influence the angular momentum $J$ of the rotational states (for more details see Sec.~\ref{p1c1:intro_dynamics}).
The $\varphi$ and $\chi$ dependencies enter only when $F^{[2]}_{M=\pm 2}\neq 0$ and $\alpha^{[2]}_{K=\pm2} \neq 0$ are fulfilled, respectively.
On the one side, $\alpha^{[2]}_{K=\pm2} \neq 0$ is only possible for asymmetric-top molecules ($A<B<C$).

\begin{figure}[h]
  \centering
  \includegraphics[width=.7\linewidth]{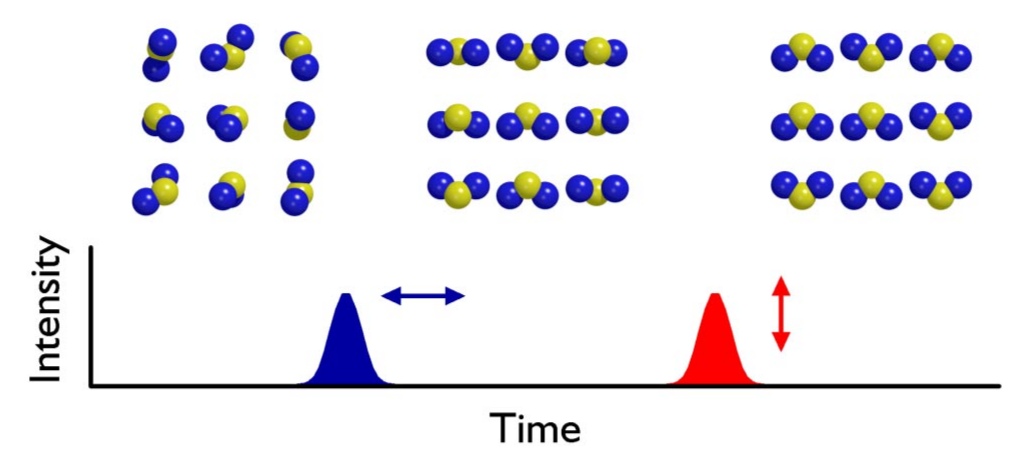}
  \caption{
    A schematic illustration of 3D alignment of SO$_2$ with two different linearly polarized pulses is shown.
    The first pulse aligns the randomly oriented molecules along the most polarizable axis.
    The second pulse aligns the second most polarizable axis leading to 3D alignment.
    This figure is taken from Ref.~\cite{LeVi-PRL-2006}.
    Copyright~\copyright~2006 American Physical Society (APS).
  }
  \label{fig1:align_3d}
\end{figure}

For symmetric-top and linear molecules, $\alpha_{aa}=\nolinebreak \alpha_{cc}$ and, therefore, $\alpha^{[2]}_{K=\pm2} = 0$.
On the other side, $F^{[2]}_{M=\pm 2}\neq 0$ can be only achieved when elliptically polarized pulses or multiple pulses with different polarization directions are used.
A schematic sketch of 3D alignment of two pulses with different linear polarizations is shown in Fig.~\ref{fig1:align_3d}.

If one of the two requirements for generating a $\varphi$- and $\chi$-dependent potential is not fulfilled, only 1D alignment is possible.
For instance, linearly polarized pulses (no $\varphi$-dependence) align asymmetric-top molecules ($\chi$-dependence) only around the most polarizable axis~\cite{PePo-PRL-2003,PePo-PRA-2004,RoGu-PRA-2006} as for linear molecules.
Panel (B) in Fig.~\ref{fig1:3dalign_potential} demonstrates this case.
Elliptically polarized pulses ($\varphi$-dependence) used in combination with symmetric-top molecules (no $\chi$-dependence) align only the symmetry axis $c$.
However, this symmetry axis can be rotated in space, synchronized with the oscillating elliptically polarized field~\cite{CrBu-PRA-2009}.

\subsection{Alignment Dynamics}
\label{p1c1:intro_dynamics}
Dynamical states in quantum mechanics are coherent superpositions of eigenstates of the system (i.e., Hamiltonian).
The discussion starts with .
Many effects can be understood with symmetric-top rotors ($A=B$), while differences and similarities to asymmetric-top rotors are pointed out as well.
The rotational kinetic operator of a symmetric-top rotor is $\hat H_\text{rot}=A\,\hat J^2 + (C-A)\,\hat J^2_c$.
Its rotational eigenstates are given by~\cite{Zare-book}
\begin{subequations}
\label{eq1:symmrotor}
\begin{align}
  \label{eq1:symmrotor_eigenstates}
  \braket{\Omega}{JKM}
  &=
  \sqrt{\frac{2J+1}{8\pi^2}}\,\rotmats{J}{M,K}(\Omega)
  ,
\\
  \label{eq1:symmrotor_spec}
  \hat H_\text{rot}
  \ket{JKM}
  &=  
  \underbrace{
  \left[\Big.
    A\,J(J+1) - (C-A) K^2
  \right]
  }_{\displaystyle E_{J,K}}
  \ket{JKM}
  ,
\end{align}
\end{subequations}
where $\rotmats{J}{M,K}$ are the well-known Wigner-D matrices~\cite{Zare-book}, and $E_{J,K}$ are the rotational energies.
$K$ and $M$ are again the angular momentum projections onto the body-fixed $c$ axis and the space-fixed $Z$ axis, respectively.
Since the rotation is defined in the molecular frame, the rotational energies are degenerate with respect to $M$.
In the case of linear molecules ($C^{-1}=0$), the rotation around the $c$ axis is energetically not possible ($K=0$) and the rotational energies simplify to $A\,J(J+1)$.
For general asymmetric-top rotors, the kinetic operator 
\begin{subequations}
\label{eq1:asymmrotor}
\begin{align}
  \label{eq1:asymmrotor_operator}
  \hat H_\text{rot}
  &=
  \frac{A+B}{2}\,\hat J^2 
  + 
  \frac{2C-A-B}{2}\,\hat J^2_0
  +
  \frac{A-B}{2}\,
  \left[
    \hat J^2_{+1}
    +
    \hat J^2_{-1}
  \right]
  ,
\\
  \label{eq1:asymmrotor_spec}
  \hat H_\text{rot}
  &
  \ket{J\tau M}
  =  
  E_{J,\tau}
  \ket{J\tau M}
  ,
\end{align}
\end{subequations}
mixes different $K$ states such that $K$ is no longer a good quantum number.
The rotational eigenstates $\ket{J\tau M}$ are superpositions of $\ket{JKM}$~\cite{ArSe-JCP-2008}.
It is not possible to give an analytic expression for the rotational energies $E_{J,\tau}$ in contrast to the rotational energies of symmetric-top rotors, $E_{J,K}$.
The angular momentum operators are rewritten as spherical tensor operators $\hat J_{0,\pm 1}$ in the body-fixed frame. 

Similar to the relation between $\hat x$ and $\hat p_x$, where the localization of the wavefunction in real space requires the wavefunction to be delocalized (coherent superposition) in momentum space, a localization of the rotational wavefunction at specific Euler angles requires a coherent superposition of rotational states, $\ket{JKM}$.
Coherent superpositions of different $J,K$, and $M$ states are needed to localize the rotational wavefunction in $\vartheta,\chi$, and $\varphi$, respectively.
Rotational wavepackets of a symmetric-top rotor can be written as
\begin{align}
  \label{eq1:wavepacket_rot}
  \ket{\Psi(t)}
  =
  \sum_{J,K,M} c_{J,K,M}(t) e^{-i\,E_{J,K}\,t} \ket{JKM}
  .
\end{align}
The time evolution of the rotational wavepacket is determined by the time-dependent Schr\"odinger equation,
\begin{align}
  \label{eq1:tdse}
  i \frac{\partial}{\partial t} \ket{\Psi(t)}
  &=
  \hat H(t) \,\ket{\Psi(t)}
  ,
\end{align}
where $\hat H(t)$ is defined in Eq.~\eqref{eq1:ham}.

\subsubsection{Alignment Revivals}
The coefficients $c_{J,K,M}(t)$ do only change in time when an aligning potential is present. 
If no external field is present, $c_{J,K,M}(t)$ are time-independent constants.
For linear molecules, all rotational energies $E_{J,K=0}=A\,J(J+1)$ are multiples of $A$.
Hence, the rotational wavepacket shows a repeating pattern in time called {\it revivals}.
The revival period is $T_\text{rev}=\pi/A$~\cite{Se-PRL-1999}.
Often the rotational period is given by $T_\text{rev}=\frac{1}{2A}$~\cite{HoVi-PRA-2007}, where $A$ is given in Hz instead of energy units.

\begin{figure}[h]
  \centering
  \includegraphics[width=.5\linewidth]{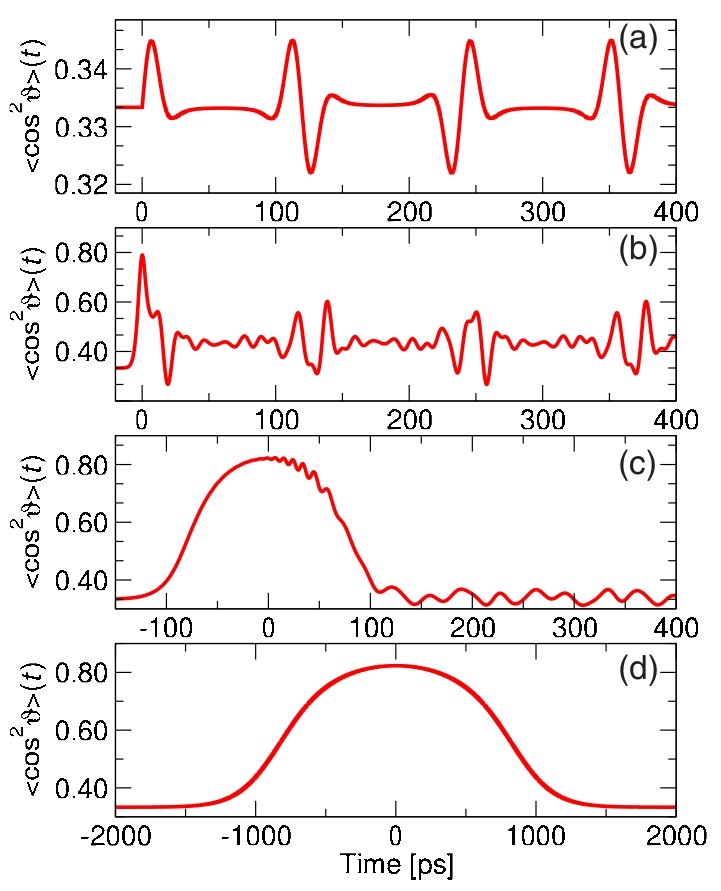}
  \caption{
    The degree of alignment $\ev{\cos^2\theta}(t)$ is shown for the symmetric-top molecule CF$_3$Br.
    The different regimes of alignment dynamics are shown ranging from purely impulsive alignment (a) via mostly non-adiabatic (b) and quasi-adiabatic alignment (c) to perfectly adiabatic alignment (d).
    The pulse duration changes from 50~fs (a) over 10~ps (b) and 95~ps (c) to 1~ns (d).
    The revival period $1/(2A)$ of CF$_3$Br seen in (a) is close to 240~ps.
    This figure is taken from Ref.~\cite{BuSa-JCP-2008}.
    Copyright~\copyright~2008 American Institute of Physics (AIP).
  }
  \label{fig1:align_dynamics}
\end{figure}

Additionally, there are features at $T_\text{rev}/2$ (half revivals) as shown in Fig.~\ref{fig1:align_dynamics}.
At $T_\text{rev}/2$ neighboring rotational states swing out of phase leading to an antialignment of molecules.
Quarter revivals at $T_\text{rev}/4$ do only occur for molecules with the appropriate spin statistics~\cite{Bunker-book,Kroto-book}.
For example, $^{16}$O$_2$ has two nuclear spins of $I=0$. 
Due to the overall symmetry of the molecular wavefunction (vibronic + rotational + nuclear spin), only rotational states with an odd $J$ are allowed and, therefore, quarter revivals can be seen~\cite{LeVi-PRL-2004}.
In the molecule $^{79}$Br-$^{81}$Br the two atoms are distinguishable and all rotational states have the same spin weight~\cite{HoMi-JCP-2009}.
Hence, no quarter revivals are visible.

The revival features do also exist for symmetric-top molecules. 
The additional degree of freedom ($K\neq 0$ is possible) does not change the nature of these features.
In asymmetric-top molecules no true revival structure can be found, since the rotational energies are non-commensurable~\cite{Zare-book}. 
However, revival-type behavior has been observed~\cite{HoVi-PRA-2007}, which lasts only for a finite amount of time till the rotational states are totally dephased.
The most common revival-like behaviors are $C$- and $J$-type revivals, where $J$-type revivals have a period of $1/(B+C)$ and occur for nearly symmetric-top molecules ($B\approx C$)~\cite{PoPe-JCP-2004,HoVi-PRA-2007}.
Revivals of the type $J$ are common in planar molecules and have a period $1/(4C)$, where $C$ is the rotational constant of the molecular $c$ axis, which points perpendicular to the molecular plane~\cite{Fe-JPC-1992}.

The repeating revival behavior of symmetric-top molecules allows one to generate the alignment first and to use/probe the alignment at a later time~\cite{Se-PRL-1999,RoVr-JCP-2002}. 
Revivals can be also used to increase the degree of alignment step-by-step by using several pulses separated by the revival period~\cite{LeAv-PRL-2003,CrBu-PRA-2009}.
Applying pulses at half or quarter revivals allows one to control the subsequent revival structure~\cite{LeVi-PRL-2004}.

\subsubsection{Adiabatic and Non-Adiabatic Alignment}
To be able to generate and to use these revivals, it is necessary that the laser pulse kicking the molecule is short compared to $T_\text{rev}$.
This type of alignment scheme is known as {\it impulsive} or {\it non-adiabatic} alignment~\cite{StSe-RMP-2003,KeDi-PRA-2000}.
Since typical rotational periods are in the picosecond to nanosecond regime, impulsively aligning pulses are typically in the femtosecond regime.
Pulses that are much longer than the rotational period lead to {\it adiabatic} alignment~\cite{Se-JCP-1995}.

In the adiabatic regime, the alignment follows the envelope of the laser pulse and returns after the pulse to its initial state.
Impulsive alignment gives the molecules a kick such that the molecular alignment dynamics persist after the pulse is over. 
Furthermore, impulsive alignment has been a very attractive technique to achieve alignment, since the molecule can be probed under field-free condition~\cite{PePo-PRL-2003}.
Figure~\ref{fig1:align_dynamics} shows the different response behaviors of the symmetric-top molecule CF$_3$Br, which ranges from impulsive alignment in panel (a) to adiabatic alignment in panel (d).

Besides the laser pulse, the degree of alignment also depends on the rotational temperature of the molecules.
As one may expect, with higher temperature the degree of alignment goes down~\cite{BuSa-JCP-2008}.
This is true in the impulsive as well as in the adiabatic case.
For both cases it is possible to derive an analytic expression for the maximum possible degree of alignment depending on the pulse parameters and the rotational temperature of the molecules~\cite{Se-JCP-2001}.
The rotational temperature of the molecules also has an impact on the characteristic rotational period, which becomes shorter for increasing temperatures.
Hence, higher temperatures lead to more adiabatic-like alignment behavior~\cite{BuSa-JCP-2008,KuBi-JCP-2006}.

Recently, quantum-state-selection techniques have been used to select only specific rotational states out of the thermally populated sea of rotational states~\cite{HoNi-PRL-2009,GeAv-JCP-2011} in order to improve the alignment by minimizing the negative impact of temperature.
An almost perfect 1D alignment of $\ev{\cos^2\vartheta_{2D}}=0.97$ was achieved with this scheme~\cite{HoNi-PRL-2009}.

Very high degrees of alignment can be also reached with a combination of adiabatic and impulsive alignment 
~\cite{GuRo-PRA77-2008}.
A very popular technique is to use a long pulse, which is turned on adiabatically and turned off abruptly such that the molecule can be probed field-free~\cite{Se-JCP-2001,UnSp-PRL-2003,PoEj-PRA-2006}.
Around the times of the turn-off, a second short pulse is used to ``kick'' the molecules into an even higher degree of alignment.
This technique has become particularly popular to align asymmetric-top molecules in 3D under field-free conditions~\cite{ViKu-PRL-2007}.

Non-adiabatic field-free 3D alignment via multiple femtosecond pulses is also possible~\cite{UnSu-PRL-2005,LeVi-PRL-2006}.
However, with such a pulse configuration it is much harder to achieve 3D alignment than 1D alignment~\cite{PaSa-PRA-2010} as it will be discussed in more detail in Sec.~\ref{p1c1:apps_ff3da}.
If it is not so important to get 3D alignment under field-free condition, adiabatic alignment schemes with an elliptically polarized pulse are quite powerful~\cite{LaHa-PRL-2000}.

\subsection{Molecular Orientation}
\label{p1c1:intro_orient}
\begin{figure}[h]
  \centering
  \includegraphics[width=.4\linewidth]{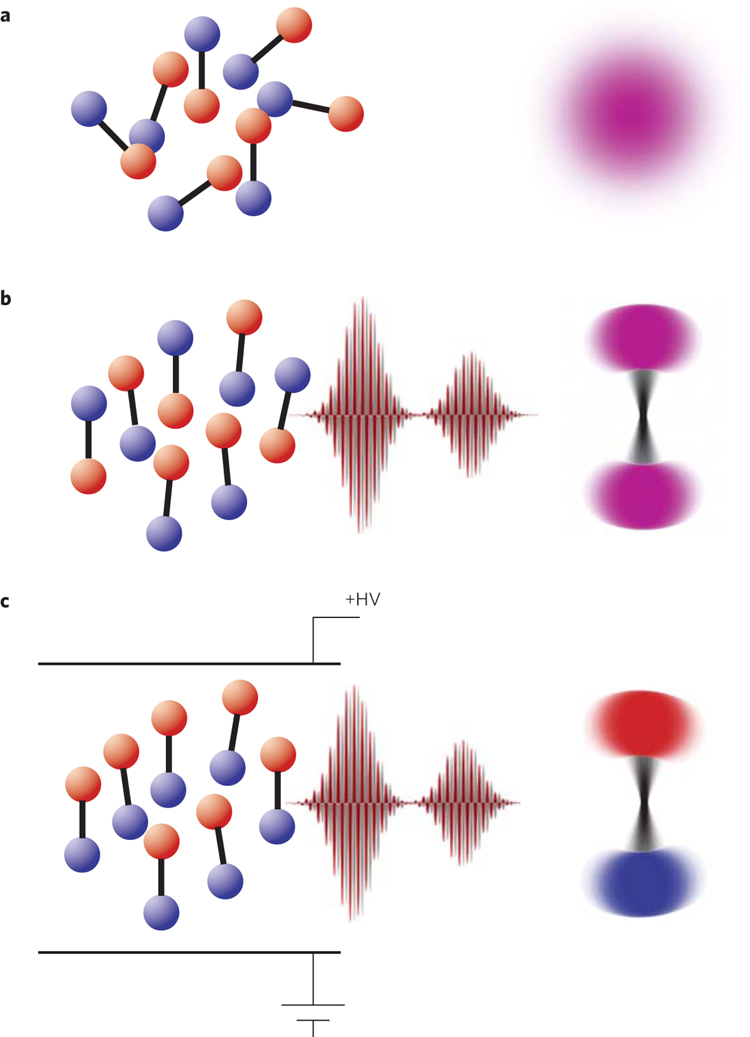}
  \caption{
    A schematic illustration of 1D orientation of a linear molecule by means of a combination of a weak static electric field and optical laser pulses.
    The coupling of the dc field to the permanent dipole moment of the diatomic molecule orients the molecule.
    Without any field the molecules are randomly oriented (a).
    Applying optical laser pulses aligns the molecules (b).
    The combination of static and optical fields (c) can be used to achieve high degrees of orientation.
    This figure is taken from Ref.~\cite{Un-NatPhys-2009}. 
    Copyright~\copyright~2009 Nature Publishing Group (NPG).
  }
  \label{fig1:orient}
\end{figure}

For molecular orientation, it is necessary that the energy corrections $E^{(1)}(\omega)$ and $E^{(3)}(\omega)$ contribute to the potential $\hat U(\omega)$ [see Eq.~\eqref{eq1:potential_rotational}].
As discussed in Sec.~\ref{p1c1:intro_align}, an optical monochromatic pulse cannot induce an orienting potential (i.e., $E^{(2n+1)}(\omega)=0$).
The reason is in the optical period, which is much shorter than the typical rotational time scale.
For static electric fields ${\cal E}_\text{static}$, where the period goes to infinity, this is not true anymore, and the coupling to the permanent dipole moment $\mu$ leads to an orienting potential 
\begin{align}
   \label{eq1:potential_orient}
   U(\vartheta,t)
   &=
   {\cal E}_\text{static}\, \mu\cos\vartheta
   .
\end{align}
Common static field strengths lie in the range of $10^2$--$10^5$~V/cm ($\approx 10^{-8}$--$10^{-5}$~a.u.)~\cite{TaMi-PRA-2005,JoHe-CPL-1997}.
Such weak fields only affect the rotational states when coupled to $\mu$ leading to the energy correction $E^{(1)}(\omega)$~\cite{FrHe-JPCA-1999}.
Since optical laser pulses have been proven to be very successful in aligning
molecules, several groups have worked on orienting molecules by combining static electric fields with optical laser pulses~\cite{BaHe-JCP-2001,TaMi-PRA-2005,SuGo-PRA-2008}.
A schematic illustration of this technique is shown Fig.~\ref{fig1:orient}.  

For non-polar molecules ($\mu=0$), multi-color pulses have to be used in order to orient molecules~\cite{KaSa-JCP-2001,TeSu-PRA-2008,MuHi-PRA-2009}.
In this case, the orientation is accomplished through the coupling of the electric field to the hyperpolarizability tensor $\beta$ (i.e., $E^{(3)}$ energy correction).
Particularly for short pulses with high electric field strengths, it was shown that it is possible to achieve orientation via the hyperpolarizability~\cite{TeSu-PRA-2008,DeZn-PRL-2009,YuKi-PRA-2011}.
In recent experiments, it has been seen that orientation substantially enhances when ionizing intensities are reached~\cite{SpPa-PRL-2012}.
Similar to the case of alignment, earlier works have focused first on accomplishing field-free 1D orientation~\cite{BaHe-JCP-2001} and later 3D orientation~\cite{NeHo-PCCP-2009}.

The quality of orientation depends highly on the rotational temperature~\cite{MaHe-PRL-2001,YuKi-PRA-2011,LaSu-PRA-2012}.
Many systematic studies have been performed to find the optimal setup for maximizing the orientation in a given molecule~\cite{LaSu-PRA-2012,DiBe-PRA-2002,TeSu-PRA-2008}.
Very high degrees of orientation can be reached when quantum-state-selection techniques are combined with orienting static fields and impulsively aligning femtosecond pulses.
This has been demonstrated with NO molecules, where an observed degree of orientation of $\ev{\cos\vartheta}=-0.74$ has been seen~\cite{GhRo-NatPhys-2009}.

Recently, progress has been made in using terahertz (THz) pulses to orient molecules.
The cycle periods is now in the picosecond regime and becomes comparable to the characteristic rotational time scale~\cite{ShYu-JCP-2010,FlWo-PRL-2011,KiIs-PRA-2011,QiTa-PRA-2012,LaSu-PRA-2012}.
As a consequence, the requirement for averaging the pulse over a period is not fulfilled anymore.
The instantaneous electric field can, now, successfully couple to the permanent dipole moment.
In contrast to static fields, the orienting potential becomes time-dependent.
Interesting new rotational motions may appear by using THz pulses together with optical pulses, which lead to high degrees of orientation.

\subsection{Applications}
\label{p1c1:apps}

In this section, several applications of molecular alignment will be discussed.
First, field-free 3D alignment studies with multiple pulses are presented in Sec.~\ref{p1c1:apps_ff3da}. 
In Sec.~\ref{p1c1:apps_xscatter}, the possibility to use 3D aligned molecules to
obtain x-ray diffraction patterns is investigated.
These diffraction patterns are used to reconstruct the molecular structure.
In Sec.~\ref{p1c1:apps_ionization} molecular alignment experiments are discussed that uses aligned molecules to study angle-dependent properties of the electronic system of the molecule.

\subsubsection{Using Multiple Pulses to 3D-Align SO$_2$ under Field-Free
Conditions}
\label{p1c1:apps_ff3da}
As discussed in Sec.~\ref{p1c1:intro_align}, the energy spacings between the
rotational eigenstates of an asymmetric-top rotor are non-commensurable, and,
therefore, no true revival pattern can be observed in contrast to linear and
symmetric-top molecules.
However, for a short time asymmetric-top molecules show revival features called
$A$-, $C$-, and $J$-revivals~\cite{RoGu-PRA-2006}.

\begin{figure}[h]
   \centering
  \includegraphics[width=.55\linewidth]{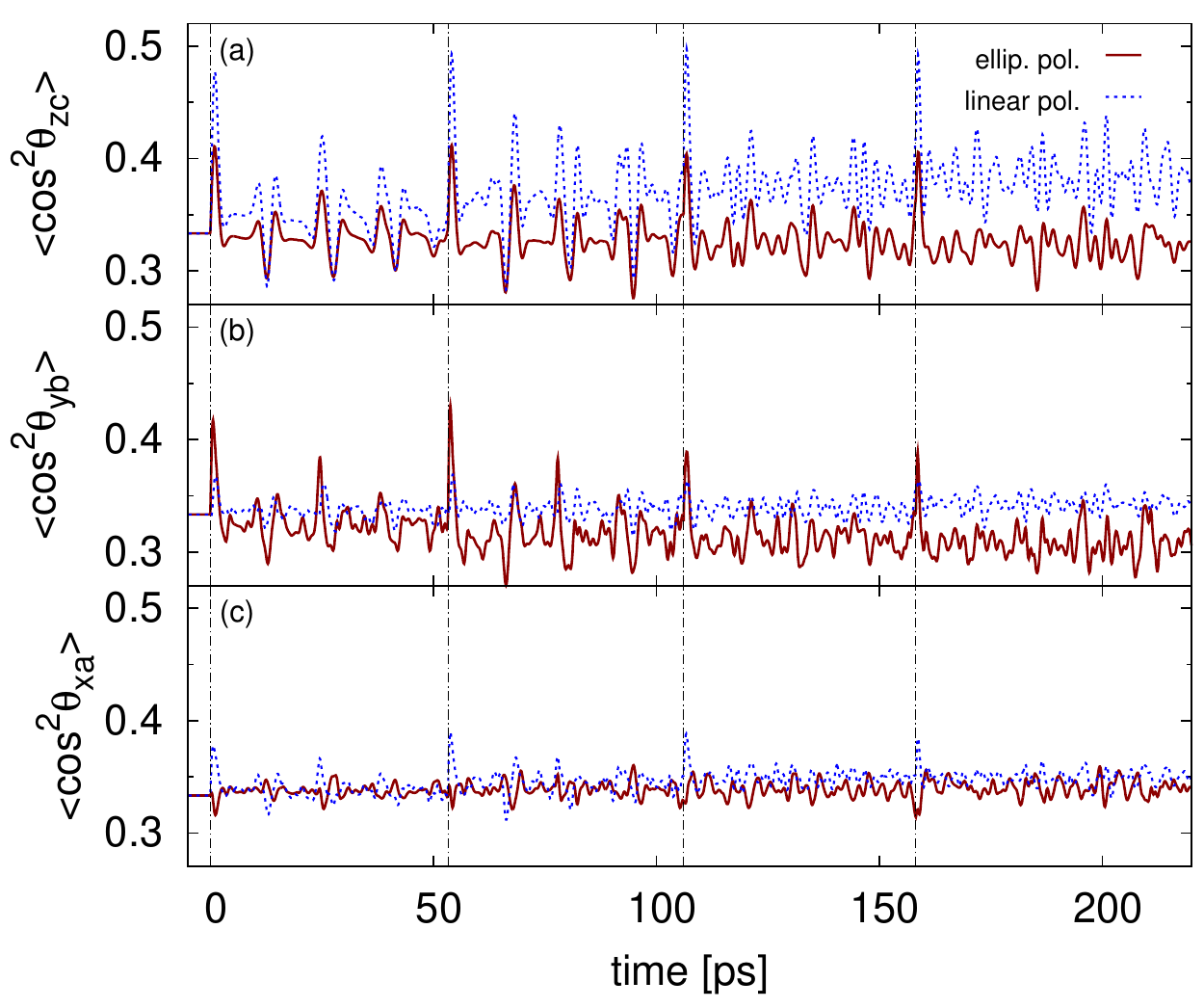}
  \caption{
    The alignment dynamics of all three molecular axes of SO$_2$ are shown.
    The short pulses (indicated by dashed lines) kick the molecules at the $J$
revivals with linearly and with elliptically polarized pulses, respectively. 
    This figure is taken from Ref.~\cite{PaSa-PRA-2010}.
    Copyright~\copyright~2010 American Physical Society (APS).
  }
  \label{fig1:so2_revival}
\end{figure}

Here, the question arises whether or not it is possible to improve field-free 3D alignment of asymmetric molecules with multi-pulse schemes similar to those used for improving 1D alignment of linear molecules~\cite{CrBu-PRA-2009}.
Field-free alignment can be only achieved with short femtosecond pulses such that the alignment dynamics is non-adiabatic.
Specifically, the asymmetric-top molecule SO$_2$ has been studied, which is close to a symmetric-top molecule $A \approx B$.
First, it was investigated to which extent pulses equally spaced by $1/(A+B)$ ($J$ revivals) improve the alignment.

In Fig.~\ref{fig1:so2_revival} the alignment dynamics of each molecular axis of
SO$_2$ is shown for linearly and elliptically polarized pulses.
As expected for linearly polarized pulses, only the most polarizable $c$ axis
gets aligned. 
With elliptically polarized pulses also the second most polarizable $b$ axis can
be aligned.
However, the third molecular axis is counter-intuitively not aligned. 
This lack of alignment is classically not possible, since the alignment of two
axes automatically aligns the third one as well.
Here it is possible, since the degree of alignment is rather weak and a
connection to a classical picture cannot be made.

\begin{figure}[h!]
  \centering
  \includegraphics[width=.55\linewidth]{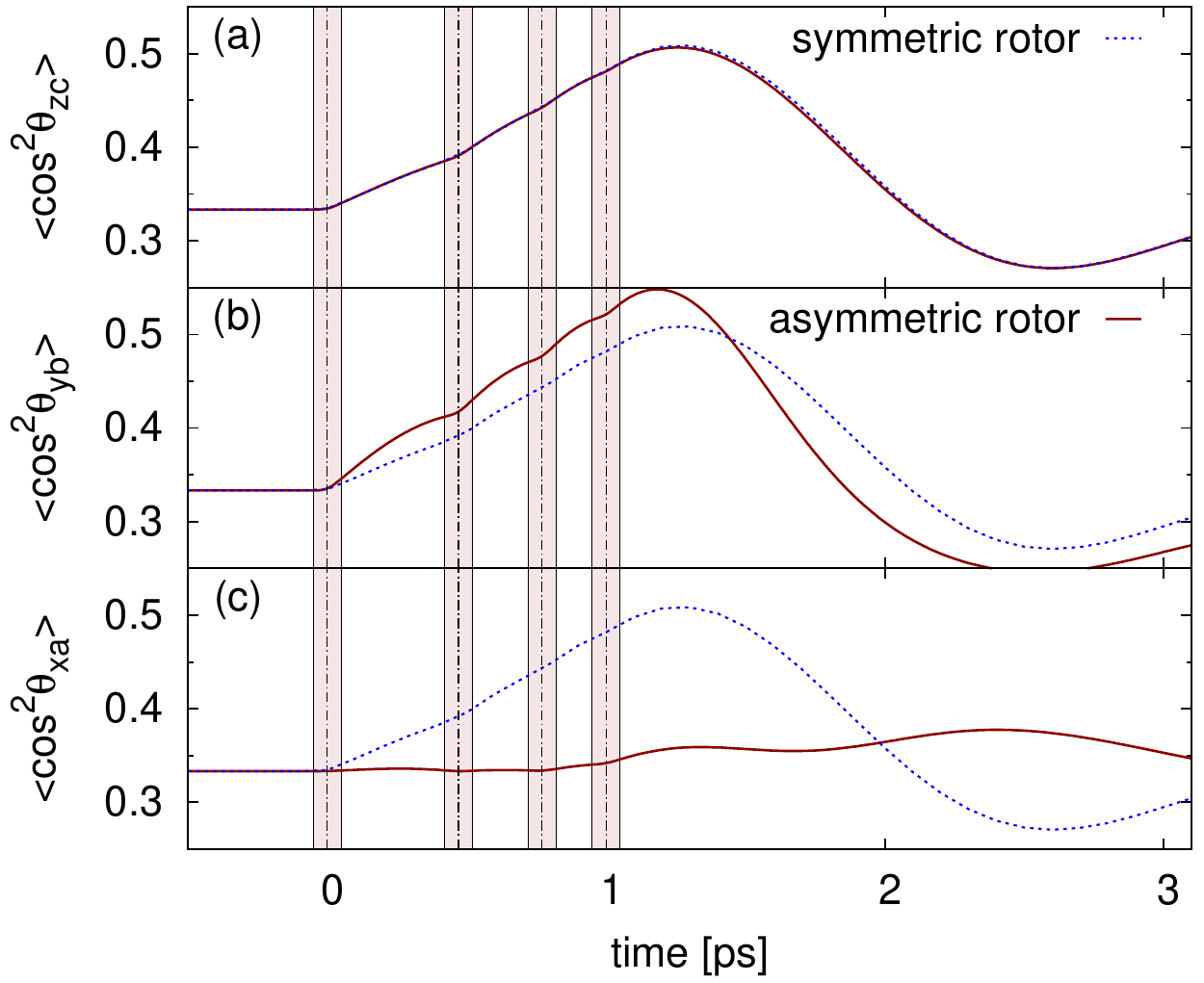}
  \caption{
    The alignment dynamics of all three molecular axes of SO$_2$ are shown.
    The short pulses kicking the molecules are highlighted (red shaded region).
    This figure is taken from Ref.~\cite{PaSa-PRA-2010}.
    Copyright~\copyright~2010 American Physical Society (APS).
  }
  \label{fig1:so2_quick}
\end{figure}

To improve on this technique, new method was proposed~\cite{PaSa-PRA-2010} of aligning asymmetric-top molecules with elliptically polarized pulses, which arrive directly after each other as discussed in Ref.~\cite{AvAr-PRL-2001} for linear molecules with linearly polarized light. 
The results are shown in Fig.~\ref{fig1:so2_quick}.
The degree of alignment is significantly enhanced for the two most polarizable axes, whereas the least polarizable axis is almost not affected.
Hence, SO$_2$ cannot be aligned in 3D with a sequence of short pulses.

Very recent studies on 3,5~difluoroiodobenzene improved on this idea by using a
combination of linearly and elliptically polarized pulses.
First results indicate it is possible to improve on 3D alignment; particularly
when the molecule is oblate\footnote{private discussion at the GRC Multiphoton
conference 2012}.

\subsubsection{X-ray Scattering from 3D-Aligned Naphthalene Molecules}
\label{p1c1:apps_xscatter}
An interesting application of aligned molecules is
electron~\cite{BlXu-Nature-2012,HeYa-PRL-2012} or x-ray~\cite{HoSa-PRA78-2008}
diffraction.
Particularly for single-molecule
imaging~\cite{NeWo-Nature-2000,HuSz-JStrucBio-2003} it is favorable to know and
to control the orientation of the molecule such that an entire 3D diffraction
pattern can be collected.
With the help of phase-retrieval algorithms it is possible to recover the phase
information of the diffraction pattern, which was lost during the
measurement~\cite{Ba-Opt-1982,MiSa-JOSAA-1998,El-JOSAA-2003,ElMi-ActaCrystA-2008}.
New approaches (namely multiwavelength anomalous diffraction) have also been
developed to determine the phase directly from the
measurement~\cite{SoCh-PRL-2011}.

A large number of molecules contributing to the scattering signal is
advantageous, since the scattering cross section is small---especially for large
scattering angles~\cite{Sa-JPB-2009}.
When molecules are periodically ordered as in a crystal, Bragg
peaks~\cite{NiMc-ElementXrayPhysics-book} occur and their scattering signal
increases quadratically with the number of molecules.
This quadratic increase is very favorable and it is at the heart of x-ray
crystallography~\cite{NiMc-ElementXrayPhysics-book}.
Periodic ordering is commonly exploited in order to determine the structure of
macromolecules~\cite{NiMc-ElementXrayPhysics-book}.
Unfortunately, not all molecules (like membrane proteins) can be grown in a
large crystal.

Recent advances have shown it is also possible to recover the molecular
structure from nanocrystals~\cite{ChFr-Nature-2011} (for a review see
Ref.~\cite{Ki-JPB-2012}).
The high x-ray intensities are needed to collect enough signal destroy the molecules
during the x-ray pulse~\cite{NeWo-Nature-2000}.
This obstacle of molecular explosion can be overcome by making the x-ray pulse
shorter than the time it takes for the molecule to
explode~\cite{NeWo-Nature-2000,GaCh-Science-2007,BoLo-Science-2012}.

Another approach for collecting x-ray diffraction patterns uses gas phase
molecules, where up to 10$^{7}$ molecules can contribute to the scattering
signal~\cite{PeBu-AppPhysLett-2008}.
These molecules are randomly positioned (not randomly aligned) in space and,
therefore, the diffraction pattern increases only linearly with the number of
molecules.
However, collecting signal over many x-ray shots reduces the required intensity
such that common 3rd generation synchrotron sources like PETRA~III at
DESY\footnote{
\url{http://hasylab.desy.de/facilities/petra_iii/index_eng.html}
} 
or the Advanced Photon Source at Argonne National Laboratory\footnote{
\url{http://www.aps.anl.gov}
} 
can be used. 

\begin{figure}[h]
  \centering
  \includegraphics[width=.4\linewidth]{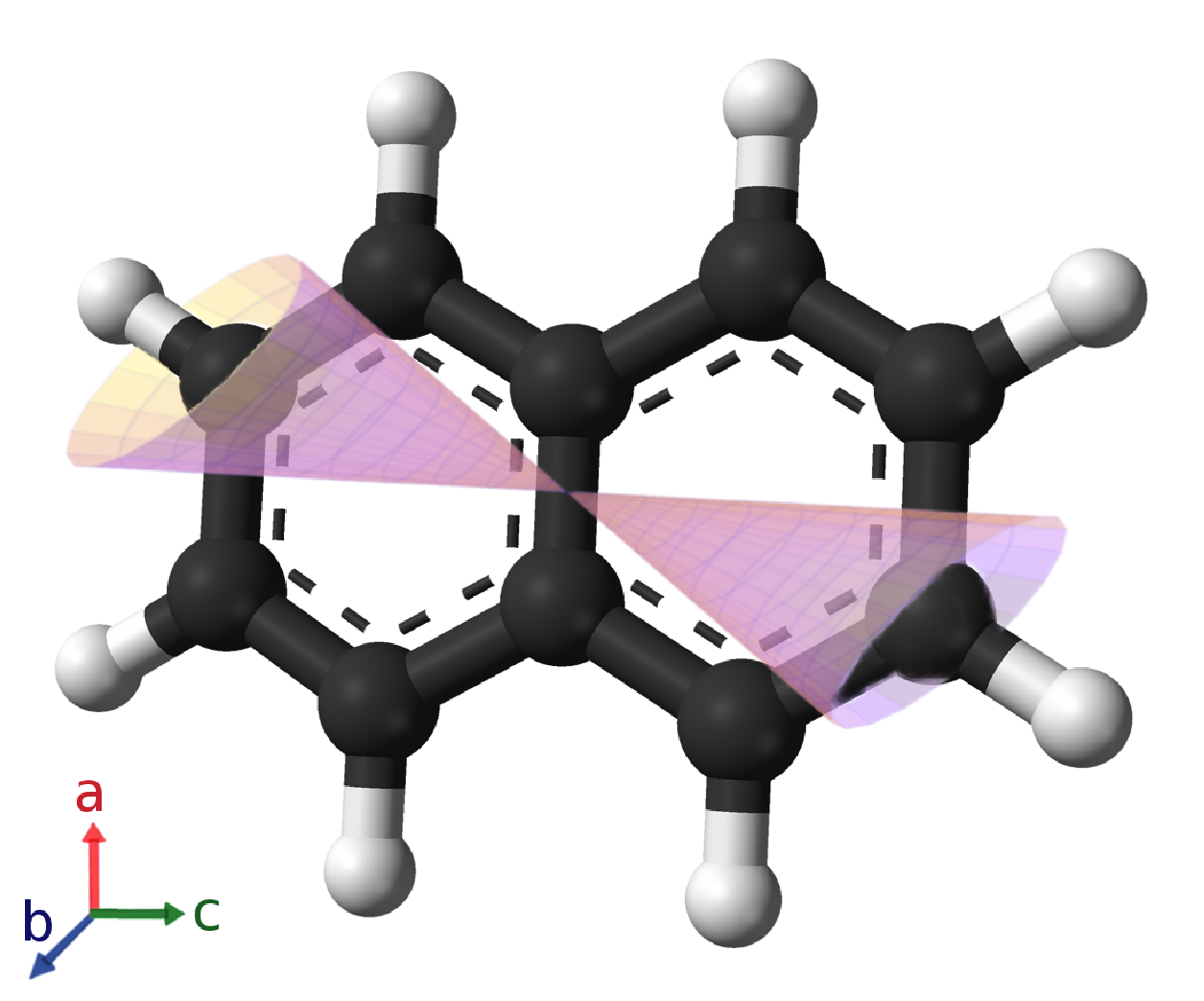}
  \caption{
  The structure of naphthalene (C$_{10}$H$_8$). 
  All atoms lie in the $ac$ molecular plane. 
  The coordinate system is shown in the lower left corner. 
  The cones symbolize the resolution limit of the carbon atoms due to imperfect
alignment.
    This figure is taken from Ref.~\cite{PaHo-PRA-2010}.
    Copyright~\copyright~2010 American Physical Society (APS).
  }
  \label{fig1:naphthalene}
\end{figure}
All molecules have to be oriented and/or aligned simultaneously to be able to
record a single-molecule scattering pattern. 
Here is where laser alignment comes into play.
Previous works have looked at x-ray scattering from 1D aligned symmetric
molecules~\cite{HoSa-PRA78-2008,HoSt-JCP-2009}.
Recently, an extension of this scheme to x-ray scattering from 3D aligned asymmetric-top molecules as been presented~\cite{PaHo-PRA-2010}.
The molecule of choice was, here, Naphthalene (see Fig.~\ref{fig1:naphthalene}), which is a planar molecule with a head-tail symmetry such that orientation and alignment become indistinguishable.
This is convenient, since different orientations add up incoherently and make it harder to reconstruct the molecular structure with phase-retrieval methods, which assume perfectly coherent diffraction patterns.

\begin{figure}[b]
  \centering
  \begin{subfigure}[b]{0.46\textwidth}
    \centering
    \includegraphics[width=\linewidth]{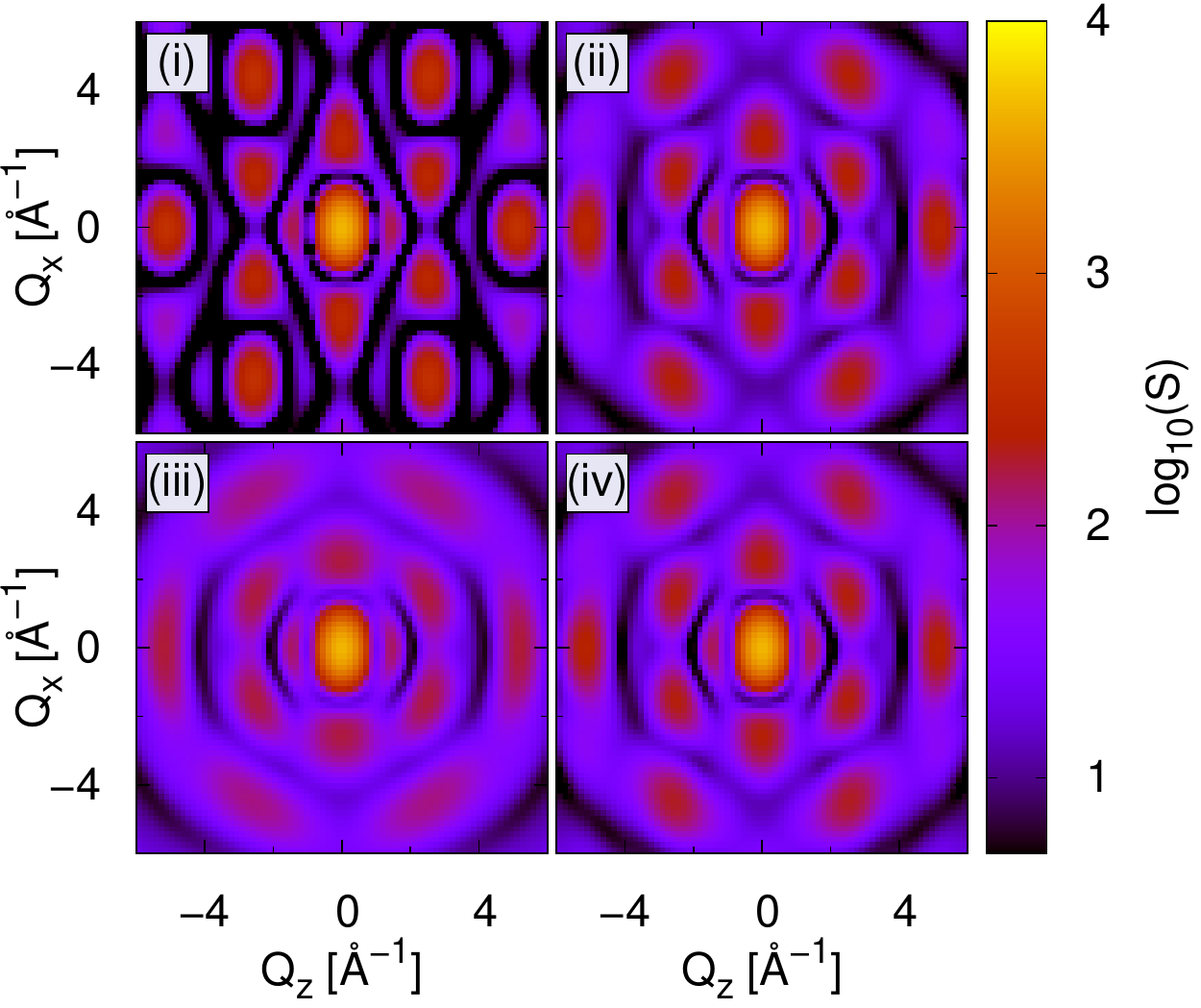}
    \caption{X-ray diffraction patterns}
    \label{fig1:xscatter_pattern}
  \end{subfigure}
  \hfill
  \begin{subfigure}[b]{0.49\textwidth}
    \centering
    \includegraphics[width=\linewidth]{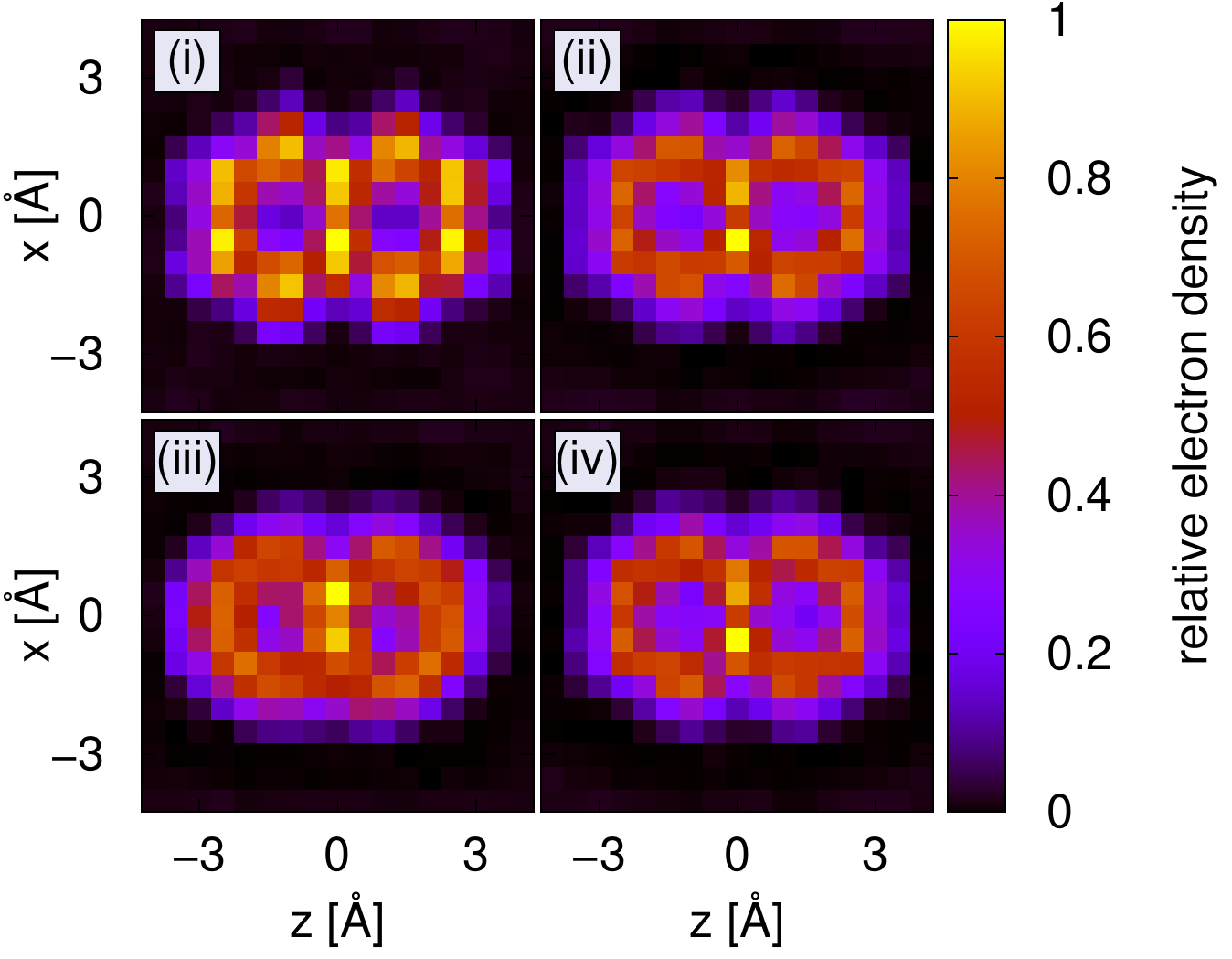}    
    \caption{Reconstructions}
    \label{fig1:xscatter_reconstruct}
  \end{subfigure}  
  \caption{
    X-ray diffraction patterns (a) and their the corresponding reconstructed
structure (b) are shown for different degrees of alignment:
    (i) perfect alignment, (ii) $T=10$~mK and a 1~ps x-ray pulse,
    (iii) $T=1$~K and a 100~ps x-ray pulse, and (iv) $T=10$~mK and a 100~ps
x-ray pulse.
    The aligning IR pulse is 100~ps long and aligns the molecule adiabatically.
    These figures are taken from Ref.~\cite{PaHo-PRA-2010}.
    Copyright~\copyright~2010 American Physical Society (APS).
  }
  \label{fig1:xscatter}
\end{figure}

The positions of the atoms in the laser frame are not well defined due to the
imperfect alignment of the molecules (illustrated by the cones in
Fig.~\ref{fig1:naphthalene}).
This uncertainty results in an incoherent average of different scattering
patterns, which blurs out the scattering pattern for large scattering angles
(i.e., large momentum transfer $Q$ of the x-ray photon) as seen
Fig.~\ref{fig1:xscatter_pattern}.
This has a direct consequence on the effective resolution in the corresponding
reconstructions shown in Fig.~\ref{fig1:xscatter_reconstruct}.
The worse the alignment the smaller is the useful part of the diffraction
patterns which can be fed into the phase-retrieval algorithm.
Hence, the effective resolution decreases with less 3D-aligned molecules.

The main contribution to the imperfect alignment is the rotational temperature
$T$ [cf.~panels (iii) and (iv) in Fig.~\ref{fig1:xscatter}].
The length of the x-ray pulse does not have a large influence as long as it is
shorter or comparable with the adiabatic alignment pulse [cf.~panels~(ii) and
(iv) in Fig.~\ref{fig1:xscatter}].
The effective resolution is only better than the typical distance between the
atoms ($\sim 1.4$~\AA) when the rotational temperature is less than 10~mK.
Quantum state selection techniques~\cite{HoNi-PRL-2009} could be used to improve
the alignment but they result in less scattering signal.

Another approach to improve on the reconstruction uses partial-coherence schemes
in the phase-retrieval algorithm~\cite{QuNu-NatPhys-2011,ThMe-Nature-2013}.
It has been shown for partially coherent x-ray pulses that such schemes improve
the quality of the reconstruction~\cite{QuNu-NatPhys-2011,WhWi-PRL-2009}.
Partial-coherence schemes make explicit use of the fact that the final
scattering pattern is an incoherent sum of coherent diffraction patterns.
For x-ray scattering from 3D-aligned molecules, the incoherent sum enters due to
the imperfect alignment.
Similarly it is expected that partial-coherence schemes can be used to build in
the effect of imperfect alignment.

\subsubsection{Strong-field Ionization from aligned molecules}
\label{p1c1:apps_ionization}
In Sec.~\ref{p1c1:apps_xscatter} the probe pulse has been an x-ray pulse to probe the ground state properties.
It is also possible to use these aligned molecules to study angle-dependent responses of molecules.
One examples, in spirit of Sec.~\ref{p1c2}, is tunnel ionization, which can depend on the alignment or even on the orientation of molecule~\cite{HoHa-NatPhys-2010,HaHo-JPB-2012}.
In heteronuclear molecules like OCS, the ionization potential is angle-dependent.
By using circularly polarized light, it is possible to mark via the vector potential from which end of the molecule the ionized electron originated from~\cite{HoHa-NatPhys-2010}. 
This experiment is an example that shows how important it is to have a good control of the rotational degrees of freedom of the molecule.

\begin{figure}[b]
\begin{center}
  \includegraphics[width=.6\linewidth]{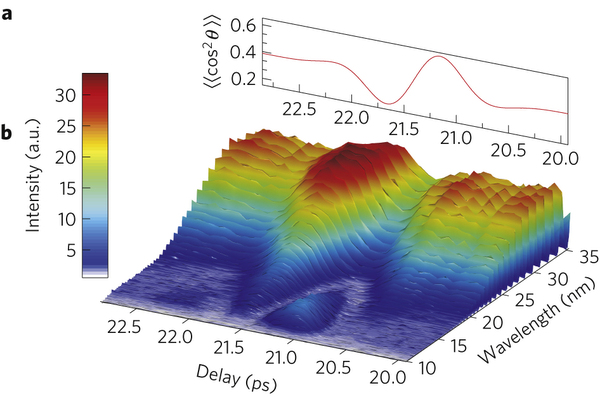}
  \caption{
    The HHG spectrum of CO$_2$ is shown as a function of pump-probe delay (and
degree of alignment in panel a).
    The pump pulse aligns the molecule and the probe pulse drives the HHG
process.
    The degree of alignment as a function of the pump-probe delay is shown in
the background.
    Structural information about the molecular orbitals can be gained from these
modifications of the HHG spectrum.    
    This figure is taken from Ref.~\cite{VoNe-NatPhys-2011}. 
    Copyright~\copyright~2011 Nature Publishing Group (NPG).
  }
  \label{fig1:hhg_co2}
\end{center}
\end{figure}

Molecular alignment can also influence the photoabsorption cross section of a
molecule.
Particularly when the absorption is favored along one specific molecular axis,
the cross section depends on the angle between the favored molecular axis and
the polarization direction of the photon~\cite{BuSa-PRA-2008}.
This has been experimentally demonstrated on CF$_3$Br, where the x-ray
absorption cross section depends on the alignment of the C-Br molecular
axis~\cite{PeBu-AppPhysLett-2008}.

HHG is another mechanism where the alignment of molecules matters.
The HHG process can be separated into 3 main parts: (1) an electron gets
tunnel-ionized by a strong laser field, (2) the same field accelerates the
electron first away and then back to the ion, (3) the electron recombines with
the ionic hole state and emits a highly energetic photon (much higher than the
photon energy of the driving laser pulse).
In Sec.~\ref{p1c2}, the physics behind HHG is discussed in more detail.

When molecules are used for the HHG process, it has been observed that the
emitted HHG spectrum depends on the alignment of the molecule---specifically the
alignment of the highest occupied molecular orbitals (HOMOs) with respect to the
returning electron~\cite{KaSa-Nature-2005}.
The angle dependence of the HHG spectrum is not just a scaling of the overall
HHG yield.
Changing the alignment direction of the molecule leads also to modification in
the substructures of the HHG
spectrum~\cite{MeCo-Science-2008,TeCh-PRA-2009,VoNe-NatPhys-2011}.

In Fig.~\ref{fig1:hhg_co2} the HHG spectrum of CO$_2$ is shown as a function of
alignment angle between the molecular symmetry axis and the polarization
direction of the linearly polarized strong field pulse.
It has been schown that structural informations about the HOMO orbitals can be
extracted from the HHG spectrum~\cite{ItLe-Nature432-2004,VoNe-NatPhys-2011}.

\section[Ultrafast Ionization Dynamics in Noble Gas Atoms]
{Ultrafast Ionization Dynamics in Noble Gas Atoms: About Femtosecond and Attosecond Dynamics}
\label{p1c2}

\subsection{Description of Light-Matter Interaction}
\label{p1c2:intro_lm}
Electron excitations and ionizations induced by ultrashort laser pulses are key mechanisms in attosecond physics~\cite{KrIv-RMP-2009}.
The energy difference between electronic states determines the time scale of the corresponding electronic motion.
Typical energy differences between electronic states of molecules and atoms lie within $0.1-100$~eV.
The corresponding time scales are 50~as to 50~fs.
The rapid technological progress in building lasers with sub-femtosecond pulses~\cite{BrKr-RMP-2000,PoCh-Science-2012} makes it possible to study fundamental mechanisms in chemical and physical processes in a time-resolved fashion~\cite{CoKr-NatPhys-2007}.
The attosecond physics community is closely tied to the strong-field physics community since, in one way or another, intense laser pulses are involved (at least in the generation of attosecond pulses).
Due to the wide interest in these fields, several review articles have been written in strong-field physics~\cite{BrKr-RMP-2000,BeLi-RMP-2012,PoBa-JMO-2008,BuRe-JPB-1993,IvSp-JMO-2005,MiPa-JPB-2006} and attosecond physics~\cite{KrIv-RMP-2009,AgDi-RPP-2004,ScIv-JPB-2006,PfAb-CPL463-2008,DaHu-JPB-2012}, which focus more on ionization processes and system dynamics, respectively. 

Pump-probe approaches are particularly advantageous for time-resolved studies~\cite{WuMe-JPB-2006}.
Each pump-probe configuration corresponds to a snapshot and a sequence of snapshots corresponds to a movie.
The time resolution is determined by the ability to accurately control the pump-probe delay $\tau$, which can be already measured to subattosecond (zeptosecond) precision~\cite{KoWo-OptExp-2011}.
Many attosecond pump-probe experiments use a femtosecond near-infrared (NIR) pulse and an attosecond ultraviolet (UV) pulse~\cite{KrIv-RMP-2009}. 
Either pulse can be used as a pump or as a probe pulse (see Sec.~\ref{p1c2:streaking} and Sec.~\ref{p1c2:tas}).
On the one hand, NIR pulses are usually strong-field pulses, which lead to non-perturbative tunnel ionization.
UV pulses, on the other hand, generally lead to electronic excitation and to perturbative multiphoton ionization. 

Pump-probe experiments can also be done with two UV pulses~\cite{RuJi-JPB-2010}.
It is, however, experimentally challenging if both UV pulses have low intensity, since the final signal strength would be quite weak.
This is not the case for UV pulses generated by free-electron lasers (FELs) such as FLASH~\cite{AcAs-NatPhot-2007}.
Here, the number of photons per pulse is quite high and UV-UV pump-probe experiments do not suffer from low statistics~\cite{RuJi-JPB-2010,MaHe-PRA-2012}.
All these different types of pump-probe schemes are complementary to each other and enable us to investigate many different aspects in the ultrafast world~\cite{KrIv-RMP-2009}.

\subsubsection{Keldysh Parameter}
\begin{figure}[ht!]
\begin{center}
  \includegraphics[width=.77\linewidth]{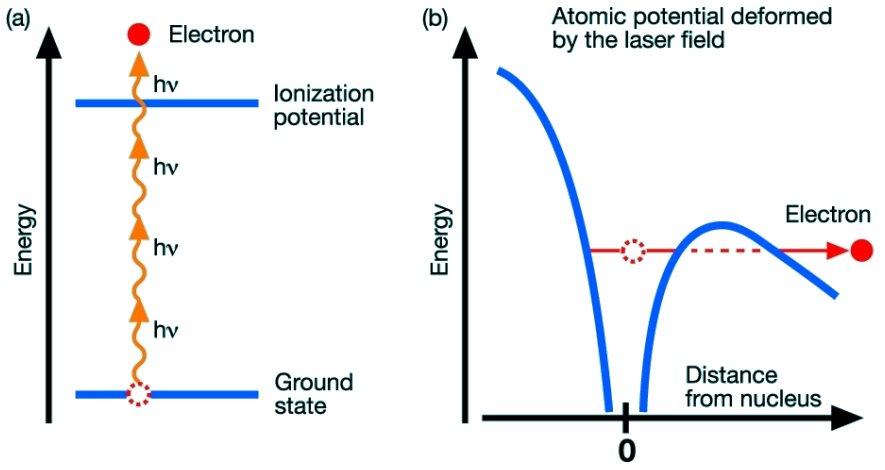}
  \caption{
    The perturbative (a) and the non-perturbative (b) multiphoton regimes are illustrated.    
    This figure is taken from \url{http://www.desy.de}.
    Copyright~\copyright~2007 DESY.
  }  
  \label{fig2:ionization}
\end{center}
\end{figure}

A common measure for characterizing the nature of field-induced ionization is the Keldysh parameter $\gamma$, which is given by~\cite{Ke-JETP-1965}
\begin{align}
  \label{eq2:keldysh}
  \gamma
  &=
  \sqrt{\frac{I_p}{2U_p}}
  ,
\end{align}
where $I_p$ is the ionization potential of the electronic state and $U_p=\frac{E^2}{4\omega^2}$ is the ponderomotive potential, which is the average energy of a free electron oscillating in the electric field with amplitude $E$ and frequency $\omega$.
The Keldysh parameter distinguishes between two ionization regimes: perturbative ($\gamma \gg 1$) and non-perturbative ($\gamma \ll 1$) multiphoton ionization.
Figure~\ref{fig2:ionization} illustrates the physical pictures behind these two regimes.

However, recent studies~\cite{ToRo-PRA-2012} have found that $\gamma$ alone is not always a rigorous measure to decide whether or not the ionization is perturbative or non-perturbative.
Particularly for $\gamma < 1$ and photon energies comparable or larger than the ionization potential ($\omega \gtrsim I_p$) the ionization process becomes more and more perturbative (contrary to what the Keldysh parameter would suggest).

In the case of $\gamma \gg 1$, the ionization lies in the perturbative multiphoton (or few-photon) regime.
The light-matter interaction can be seen as a correction to the field-free system.
The system gets ionized by absorbing a certain number of photons  (see Fig.~\ref{fig2:ionization}(a)).
In other words, the final state of the system is a result of a well-defined (and countable) number $n$ of light-matter interactions.
These $n$ light-matter interactions are exactly captured by the $n^\text{th}$-order perturbation correction.
Therefore, this regime is the perturbative multiphoton regime.

The ionization is considered to be in the non-perturbative multiphoton (or tunneling) regime\footnote{$\gamma \ll 1$ is also commonly referred to as the quasistatic limit, since $\gamma$ goes to zero for a given field strength and $\omega\rightarrow 0$.} for $\gamma\ll 1$.
Here, the Coulomb potential of the system rather than the field is considered to be the perturbation.
Furthermore, the light-matter interaction is pictured as a local potential that strongly distorts the Coulomb potential, and already after a few Bohr radii the field-induced potential starts to dominate the motion of the electron (see Fig.~\ref{fig2:ionization}(b)).
This distortion creates a potential barrier, which can be overcome by the electron by {\it tunneling} through this barrier and, consequently, out of the system.
Once the electron is tunneled to the outer side of the barrier, the Coulomb potential becomes negligible and the dynamics is govern by the field-induced potential.
Note, not just quantized photon picture has been dropped but also the number of photons an electron absorbs during ionization becomes an ill-defined quantity due to the non-perturbative nature of tunnel ionization.

To distinguish experimentally these two ionization regimes, it is quite convenient to study the photoelectron spectrum, specifically the energy distribution~\cite{ToRo-PRA-2012}.
In the tunneling limit, no characteristic photoelectron energies are observed whereas in the perturbative regime the photoelectron spectrum becomes discrete and the positions of the energy peaks reveal directly the number of photons the electron has absorbed during ionization.

\subsubsection{Minimal Coupling Hamiltonian}
By solving the full time-dependent Schr\"odinger equation (TDSE),
\begin{subequations}
\begin{align}
  \label{eq2:tdse}
  i\frac{\partial}{\partial t} \ket{\Psi(t)}
  &=
  \hat H(t) \ \ket{\Psi(t)}
  ,
\end{align}
\end{subequations}
both ionization limits as well as everything in between can be fully described.
However, making approximation to the Hamiltonian often helps to understand/identify better the dominant processes (e.g., see Sec.~\ref{p1c2:tunnel}).
Following the principle of {\it minimal coupling}~\cite{Jackson-book,Peksin-QFT-book}, the interaction between light and matter is captured by replacing the canonical momentum ${\bf \hat p}$ by the kinetic momentum ${\bf \hat v}(t):={\bf \hat p} - \alpha\, q_e\,{\bf A}({\bf r},t)$ and adding a scalar potential $q_e\,\phi({\bf r},t)$ to the Hamiltonian of field-free, non-interaction particles, i.e., $\hat H = \sum_n \hat{\bf p}^2_n/2$.
Here, $q_e$ is the charge of the electron and the speed of light in atomic units is given by $c=\alpha^{-1}\approx \frac{1}{137}$, where $\alpha$ is the fine-structure constant.

In the Coulomb gauge, $\hat {\bf \nabla}\cdot {\bf A}({\bf r},t)=0$, the scalar potential $\phi({\bf r})$ represents the instantaneous Coulomb potential generated by the charged particles, and the vector potential ${\bf A}({\bf r},t)$ represents the propagating light field.
The minimal coupling Hamiltonian for the electronic system reads
\begin{align}
  \label{eq2:ham_minicoupl}
  \hat H(t)
  &=
  \sum_n \frac{\big[ \hat{\bf p}_n - \alpha\,q_e\,{\bf A}(\hat{\bf r}_n,t) \big]^2}{2}
  + \sum_n \frac{Z\,q_e}{\hat{\bf r}_n}
  + \sum_{n \neq m} \frac{q_e^2}{|\hat{\bf r}_n-\hat{\bf r}_m|}
  ,
\end{align}
where $n$ runs over all electrons in the system, and $Z$ is the nuclear charge.
The Coulomb interaction between the charged particles, which is due to the exchange of virtual photons, is commonly not considered as a part of the light-matter interaction. 
Here, there term light-matter interaction refers to the interaction with external light fields---specifically the electric field component.
Consequently, the terms describing the light-matter interaction are
\begin{align}
  \label{eq2:lm_minicoupl}
  \hat H_\text{int}(t)
  &=  
  -\alpha\,q_e
  \sum_n 
    {\bf A}(\hat{\bf r}_n,t) \cdot \hat{\bf p}_n
  +
  \sum_n
  \frac{\alpha^2\,q_e^2}{2} \, {\bf A}^2(\hat{\bf r}_n,t)
  , 
\end{align}
where we have used the Coulomb gauge to rewrite the first term in Eq.~\eqref{eq2:lm_minicoupl}.
Often the space dependence of the light pulse can be dropped.
This is known as the dipole approximation, i.e., ${\bf A}({\bf r},t)\rightarrow {\bf A}(t)$.
This approximation is valid as long as the wavelength is large compared to the system size.
Typical wavelengths in strong-field and ultrashort pulses lie within tens to thousands of nanometers (UV to IR light) which are large compared to the size of atoms and molecules (sub-{\AA}ngstr\"om to a few nanometers).

The ${\bf A}^2(t)$ term is a global energy shift and is normally dropped, since it has no affect on the electronic dynamics (as long as the dipole approximation is valid).
For scattering experiments, where the wavelength is smaller than the object size, the dipole approximation cannot be made.
In fact, the non-dipole terms of ${\bf A}^2({\bf r},t)$ are essential for describing (x-ray) scattering~\cite{Sa-JPB-2009}.
Without them no scattering pattern would occur.

\subsubsection{Equivalent Forms of the Light-Matter Interaction}
By performing a unitary transformation of the the wavefunction, alternative forms of the Hamiltonian can be obtained, which are all equivalent to each other~\cite{Mi-book}.
The transformed wavefunction and the transformed Hamiltonian read
\begin{subequations}
\label{eq2:gauge-trafo}
\begin{align}
  \label{eq2:wfct_gauge-trafo}
  \Psi({\bf r},t)
  &\rightarrow 
  e^{-i\chi({\bf r},{\bf p},t)}\, \Psi({\bf r},t),
\\
  \label{eq2:ham_gauge-trafo}
  \hat H(t)
  &\rightarrow
  e^{i\chi({\bf r},{\bf p},t)}  \hat H(t)  e^{-i\chi({\bf r},{\bf p},t)}
  -
  \frac{ \partial\, \chi({\bf r},{\bf p},t) }{\partial t}
  ,
\end{align}
\end{subequations}
where $\chi({\bf r},{\bf p},t)$ can be any real function.
Hence, there exist an infinite number of possibilities~\cite{Mi-book}. 
These transformations are often called gauge transformations as well~\cite{Mi-book,BaFi-arxiv-2013}.
However, we will refer to these different forms as {\it frames} so that we do not confuse them with the gauge transformations (like the Coulomb gauge) that are based on fundamental invariances of nature\footnote{
Here $\chi$ is only a function of time and space and transforms also the vector potential ${\bf A}\rightarrow{\bf A}+\alpha^{-1}\nabla\chi$ and the scalar potential $\phi\rightarrow\phi-\partial_t \chi$ such that the minimal coupling Hamiltonian stays invariant~\cite{Peksin-QFT-book}.}.

Rather than just dropping the ${\bf A}^2(t)$ in Eq.~\eqref{eq2:lm_minicoupl} it can be also transformed away with $\chi_A(t)=\frac{\alpha^2\,q_e^2}{2}\int_{-\infty}^t\!dt'\, {\bf A}^2(t')$.
There are three popular choices for representing the light-matter interaction.
The light-matter interaction in the {\it velocity} frame with $\chi_V({\bf r},t)=0\ [+ \chi_A(t)]$ is already shown in Eq.~\eqref{eq2:lm_minicoupl}.
The light-matter interaction in the {\it length} frame~\cite{Go-AdP-1931} and the acceleration (or {\it Kramers-Henneberger}) frame~\cite{He-PRL-1968,ReBu-PRA-1990} is given by
\begin{subequations}
\label{eq2:frames}
\begin{align}
  \label{eq2:frames_l}
  \hat H_\text{int}(t)
  &=
  -q_e\,{\bf E}(t)\cdot \sum_n \hat{\bf r}_n
  \  &\text{with}&& \
  \chi_L &= -\alpha\,q_e\,{\bf A}(t)\cdot \sum_n \hat{\bf r}_n + \chi_A
  , \\
  \label{eq2:frames_kh}
  \hat H_\text{int}(t)
  &=
  \sum_n 
    \frac{Z\,q_e}{|\hat{\bf r}_n-{\bf a}(t)|}
    -
    \frac{Z\,q_e}{|\hat{\bf r}_n|}
  \  &\text{with}&& \
  \chi_\text{KH} &= -\alpha\,q_e \int_{-\infty}^t\hskip-1.5ex dt'\, 
    {\bf A}(t')\cdot \sum_n \hat{\bf p}_n
    +\chi_A
  .
\end{align} 
\end{subequations}
The Kramers-Henneberger frame [see Eq.~\eqref{eq2:frames_kh}] corresponds to a coordinate transformation into the acceleration frame of the electron, where the electronic position is displaced by ${\bf a}(t)=-\alpha\,q_e \int_{-\infty}^t\!dt'\,{\bf A}(t')$.
The subtraction of the field-free nuclear potential in Eq.~\eqref{eq2:frames_kh} means the resulting overall Hamiltonian is identical to the field-free Hamiltonian just with a time-dependent nuclear potential $\sum_i\frac{Z\,q_e}{|\hat{\bf r}_i-{\bf a}(t)|}$.
The electron-electron interaction is not affected by this transformation, since the relative distances between electrons is unchanged by this transformation.
In the Kramers-Henneberger frame, the coordinates are transformed such that one would stay in the rest frame of the electron if it is freely moving in the field (without any Coulomb potentials). 
As a result, the nuclei are moving and not the electrons as in the other two frames.
Therefore, the Kramers-Henneberger frame is also known as the acceleration frame.

Equation~\eqref{eq2:frames_l} is the most popular choice for the light-matter interaction, where ${\bf E}(t)=-\alpha \partial_t {\bf A}(t)$ is the electric field.
The intuitive picture that comes with treating the light-matter interaction as a local potential, ${\bf E}(t) \cdot \hat{\bf r}$, is one reason for its popularity.

From now on, we explicitly set the electronic charge $q_e=-1$.

\subsection{Photoionization}
\label{p1c2:photo}
One of the best studied light-matter interaction processes is one-photon ionization or simply photoionization~\cite{Ha-RMP-1936,FaCo-RMP-1968}.
Photoionization is often discussed in terms of cross sections, which are measures of how likely a photon interacts with the object.
The photoionization cross section of an $N$-electron system in the dipole approximation is given by~\cite{St-Springer-1980,Sa-JPB-2009}
\begin{align}
  \label{eq2:cs_vel}
  \sigma(\omega)
  &=
  \frac{4\pi^2\,\alpha}{\omega} 
  \sum_F \left|
    \sum_{n=1}^N \bra{\Psi_F}  \bm{\epsilon} \cdot \hat{\bf p}_n \ket{\Psi_{I}}
  \right|^2  
  \delta(E_{F}-E_I-\omega)
  ,
\end{align}
where $\omega$ is the photon energy, $E_I$ is the energy of the initial state $\Psi_I$, and $E_{F}$ is the energy of the final state $\Psi_F$.
The delta distribution enforces the condition $E_{F}-E_I=\omega$, which ensures energy conservation.
The polarization direction of the absorbed photon is given by $\bm{\epsilon}$.
The continuum states $\Psi_F$ are energy normalized, i.e., $\braket{\Psi_F}{\Psi_G}=\delta(E_F-E_G)$.
Equation~\eqref{eq2:cs_vel} characterizes the absorption of one photon and can be derived from $1^\text{st}$-order perturbation theory.

The photoionization cross section in Eq.~\eqref{eq2:cs_vel} is given in terms of the momentum operator, ${\bf \hat p}_n$.
Alternative and equally exact expressions for the electronic dipole transitions can be found.
In order to do so, the following commutator relations involving the exact Hamiltonian are used:
\begin{subequations}
  \label{eq2:dipoleforms}
  \begin{align}
    \label{eq2:dipoleforms_l}
    \hat {\bf p}_n &= -i[\hat {\bf r}_n,\hat H]
    , \\
    \label{eq2:dipoleforms_a}
    [\hat {\bf p}_n,\hat H] &=-i (\hat{\bf \nabla}_n \hat V)
      =-i\,Z\, \frac{\hat{\bf r}_n}{|\hat{r}_n|^3}
    ,
  \end{align}
\end{subequations}
with $\hat V=-\sum_n Z/|\hat{\bf r}_n| + \sum_{n\neq m} 1/|\hat{\bf r}_n-\hat{\bf r}_m|$ being the exact electronic potential consisting of the nuclear-electron and electron-electron Coulomb interactions.
If the initial and the final states are energy eigenstates, the commutators in Eqs.~\eqref{eq2:dipoleforms} can be easily evaluated. 
The relationship between the corresponding matrix elements are
\begin{subequations}
  \label{eq2:matrixforms}
  \begin{align}
  \label{eq2:matrixforms_l}
   (E_F-E_I)
   \bra{\Psi_F}\sum_n \hat {\bf p}_n\ket{\Psi_I} 
   &= 
   i\bra{\Psi_F} Z \sum_n \frac{\hat{\bf r}_n}{|\hat r_n|^3} \ket{\Psi_I}
   , \\
   \label{eq2:matrixforms_v}
   \bra{\Psi_F}\sum_n \hat {\bf p}_n\ket{\Psi_I} 
   &= 
   i(E_F-E_I)\bra{\Psi_F}\sum_n\hat{\bf r}_n \ket{\Psi_I}
  \end{align}
\end{subequations}
The expressions involving the matrix elements of ${\bf \hat r}_n$, ${\bf \hat p}_n$, and $Z\,{\bf \hat r}_n/r_n^3$ are known as ``length'', ``velocity'', and ``acceleration'' forms of the electric dipole matrix~\cite{FaCo-RMP-1968,St-Springer-1980}.
The equality of the different expressions for the electronic dipole matrix has been first discussed by Chandrasekhar~\cite{Chandra-Astrophys-1945}.
For approximations to the exact eigenstates of the system, these forms start to differ and tend to emphasize different spatial parts of the wavefunction. 
The length form stresses the large distances, the velocity form stresses the intermediate ones, and the acceleration form stresses the short distances (for more details see Refs.~\cite{St-AdvAMO-1968,Cr-AdvAMO-1969}).

\subsubsection{Independent Particle Picture}
It is not possible to determine the exact states $\Psi_I$ and $\Psi_F$ for many-electron systems due to the Coulomb interaction between the electrons~\cite{SzOs-book}.
Even the exact ground state of Helium, a two-electron atom, has not yet been found~\cite{Dr-PS-1999}.
This discussion focuses on noble gas atoms, which have a closed-shell electronic structure. 
The big advantage of closed-shell atoms is that their electronic ground state $\Psi_0$ can be well approximated by the Hartree-Fock (HF) ground state $\Phi_0$,
\begin{align}
  \label{eq2:GS_HF}
  \ket{\Psi_0}
  &\approx
  \ket{\Phi_0}
  =
  \prod_{n=1}^N \hat c_i^\dagger \ket{\text{vacuum}}
  ,
\end{align}
where the operator $\hat c_i^\dagger$ creates an electron in orbital $\varphi_i$ and $\hat c_i$ destroys it~\cite{Ro-RMP-1951}.
A very nice side effect of HF, the $(N-1)$-electron states, $\ket{\Phi_i}=\hat c_i \ket{\Phi_0}$, with a missing electron in orbital $\varphi_i$ correspond quite well to the eigenstates of the ionic system.
Hence, the ionization potential of an electron in orbital $\varphi_i$ is given by $-\varepsilon_i$, where $\varepsilon_i$ is the corresponding HF orbital energy. 
This statement is known as Koopmans' theorem~\cite{SzOs-book}.

Assuming the absorbed photon affects only one electron (active electron) and all the other electrons are spectators (passive electrons), the final $N$-body state can be simplified to $\ket{\Psi_F}\approx\ket{\Phi^a_i}=\cre_a \ann_i \ket{\Phi_0}$, where the index $i$ refers to occupied orbitals and $a$ refers to unoccupied orbitals in the HF ground state. 
This is also known as the sudden approximation~\cite{Stoehr-book}.
This assumption holds when the relaxation process of the remaining electrons is slow compared to the ionization dynamics.
Hence, the electronic transition elements in Eq.~\eqref{eq2:cs_vel} reduce to one-body transition elements
\begin{align}
  \sum_{n=1}^N \bra{\Psi_F} \bm{\epsilon} \cdot \hat{\bf p}_n \ket{\Psi_{I}}
  =
  \bra{\varphi_a} \bm{\epsilon} \cdot \hat {\bf p}\ket{\varphi_i}
  \underbrace{\braket{\Phi_i}{\Phi_i}}_{=1}
  ,
\end{align}
where the initial state is the Hartree-Fock ground state $\Phi_0$.
%
The photoionization cross section simplifies to~\cite{Sa-JPB-2009}
\begin{align}
  \label{eq2:cs_total}
  \sigma(\omega) 
  &= 
  \frac{4\pi^2\, \alpha }{\omega}
  \sum_a \left|
    \bra{\varphi_a} \bm{\epsilon} \cdot \hat {\bf p} \ket{\varphi_i}
  \right|^2  
  \delta(\varepsilon_a-\varepsilon_i-\omega)
  .
\end{align}
From $\sigma(\omega)$ the ionization rate $\Gamma$ of an atom can be easily calculated by $\Gamma=J_\text{EM}\,\sigma(\omega)$, where $J_\text{EM}$ is the photon flux of the ionizing pulse hitting the atom.
Note that the cross section itself is not field-dependent.
The photoionization rate increases linearly with the photon flux (intensity) for a fixed photon energy.
For tunnel ionization, the ionization rate does not follow a power law and depends rather exponentially on the intensity (for details see Sec.~\ref{p1c2:tunnel}). 

%
\begin{figure}[ht!]
\begin{center}
  \includegraphics[width=.9\linewidth]{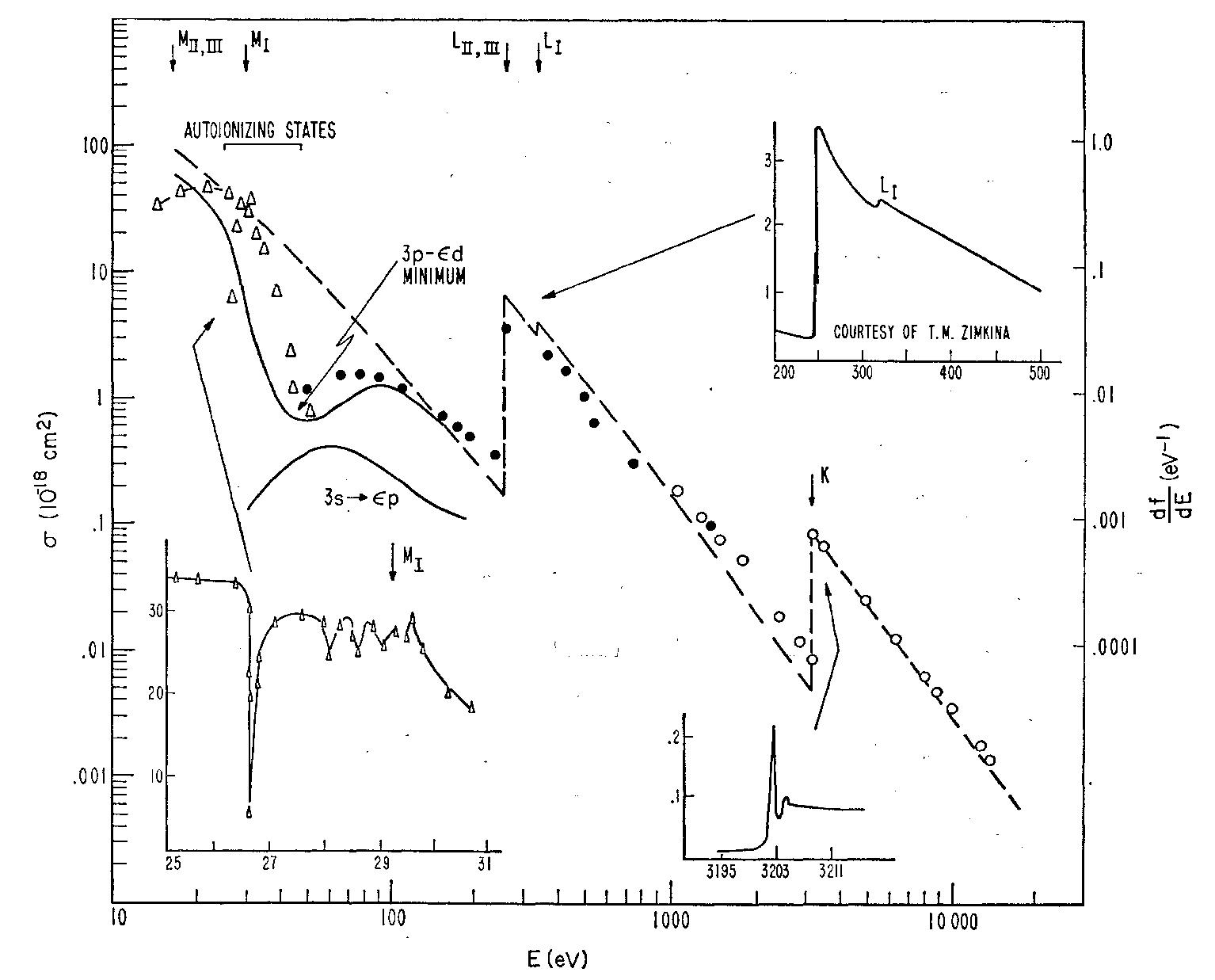}
  \caption{
    The total cross sections $\sigma(\omega)$ for argon: hydrogenic approximation (dashed line), one-electron model (solid line), experimental data (points).
    Insets show details of $\sigma(\omega)$ around certain energy regions.
    This figure is taken from Ref.~\cite{FaCo-RMP-1968}. 
    Copyright~\copyright~1968 American Physical Society (APS).
  }  
  \label{fig2:cs_c}
\end{center}
\end{figure}

\subsubsection{Structural Information in Photoionization Cross Sections}
The cross section also contains a wide range of electronic structural information about the system.
A typical cross section behavior as a function of photon energy is shown in Fig.~\ref{fig2:cs_c}, where experimental values (points) as well as theoretical predictions (lines) for argon are shown.
Very distinct jumps in $\sigma(\omega)$ occur when new ionization channels open, i.e., when the photon energy is high enough to ionize electrons out of the next energetically deeper lying subshell.
In Fig.~\ref{fig2:cs_c} the $K$ ($1s$ shell) and $L$ ($2s,2p$ shells) edges are clearly visible at 3200~eV and 250~eV, respectively.
The $L$ edge consists actually of 3 edges corresponding to the subshells $2p_{1/2},2p_{3/2}$ and $2s$.
The energy difference between $2p_{1/2}$ ($L_\text{II}$ edge) and $2p_{3/2}$ ($L_\text{III}$ edge) is $\approx 2$~eV~\cite{XDB} and, therefore, not visible in Fig.~\ref{fig2:cs_c}.
The $2s$ ($L_\text{I}$) edge can be singled out, since it is $\approx 25$~eV~\cite{XDB} apart (see upper right inset in Fig.~\ref{fig2:cs_c}).

Another interesting feature appears shortly before the ionization edges and is called XANES (x-ray absorption near edges structure)~\cite{ReAl-RMP-2000}.
A typical XANES profile is shown in the lower right inset in Fig.~\ref{fig2:cs_c}.
The peak structure around the edge corresponds to electron excitations; more specifically, excitations into the Rydberg states of the atom.
With increasing energies the spacing becomes smaller between the Rydberg peaks, which are broadened by the lifetime of the corresponding state.
When the spacing is smaller than their line widths, individual peaks are no longer visible and the cross section goes smoothly over into a continuum structure~\cite{ReAl-RMP-2000}. 

\begin{figure}[ht!]
\begin{center}
  \includegraphics[width=.7\linewidth]{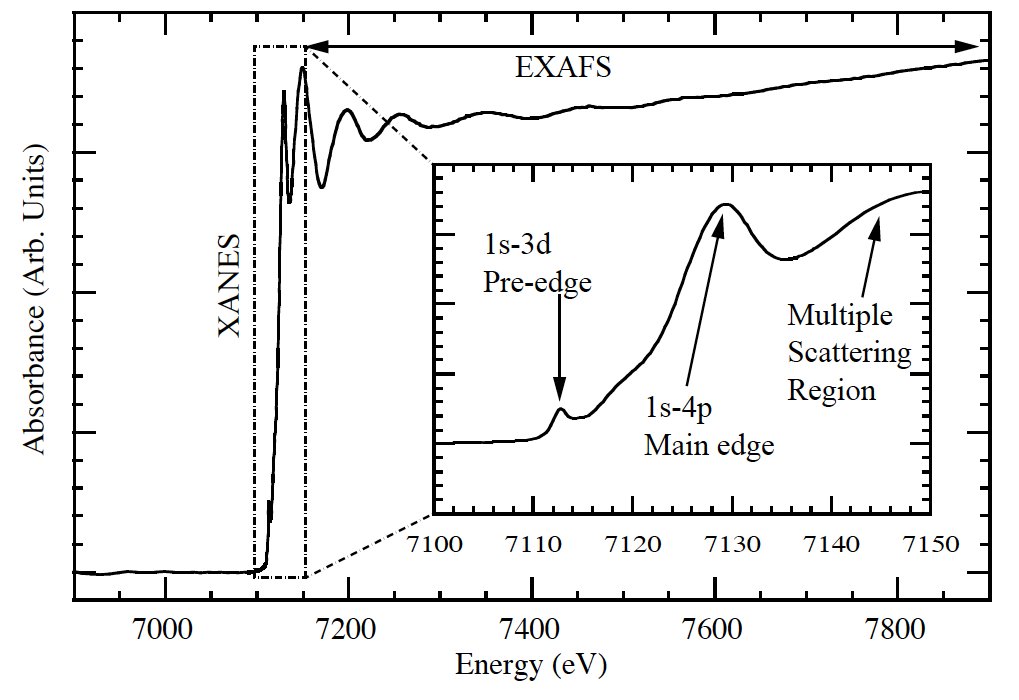}
  \caption{
    The cross section around the K-edge of Fe is shown.
    The XANES and EXAFS regions are highlighted. 
    This figure is taken from Ref.~\cite{Carp-thesis-2010}.
    \copyright~2010 Matthew H. Carpenter.
  }
  \label{fig2:xanes_exafs}
\end{center}
\end{figure}

Above the ionization threshold, the XANES features slowly transform into EXAFS (extended x-ray absorption fine structure) features~\cite{BuSt-PRL-1984}. 
EXAFS originates from the interference with the photoelectron that has been backscattered by neighboring atoms~\cite{ReAl-RMP-2000,NiMc-ElementXrayPhysics-book}.
Hence, EXAFS can be used as a probe to study the vicinity of specific atoms, which can be chosen by tuning the photon energy close to an atom-specific absorption edge~\cite{Sa-JPB-2009}.
In Fig.~\ref{fig2:xanes_exafs} the XANES and EXAFS regions are shown for the $K$-edge of iron. 
%

Besides the characteristic absorption edges, the cross section contains more information about the electronic structure, particularly about the continuum structure~\cite{St-Springer-1980}. 
A famous example is the Cooper minimum~\cite{Co-PR-1962}, which appears at 48~eV in argon~\cite{SaSt-JESRP-2002} (see Fig.~\ref{fig2:cs_c}).
It is a consequence of a sign change in the transition elements $\bra{\varphi_{3p}} r\ket{\varepsilon d}$ from negative ($\omega<48$~eV) to positive ($\omega>48$~eV). 
The term $\varepsilon$ is the energy of the ionized electron.
This sign change results in a vanishing contribution from the $3p\rightarrow \varepsilon d$ ionization channel.
The total cross section is, however, not really zero, since the much weaker $3p\rightarrow \varepsilon s$ and $3s\rightarrow \varepsilon p$ ionization channels have nonzero contributions~\cite{SaSt-JESRP-2002}.
Note that the channel $nl\rightarrow \varepsilon (l+1)$ is usually the dominant signal and much larger than $nl\rightarrow \varepsilon (l-1)$~\cite{FaCo-RMP-1968}, where $n$ is the principal quantum number of the orbital.
Cooper and Fano~\cite{FaCo-RMP-1968} have given a rule determining when a Cooper minimum can occur in the photoionization cross section of a subshell.
The rule excludes a Cooper minimum in the noodless subshells $1s, 2p, 3d,$ and $4f$ as well as in the transitions $nl\rightarrow \varepsilon(l-1)$.

The occurrence of a Cooper minimum can be well-explained with a one-particle picture.
However, the details like the exact position and the form of the minimum does depend on many-body physics~\cite{ChFa-PRA-1976}.
For the Cooper minimum in Ar, doubly excited configuration corrections to the ground state, as included in the random phase approximation (RPA), are needed to predict the correct position~\cite{ChFa-PRA-1976}.

\begin{figure}[b!]
  \centering
  \begin{subfigure}[b]{0.49\textwidth}
    \centering
    \includegraphics[width=\linewidth]{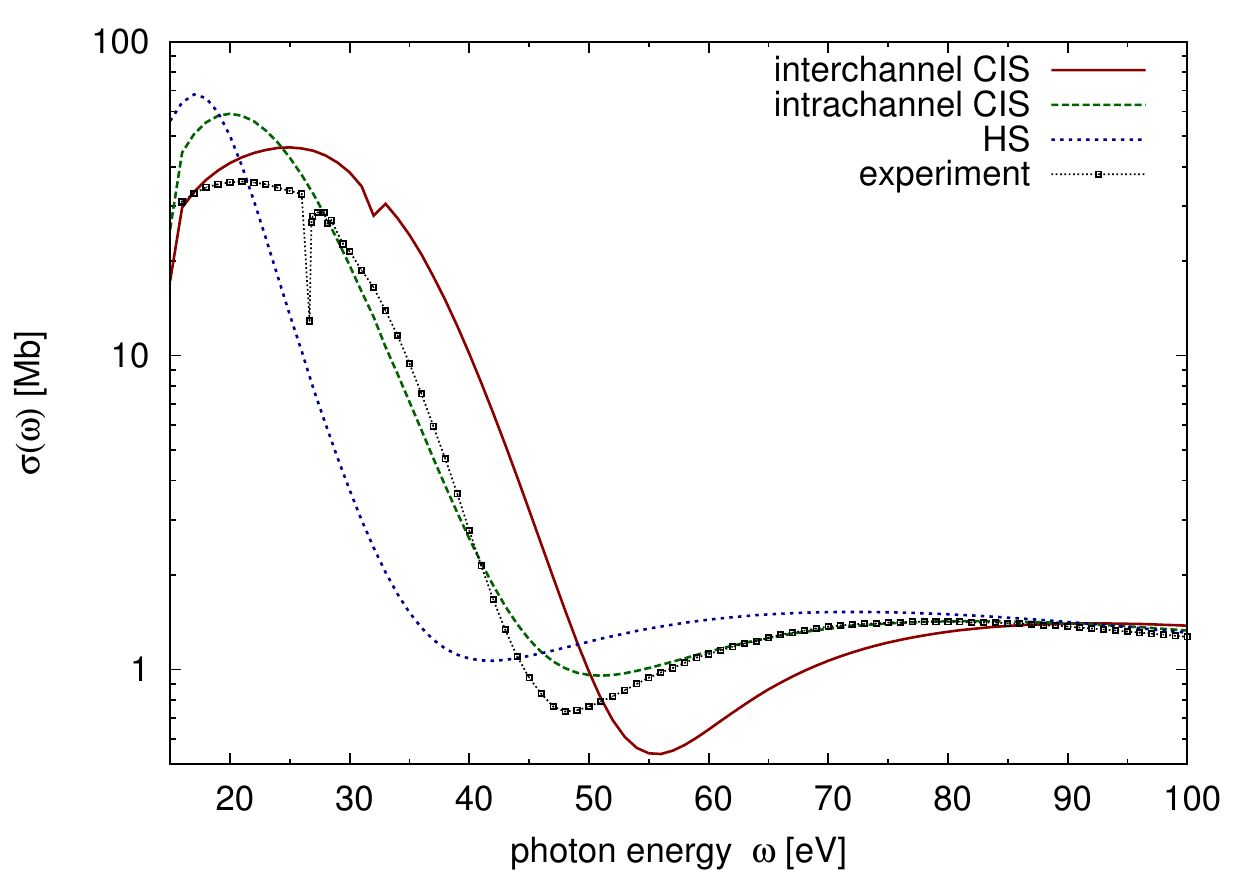}
    \caption{Argon}
    \label{fig2:cs_total_ar}
  \end{subfigure}
  \begin{subfigure}[b]{0.49\textwidth}
    \centering
    \includegraphics[width=\linewidth]{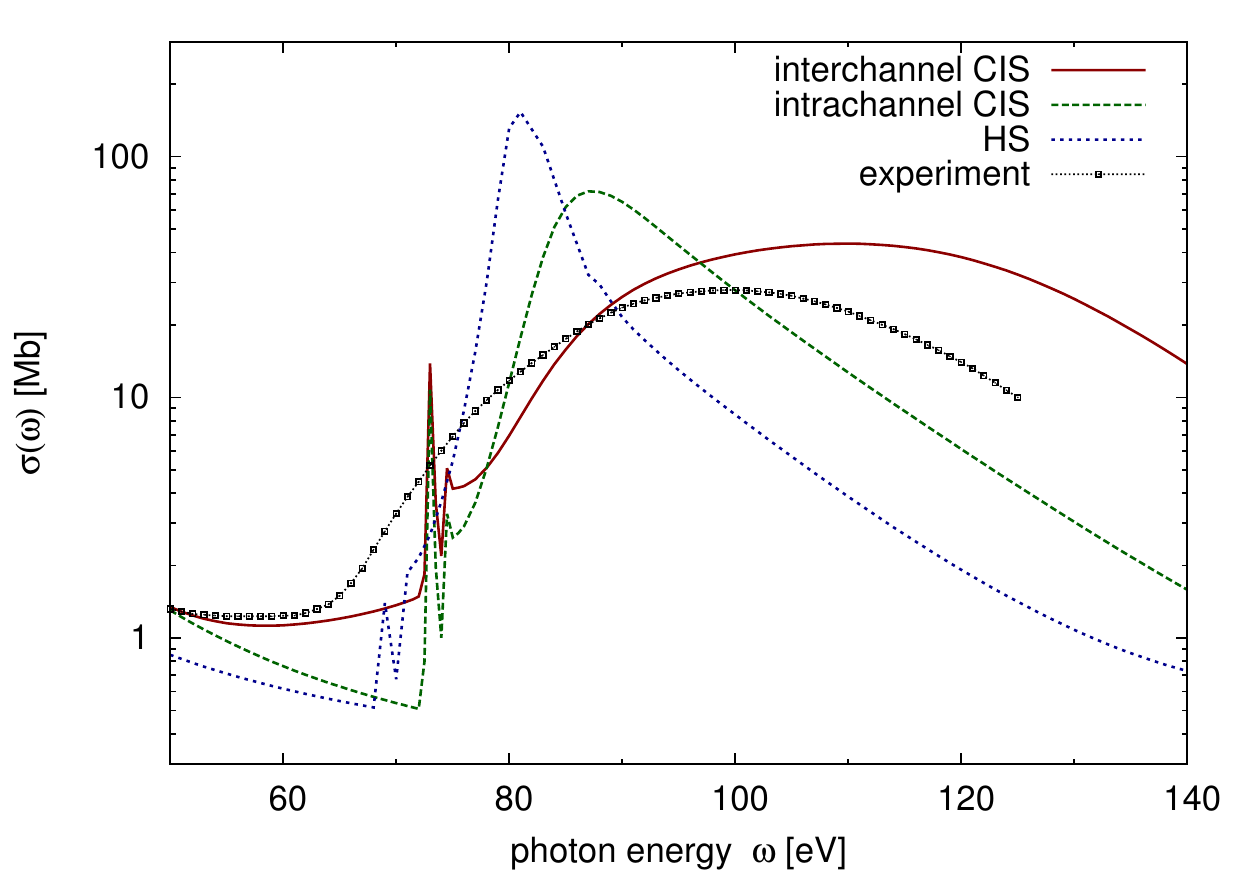}
    \caption{Xenon}
    \label{fig2:cs_total_xe}
  \end{subfigure}
  \caption{
    The absorption cross sections of Ar (a) and Xe (b) are shown for the Hartree-Fock-Slater (blue), the interchannel CIS (red), and the intrachannel CIS (green) models.
    Note that the interchannel CIS model also includes intrachannel interactions.
    The experimental data (line and points) are taken from Ref.~\cite{SaSt-JESRP-2002}. 
  }
  \label{fig2:cs_total}
\end{figure}

In Fig.~\ref{fig2:cs_total}, the total photoionization cross sections of Ar and Xe are shown for the Hartree-Fock-Slater (HS) model~\cite{SoYo-PRA-2011,Sl-PR-1951}, an intrachannel CIS model, and an interchannel CIS model (explained below).
CIS stands for the configuration-interaction singles method described in Sec.~\ref{p1c2:tdcis}.
All CIS results presented in this section are calculated with the {\sc xcid} package described in Ref.~\cite{GrSa-PRA-2010}.
The basic differences between the models lie in the approximations made to the Hamiltonian and to the wavefunction.
In HS, a single Slater determinant is used for the wavefunction as in HF whereas the CIS wavefunction is a sum of Slater determinants containing the HF ground state and all singly-excited (one-particle-one-hole) configurations $\Phi^a_i$.

For HS, a spherical model potential is used, where the exchange potential between electrons is approximated by a free electron gas model~\cite{Sl-PR-1951} and the long range $1/r$ potential is enforced via the Latter-type correction~\cite{La-PR-1955}.
The adjectives {\it interchannel} and {\it intrachannel} in the CIS model refer to the allowed electron-electron interaction.
Intrachannel coupling considers only electron-electron interactions where the ionic state does not change.
In this case, the ionization channels are independent of each other. 
The intrachannel CIS model includes only intrachannel interactions.
Interchannel coupling refers to electron-electron interactions where the excited/ionized electron changes the ionic state.
The interchannel CIS model includes both intrachannel and interchannel interactions.


In Fig.~\ref{fig2:cs_total}, the total cross sections of argon (a) and xenon (b) are shown for different theoretical models.
The Cooper minimum in Ar and the giant dipole resonance in Xe do exist in $\sigma(\omega)$ for all three models.
It shows that the Cooper minimum as well as the giant dipole resonance are not results of many-body effects.
They can be solely explained by one-particle physics.
The position as well as the form of these features do, however, strongly differ between the different models.

In the HS model, these features occur at too low photon energies and are too narrow and too high in their shape.
The potential resulting from intrachannel and interchannel interactions, which are large and repulsive, corrects the HS model too much and shifts the features to too high energies.
Nevertheless, models including interchannel and intrachannel interactions generally yield results that are closer to the experimental values.
Particularly in xenon, the shape and the position of the giant dipole resonance changes significantly when intrachannel and interchannel interactions are included (see Fig.~\ref{fig2:cs_total_xe}).

\begin{figure}[ht!]
  \centering
  \begin{subfigure}[b]{0.49\textwidth}
    \centering
    \includegraphics[width=\linewidth]{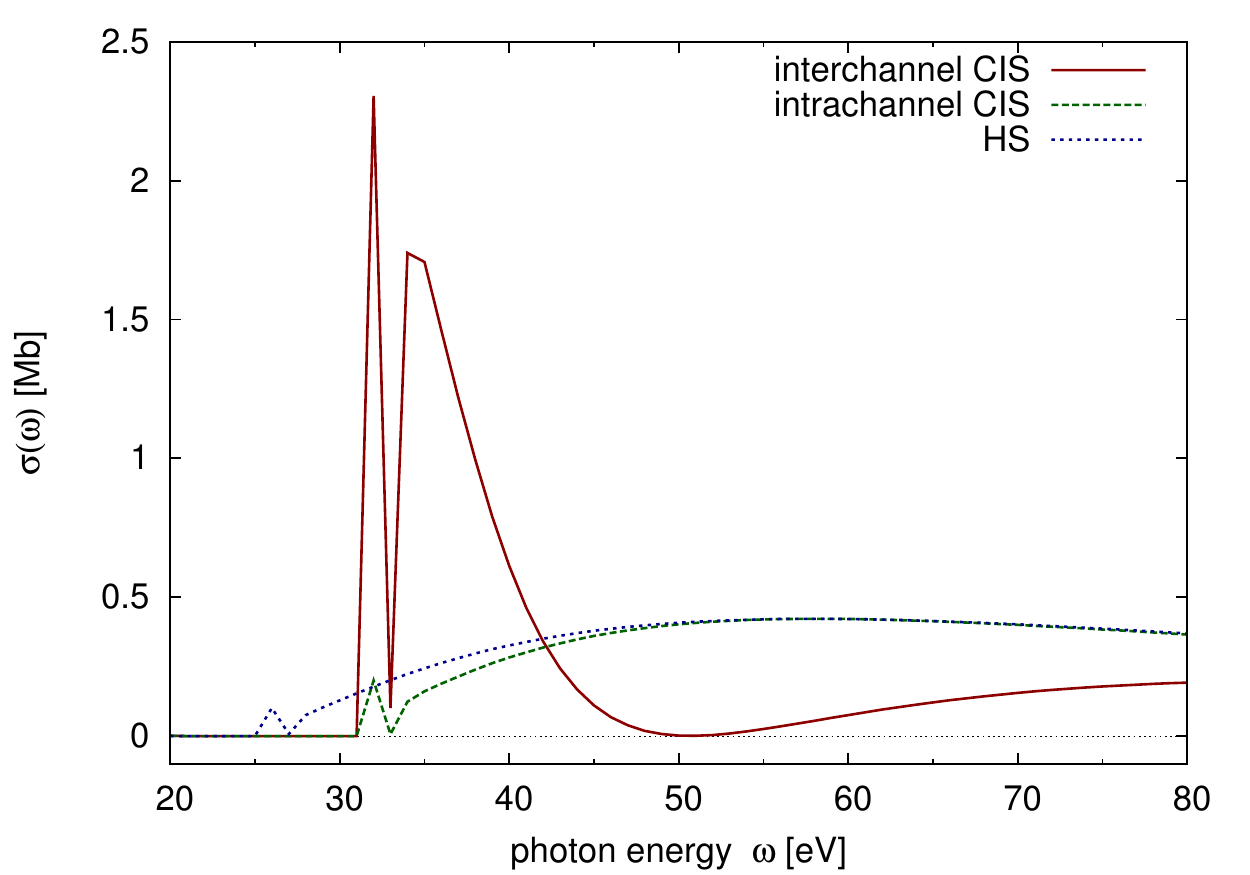}
    \caption{Argon}
    \label{fig2:cs_subshells_ar}
  \end{subfigure}
  \begin{subfigure}[b]{0.49\textwidth}
    \centering
    \includegraphics[width=\linewidth]{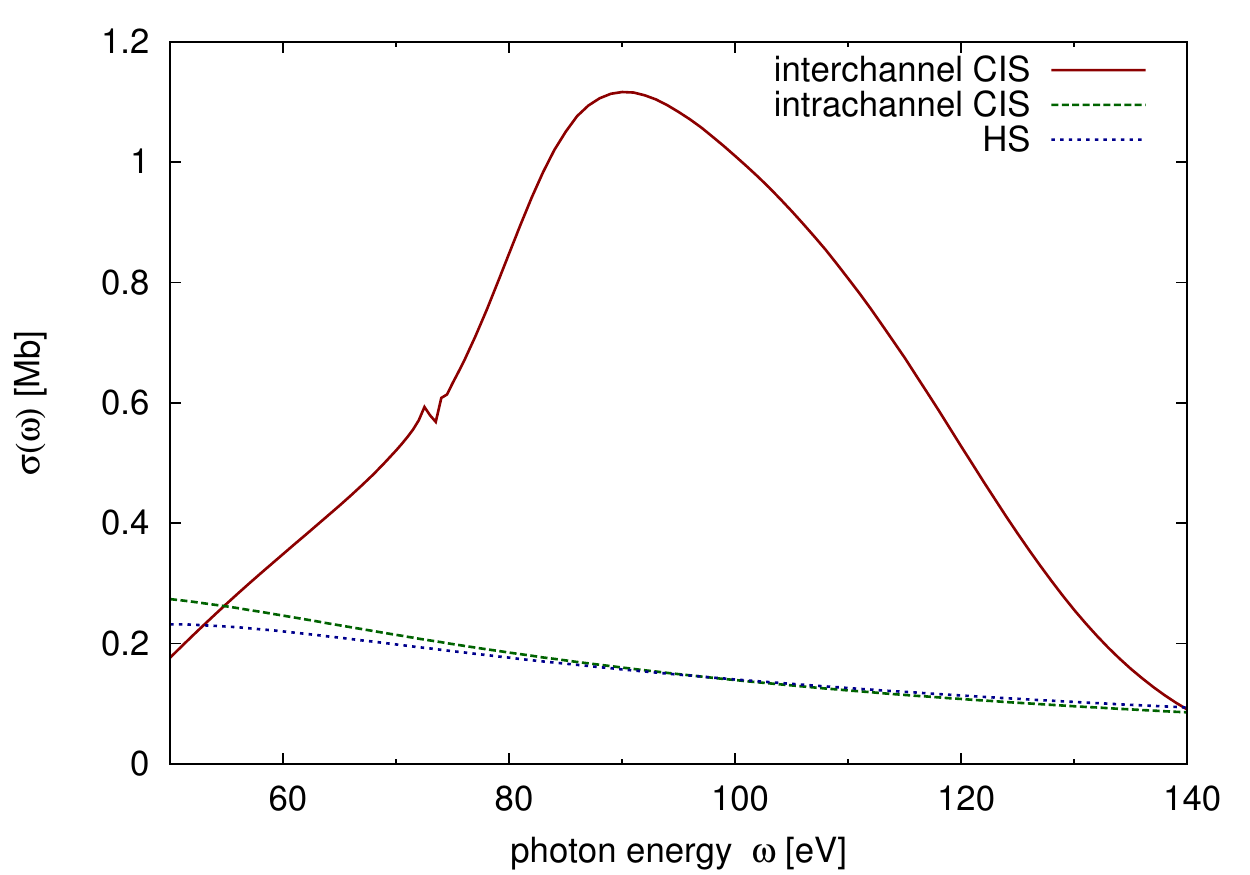}
    \caption{Xenon}
    \label{fig2:cs_subshells_xe}
  \end{subfigure}
  \caption{
    The absorption cross sections of Ar (a) and Xe (b) of the subshells $3s$ and $5s$, respectively, are shown for the Hartree-Fock-Slater (blue), the interchannel CIS (red), and the intrachannel CIS (green) models.
    Note that the interchannel CIS model also includes intrachannel interactions.
  }
  \label{fig2:cs_subshells}
\end{figure}

Interchannel coupling is particularly important for partial cross sections of deeper lying subshells.
In Fig.~\ref{fig2:cs_subshells}, the photoionization cross sections of the $3s$ and $5s$ subshells of Ar and Xe, respectively, are shown. 
For argon, the Cooper minimum of the $3p\rightarrow \varepsilon d$ ionization channel affects the $3s$ cross section. 
Without interchannel interactions the effect is gone.
Similarly for Xe, the giant dipole resonance in the $4d$ cross section at 100~eV also affects the cross section of the $5s$ (shown) and $5p$ (not shown) subshells.
Switching off interchannel interactions leads to a dramatic change in the $5s$ cross section shown in Fig.~\ref{fig2:cs_subshells_xe}.
Intrachannel interactions alone only slightly affect the cross sections.

\subsubsection{Angular Distribution}
The electronic structure features are also imprinted in the angular distribution of the differential cross section, $\frac{d\sigma}{d\Omega}$~\cite{DeDi-PRL-1976,DiSt-PRA-1975}.
Studying the angular distribution of the photoionized electron is known as angle-resolved photoemission spectroscopy (ARPES)~\cite{BeSh-book}. 
Changes in the angular distribution contain a wide range of information about the interaction between the electron and the ion, and about the orbital structure from which the electron came~\cite{FaDi-PRA-1972,DiSt-PRA-1975}.
This is particularly interesting in molecules~\cite{HeGe-PRL-1997,BoLi-PRL-2001,ScTi-Science-2008} and solids~\cite{DaHu-RMP-2003,LuVi-ARCMP-2012}.
In atomic systems, the differential photoionization cross section averaged over all initial magnetic quantum numbers $m_i$ reduces to~\cite{BlBi-RMP-1952}
\begin{align}
  \label{eq2:dcs_beta}
  \frac{d\sigma_i(\omega)}{d\Omega}
  &=
  \frac{\sigma_i(\omega)}{4\pi}
  \left[
  1 + \beta(\omega)\, P_2(\cos\vartheta) 
  \right]
  ,
\end{align}
where $P_2(x)=\frac{3}{2}x^2-\half$ is the $2^\text{nd}$-Legendre polynomial, and $\vartheta$ is the angle between the photoelectron and the polarization direction of the light.
The term $\sigma_i(\omega)$ is the partial photoionization cross section of the subshell $n_il_i$, where $n_i$ is the principal quantum number and $l_i$ is the angular momentum quantum number.

\begin{figure}[ht!]
  \centering
  \begin{subfigure}[b]{0.49\textwidth}
    \centering
    \includegraphics[width=\linewidth]{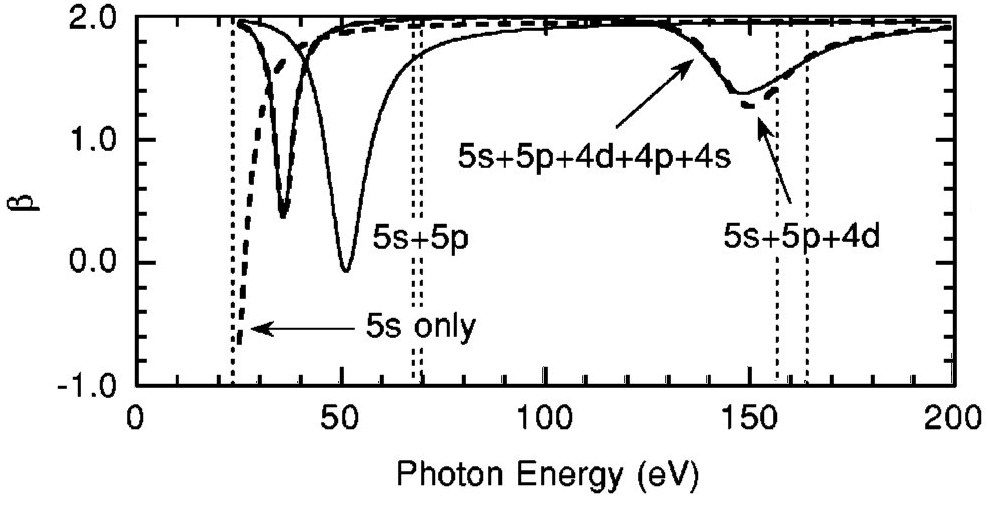}
    \caption{Xenon $5s$}
     \label{fig2:cs_beta_5s}
  \end{subfigure}
  \begin{subfigure}[b]{0.49\textwidth}
    \centering
    \includegraphics[width=\linewidth]{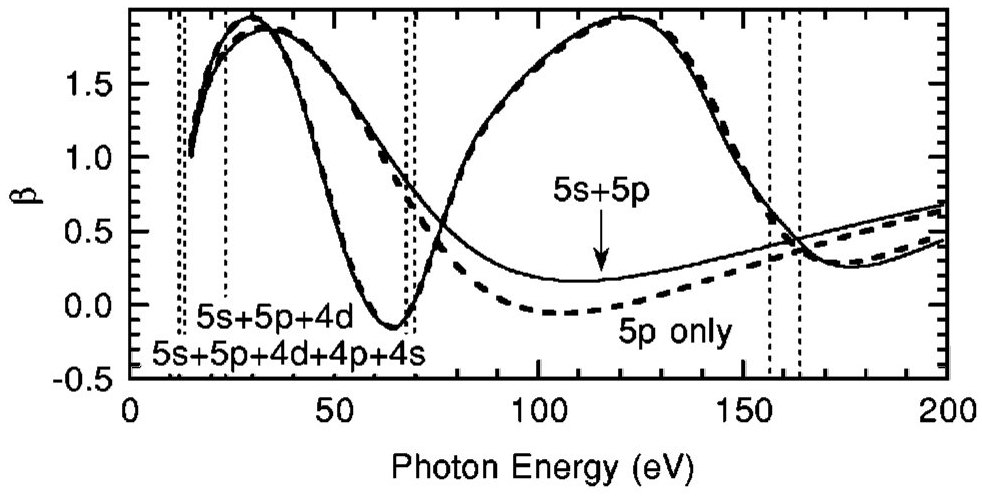}
    \caption{Xenon $5p$}
    \label{fig2:cs_beta_5p}
  \end{subfigure}
  \caption{
    The $\beta(\omega)$ parameter is shown for the $5s$ subshell (a) and the $5p$ subshell (b) of Xe including various interactionss with other orbitals (see labels in the figures).
    These figures are taken from Ref.~\cite{JoCh-PRA-2001}. 
    Copyright~\copyright~2001 American Physical Society (APS).
  }
  \label{fig2:cs_beta}
\end{figure}

The angular distribution of $\frac{d\sigma_i}{d\Omega}$ is characterized by a single quantity: the $\beta$ parameter~\cite{BlBi-RMP-1952}.
Since $\frac{d\sigma_i}{d\Omega}$ must be positive, the range of $\beta$ is restricted to $-1\leq \beta \leq 2$.
In Fig.~\ref{fig2:cs_beta}, the beta parameter is shown for the $5s$ (Fig.~\ref{fig2:cs_beta_5s}) and the $5p$ (Fig.~\ref{fig2:cs_beta_5p}) subshells of Xe with and without various interchannel interactions to neighboring orbitals.
In both cases, the behavior of $\beta$ is significantly altered when interactions with other subshells are ignored.
Hence, $\beta$ is an ideal quantity for studying these interchannel effects~\cite{Ilchen-Phd}.

\subsubsection{Perturbative Few-Photon Ionization}
The probability of absorbing a second or even a third photon increases with higher intensities.
Recent experiments~\cite{YoKa-Nature-2010} at the \href{http://lcls.slac.stanford.edu}{LCLS} in the x-ray regime have impressively shown that up to eight photons can be easily absorbed by neon producing Ne$^{+8}$.
In the x-ray regime, multiphoton ionization is dominated by {\it sequential} photoionization meaning one photon is absorbed at a time (i.e., multiple one-photon ionizations).
This can be well understood by looking at the Keldysh parameter $\gamma$ defined in Eq.~\eqref{eq2:keldysh}.
The high photon energy of x-rays leads to very small ponderomotive potentials and, therefore, to very large $\gamma$ values indicating that x-ray ionization is far in the perturbative regime.

The ionization of an electron by simultaneously absorbing two photons is the simplest ``true'' multiphoton step. 
The absorption of several photons at the same time is also called {\it non-sequential} multiphoton ionization (see Fig.~\ref{fig2:ionization}).
It is particularly favored if the photon energy is close to a resonance.
Non-sequential two-photon absorption is characterized by the two-photon ionization cross section $\sigma^{(2)}(\omega)$, which reads in the length form in lowest-order perturbation theory~\cite{SaLa-JPB-1999,SyPa-PRA-2012},
\begin{align}
  \label{eq2:cs_2photon}
  \sigma^{(2)}(\omega)
  &=
  \pi (4\pi\,\alpha\,\omega)^2
  \sum_F \left|
    \sum_{H\neq I}  \sum_{n,n'}
      \frac{ \bra{\Psi_F}\bm{\epsilon}\cdot\hat{\bf r}_n\ket{\Psi_{H}} 
             \bra{\Psi_H}\bm{\epsilon}\cdot\hat{\bf r}_{n'}\ket{\Psi_{I}} }
      {E_{H}-E_I-\omega}
  \right|^2 \
  \delta(E_{F}-E_I-2\omega)
  ,
\end{align}
where $E_H$ is the energy of the intermediate state $\Psi_H$.

In another experiment~\cite{DoRo-PRL-2011} also performed on Ne at the LCLS, it has been seen that it was possible to produce Ne$^{+9}$ even though it was energetically not possible to ionize Ne$^{+8}$ further with a single photon.
The ratio between the Ne$^{+9}$ and Ne$^{+8}$ productions showed a quadratic behavior in intensity pointing towards a two-photon ionization.
Theoretical investigations showed that the production of Ne$^{+9}$ is a combination of sequential and non-sequential two-photon ionization.
To fully explain the production of Ne$^{+9}$, it is necessary to include coherence properties of the x-ray pulse~\cite{SyPa-PRA-2012}.

\subsection{Tunnel Ionization}
\label{p1c2:tunnel}
For very high field strengths, the perturbative approach for describing multiphoton ionization breaks down.
In the UV and x-ray regimes, it is quite difficult to reach the non-perturbative regime ($\gamma \ll 1$) with current light sources due to the high intensity needed.
Since the Keldysh parameter goes linearly with $\omega$ at a given intensity, it is much easier to reach the non-perturbative regime with optical frequencies.
In the non-perturbative multiphoton regime or strong-field regime, it is more favorable to picture the light-matter interaction in terms of a potential that distorts the electronic system (see Fig.~\ref{fig2:ionization}).
If the deformation of the Coulomb potential is large enough, the electron can tunnel out of the system into the continuum.
This ionization process is, therefore, also known as tunnel ionization.
In the following, two prominent strong-field theories, i.e., the ADK and the SFA models, are discussed.

\subsubsection{ADK Model}
The pioneering work to describe tunnel ionization rates has been done by Landau and Lifshitz for the ground state of atomic hydrogen exposed to a static electric field~\cite{LaLi-QM-book}.
Perelomov, Popov, and Terent'ev~\cite{PPT-JETP-1966} extended the theory to any Coulomb wavefunction and to electromagnetic fields of low frequency (quasistatic limit).
A generalized version, which can also describe many-electron atoms, is the popular Ammosov-Delone-Krainov (ADK) theory~\cite{ADK-JETP-1986}.
Recently, the ADK theory has been extended to molecular systems, where additionally the orientation of the molecule influences the ionization rates~\cite{ToZh-PRA-2002}.
The molecular ADK model has become a quite popular tool for understanding strong-field ionization of molecules~\cite{Co-PhysTod-2011,ZhLi-PRA-2009}.
The popularity is also supported by the fact that numerically solving the TDSE (even within the single-active-electron approximation) for molecular systems is very challenging.

The basic idea behind the ADK theory is that the solution of a pure Coulomb problem is matched to a semiclassical solution, which describes the electron in the classically forbidden region below the barrier and in the classically allowed region outside the barrier (for a review see Ref.~\cite{BiMa-AJPhys-2004}).
It is assumed that the bound electronic state (near the nucleus) behaves under the barrier like a Coulomb wavefunction for large distances, which reads
\begin{align}
  \label{eq2:tunnel_adk}
  \psi_c(r,\Omega)
  &=
  D\, r^{Z/\kappa-1}\, e^{-\kappa\,r}\,  Y_{l,m}(\Omega)
  ,
\end{align}
where $r$ is the radius, $\Omega=(\theta,\phi)$ are the spherical angles, $Y_{l,m}(\Omega)$ are spherical harmonics~\cite{Zare-book} with the angular momentum $l$ and its projection $m$, and $Z$ is the nuclear charge.
The ionization potential is given by $I_p=\kappa^2/2$.
In the classically forbidden region under the barrier, the behavior of $\psi_c$ has to be matched to a semiclassical.
Ammosov, Delone, and Krainov~\cite{ADK-JETP-1986} derived an analytical expression to match the Coulomb wavefunction to the semiclassical solution.

Note that the angular momentum $l$ is not a conserved quantity anymore when an electric field is applied and the wavefunction does not factorize with respect to the spherical coordinates $r,\Omega$ as in the field-free case.
The solution of the Hamiltonian including the electric field (within the dipole approximation) factorizes for parabolic coordinates:\linebreak $\zeta=r+z,\ \eta=r-z$ and $\phi$.
Hence, the wavefunction reads $\psi=f_1(\zeta)\,f_2(\eta)\,e^{im\phi}/\sqrt{2\pi}$~\cite{BiMa-AJPhys-2004}.

\begin{figure}[ht!]
\begin{center}
  \includegraphics[width=.67\linewidth]{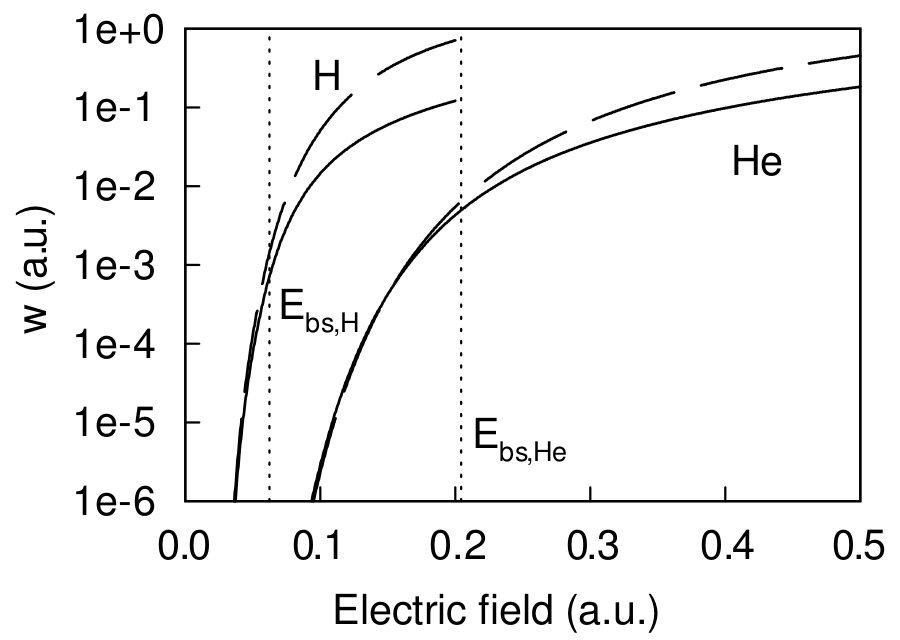}
  \caption{
    The static field ionization rates obtained with the ADK model (dashed) and numerically solving the Schr\"odinger equation (solid) are compared for H and He.
    The vertical dotted lines indicate the barrier suppression field strength for H and He beyond which ABI starts to arise.
    These figures are taken from Ref.~\cite{ScGe-PRL-1999}.
    Copyright~\copyright~1999 American Physical Society (APS).
  }
  \label{fig2:adk}
\end{center}
\end{figure}

For very strong fields, the Coulomb barrier is suppressed below the field-free binding energy.
Consequently, the electron can be ionized directly over the barrier (above-barrier-ionization or ABI) without the need for tunneling~\cite{AuSt-PRL-1989}.
In Fig.~\ref{fig2:adk} the ADK tunnel ionization rates for H and He are compared to results obtained by numerically solving the Schr\"odinger equation (see Sec.~\ref{p1c2:sae}).
Up to the field strength $E_\text{bs}$, the ADK rates are identical to the numerical results. 
Above $E_\text{bs}$ the ABI starts to kick in and the ADK model overestimates the ionization rate.

\subsubsection{SFA Model --- Path-Integral Approach}
Another approach to describe tunnel ionization is based on path-integral techniques~\cite{BeFa-JPB-2005,SaCa-Science-2001} pioneered by Keldysh, Faisal, and Reiss~\cite{Ke-JETP-1965,Fa-JPB-1973,Re-PRA-1980} (also called the KFR theory).
Since the impact of the atomic potential is neglected and only the influence of the strong laser field is considered, this approach is also referred to as the strong-field approximation (SFA)~\cite{PoBa-JMO-2008}.
The formally exact solution of the TDSE [cf. Eq.~\eqref{eq2:tdse}] reads
\begin{align}
  \label{eq2:}
  \ket{\Psi(t)}
  &=
  {\cal T}\left[
    \exp\left(-i\int_{t'}^t\!\!\!dt''\ \hat H(t'') \right)
    \right]
  \ket{\Psi(t')}
  =
  \hat U(t,t')
  \ket{\Psi(t')}
  ,
\end{align}
where $\hat U(t,t')$ is the exact time propagator, and ${\cal T}$ stands for the time ordering of the operator.
The probability of finding an electron with velocity ${\bf v}$ at time $t$ is given by $|a_{\bf v}(t)|^2$ with
\begin{align}
  \label{eq2:trans_amp}
  a_{\bf v}(t)
  &=
  -i \int_{t'}^t \!\!\! dt'' \
  \bra{\bf v(t)}    
    \hat U(t,t'') \   
    \hat H_\text{int}(t'')\
    \hat U_\text{elec}(t'',t')
  \ket{\Psi_0}
  .
\end{align}
$\hat U_\text{elec}(t'',t')$ is the field-free propagator, and $\ket{\bf v}$ describes the outgoing electron which turns asymptotically into a plane wave~\cite{IvSp-JMO-2005}.
Physically, Eq.~\eqref{eq2:trans_amp} means the electron starts in state $\Psi_0$.
At time $t''$, the electron is promoted into a continuum state where it remains till the time $t$ is reached.

In the SFA, several approximations are made to Eq.~\eqref{eq2:trans_amp}: 
(1) the final state $\ket{\bf v}$ is approximated by plane waves or Volkov states~\cite{Wo-ZPhys-1935}; 
(2) the exact propagator $\hat U$ is approximated by $\hat U_\text{SFA}$, which is the time propagator of the simplified SFA Hamiltonian\linebreak $\hat H_\text{SFA}(t) = [{\bf \hat p + \alpha A}(t)]^2/2$, where all Coulomb interactions are neglected.
The advantage of using Volkov states instead of plane waves is that Volkov states are the eigenstates of $\hat H_\text{SFA}(t)$.
Since $\hat H_\text{SFA}(t)$ does not depend on the position ${\bf \hat q}$, the {\it canonical} momentum ${\bf p}$ is a conserved quantity. 
For any time $t'$ the {\it kinetic} momentum (velocity) ${\bf v}(t')={\bf p}+\alpha {\bf A}(t')$ is given by 
\begin{align}
  \label{eq2:sfa_velocity}
  {\bf v}(t')
  &=
  {\bf v}(t)-\alpha {\bf A}(t)+\alpha {\bf A}(t')
  .
\end{align}
It is not surprising that the favorite form of the laser-matter interaction in the SFA theory is the velocity form\footnote{In the length form, a potential energy shift appears whose physical meaning is not clear~\cite{ChLe-PRA-2006,FiAu-PRA-2010}.}.

Both propagators appearing in Eq.~\eqref{eq2:trans_amp} can be analytically expressed and the resulting phase factor $S(t,t')$ reads~\cite{IvSp-JMO-2005}
\begin{align}
  \label{eq2:sfa_saddlepoint}
  S(t,t')
  &=
  \frac{1}{2}
  \int_{t'}^{t}  
    \big[ v_z(t)-\alpha\,A(t)+\alpha\,A(t'') \big]^2 
    dt''
    -
    I_p\, t'
    +
    \frac{v_\perp^2}{2}(t-t')
  ,
\end{align}
where the vector potential is linearly polarized in the $z$ direction.
The terms $v_z$ and $v_\perp$ are the velocity components along and perpendicular to the vector potential.
Note that the initial ground state energy of $\Psi_0$ is the negative ionization potential $E_0=-I_p$. 
It is used to evaluate the propagator $\hat U_\text{elec}(t'',t')$ in Eq.~\eqref{eq2:trans_amp}.

The main contribution to $a_{\bf v}(t)$ comes from the time $t'=t_0$, where $\partial_{t'}\,S(t,t')\Big|_{t'=t_0}\!\!=0$. 
For other times $t'$, the phase of $S(t,t')$ oscillates too rapidly such that these times do not contribute to the overall signal in first order~\cite{IvSp-JMO-2005}.
This approach is known as the saddle point method.
The derivative condition leads to an equation for $t_0$.
It reads
\begin{align}
  \label{eq2:sfa_simple}
  \sin^2(\omega_L\,t_0)
  +
  \gamma^2
  &=
  0
  ,
\end{align}
where $\gamma$ is the Keldysh parameter, and the final time has been conveniently set to $t=n\pi$ such that the vector potential vanishes.
Any other final time is also possible, since the result should not depend on the final time, which may be defined by the measuring process.

The electric field is taken as a plane wave with frequency $\omega_L$ and amplitude $E_0$.
Furthermore, setting $v_\perp=v_z=0$ means the initial velocity of the electron after tunneling is 0.
The time which solves Eq.~\eqref{eq2:sfa_simple} is pure imaginary $t_0=i\tau$ indicating a classically forbidden region.
For $\gamma\ll 1$, the relation $\gamma=\omega_L\,\tau$ emerges, which connects the Keldysh parameter with the so called tunneling time $\tau$~\cite{IvSp-JMO-2005}.
For $t_0=i\tau$, the phase is purely imaginary as well and reads $S=-i2/3\,I_p\,\tau$.
Hence, one finds the ionization rate 
\begin{align}
  \label{eq2:rate_tunneling}
  \Gamma \propto \left|e^{-i\,S} \right|^2
  =
  e^{2\,\text{Im}[S]} = e^{-2/3\sqrt{2I_p}^3 /E_0}
\end{align}
decreases exponentially with a higher ionization potential and a weaker electric field.
The prefactor of the ionization rate can be approximated by analytical results for a Coulomb barrier~\cite{LaLi-QM-book,DeKr-book}.

In the perturbative multiphoton regime ($\gamma\gg1$), Eq.~\eqref{eq2:sfa_simple} interestingly yields\linebreak $\tau \approx \ln(2\gamma)/\omega_L$.
Consequently, the ionization rate becomes proportional to the electric field, i.e., 
\begin{align}
  \label{eq2:rate_photoionization}
  \Gamma \propto E_0^{2N}=I^N \quad \text{with} \quad N=I_p/\omega_L
\end{align}
and $I=E^2_0$ the field intensity.
The exponent $N$ gives rise to a photon order and is a good estimate how many photons are required to ionize the system.

\begin{figure}[b!]
  \centering
  \includegraphics[width=.7\linewidth]{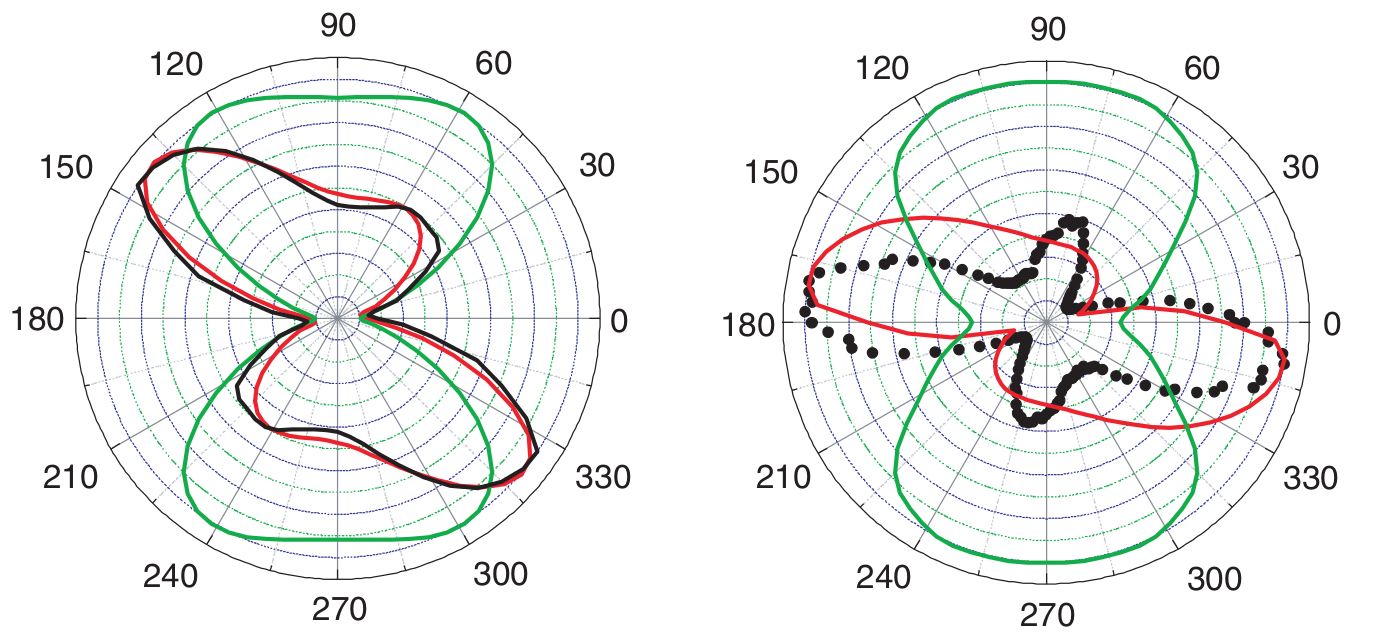}
  \caption{
    The angular momentum distributions of the photoelectron are shown for Ne (left) and Ar (right).
    The SFA model (green line), the CCSFA (red line), the TDSE results with SAE approximation (black line), and the experimental data (black dots) are compared.
    These figures are taken from Ref.~\cite{PoBa-JMO-2008}.
    Copyright~\copyright~2008 Taylor \& Francis Group.
  }
  \label{fig2:sfa_angular}
\end{figure}

Despite the success of the SFA, it is known that neglecting the long-range Coulomb potential leads to disagreement with experimental~\cite{BaBu-PRL-1988,RuZr-JPB-2005} and numerical~\cite{BaMi-JPB-PRA-2005,BaMi-JMO-2006} results in the total ionization rates and especially in the angular photoelectron distribution.
Groups have worked on including Coulomb corrections by eikonal-like approximations for the Coulomb-Volkov continuum~\cite{SmSp-JPB-2006,SmSp-PRA-2008} or by semiclassical perturbation theory~\cite{PoBa-JMO-2008}. 
The latter method is commonly called Coulomb-corrected SFA (CCSFA)~\cite{PoBa-JMO-2008,HuRo-Science-2011}. 
In Fig.~\ref{fig2:sfa_angular}, the angular photoelectron distributions are shown for the SFA (green line), the CCSFA (red line), the TDSE results with the single-active-electron (SAE) approximation (black line), and the experimental data (black dots).
The SFA does not correctly describes the angular distribution. 
The CCSFA significantly improves the results towards the SAE and experimental results.

Other widely used techniques to determine the tunnel ionization rates are numerical based  Floquet theory~\cite{Sh-PR-1965}, complex scaling~\cite{RiMe-JPB-1993,GoSa-PRA-2005}, and explicit time integration methods~\cite{Ch-PRA-2010,AbMa-PRA-2010,TaBe-PRL-2010} discussed in Sec.~\ref{p1c2:theory}.

\subsection{High Harmonic Generation}
\label{p1c2:hhg}
One of the most fundamental processes in attosecond physics is high harmonic generation (HHG) (for a review see Ref.~\cite{KoPf-AAMO-2012}).
HHG is used to generate sub-femtosecond pulses with photon energies in the EUV range from NIR femtosecond pulses.
HHG was first observed in the late 1980s in rare gas atoms~\cite{McGi-JOSAB-1987,FeHu-JPB-1988}.
Rapid developments have now made it possible to generate isolated attosecond pulses shorter than 100~as~\cite{GoSc-Science-2008,ZhZh-OptLett-2012}, and with photon energies up to the x-ray regime~\cite{PoCh-Science-2012}.
These x-ray pulses can in principle be used to generate subattosecond (zeptosecond) pulses~\cite{PoCh-CLEO-2011,XiNi-JMO-2010}.

\begin{figure}[ht!]
  \centering
  \begin{subfigure}[b]{0.5\textwidth}
    \centering
    \includegraphics[width=\linewidth]{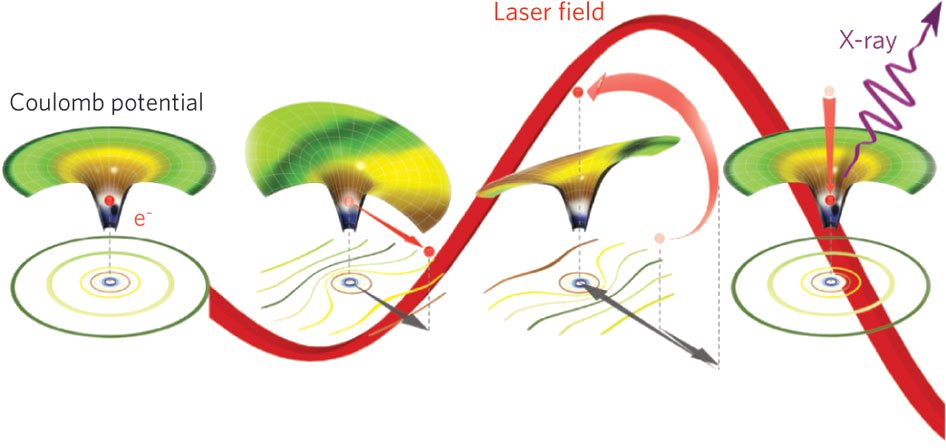}    
    \caption{Three-step model}
     \label{fig2:hhg_3step}
  \end{subfigure}
  \hspace{5mm}
  \begin{subfigure}[b]{0.39\textwidth}
    \centering
    \includegraphics[width=\linewidth]{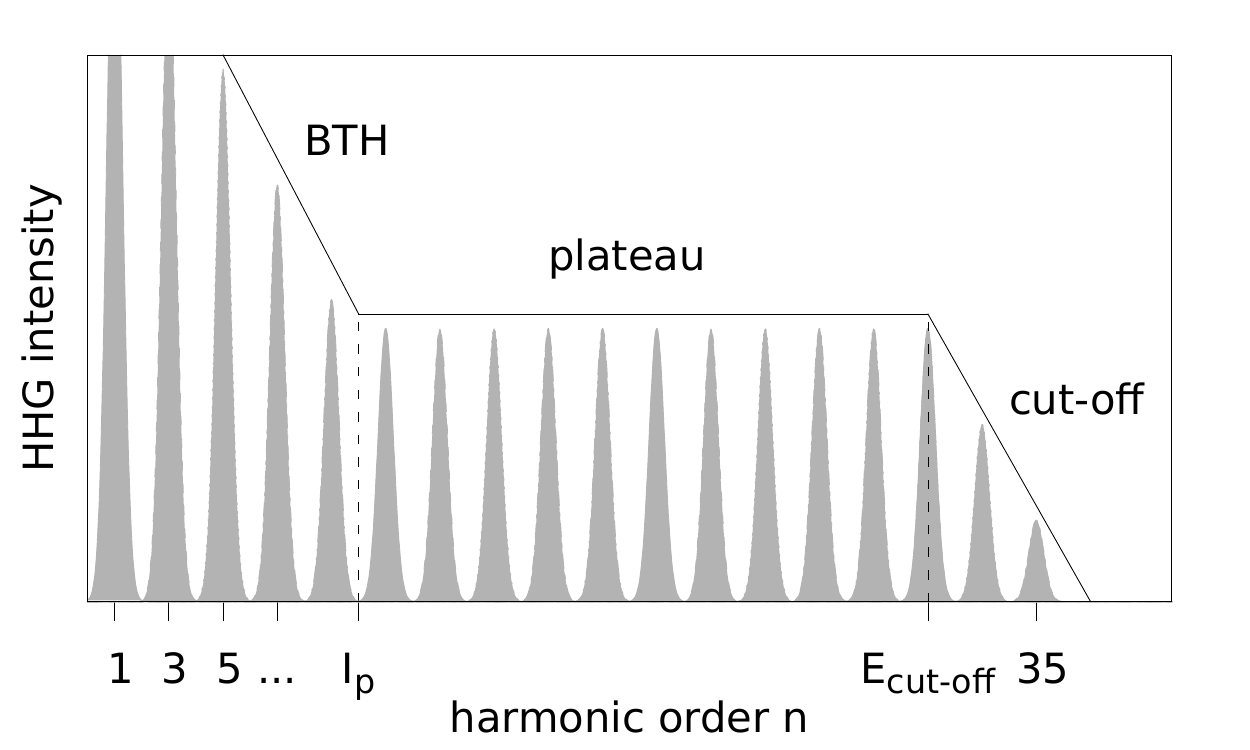} 
    \caption{HHG spectrum}
    \label{fig2:hhg_spectrum}
  \end{subfigure}
  \caption{
    (a) Illustration of the semiclassical three-step model for HHG.
    The figure is taken from Ref.~\cite{PoCh-NatPhoton-2010}.
    Copyright~\copyright~2010 Nature Publishing Group (NPG).
    (b) A sketch of a typical HHG spectrum showing the below-threshold harmonics (BTH), the plateau region, and the cut-off region.
  }
  \label{fig2:hhg}
\end{figure}

The mechanism behind HHG is well explained by a semiclassical model called the three-step model~\cite{Co-PRL-1993,ScKu-PRL-1993}.
It factorizes the HHG mechanism into three separate steps.
An illustration of the three-step process is shown in Fig.~\ref{fig2:hhg_3step}.
In the first step the outer-most electron gets tunnel-ionized by the NIR field.
In step two, the electron moves in the presence of the electric field and due to the short cycle period of the NIR pulse, the electric field drives the electron back towards the ion.
In the third step, the electron can recombine with the ion via emitting a high energy photon.
The photon energy is determined by the ionization potential $I_p$ plus the amount of energy that the electron gained in the NIR field.
The maximum emitted photon energy (commonly referred to as the cut-off energy) is given by~\cite{Co-PRL-1993}
\begin{align}
  \label{eq2:hhg-cutoff}
  E_\text{cutoff}
  &=
  I_p
  +
  3.17\ U_p
  , 
\end{align}
where $U_p=\frac{E^2}{4\omega^2}$ is the ponderomotive potential, i.e., the cycle-averaged quiver energy of a free electron in an electric field with amplitude $E$ and frequency $\omega$. 
Characteristic for HHG is the plateau region, where the harmonics extend up to the cut-off energy without decreasing in strength (see Fig.~\ref{fig2:hhg_spectrum}).

A thorough theoretical discussion of HHG based on the SFA has been given in Ref.~\cite{LeCo-PRA-1994}, which relates the analytically derived dipole moment, $\ev{x}(t)$, to the HHG spectrum.
Particularly the saddle-point approximations, which are made in the derivation, provide a physical picture of the underlying process.
Whereas in Sec.~\ref{p1c2:tunnel} the saddle-point approximation has been made only with respect to the tunneling time $\tau$ [cf. Eq.~\eqref{eq2:sfa_simple}], the saddle-point approximation of the classical action of $\ev{x}(t)$ is done in HHG with respect to three variables: (1) canonical momentum ${\bf p}$;
(2) the tunneling time $\tau$; 
and (3) the return time $t$.
As a result, three saddle-point equations appear:~\cite{LeCo-PRA-1994}
\begin{subequations}
\label{eq2:hhg_saddlepoint}
\begin{align}
  \label{eq2:hhg_saddlepoint_p}
  \partial_{\bf p} S({\bf p},t,\tau)
  &=
  {\bf x}(t) - {\bf x}(\tau)
  =
  0
  ,
\\
  \label{eq2:hhg_saddlepoint_tau}
  \partial_{\tau} S({\bf p},t,\tau)
  &=
  \frac{[{\bf p} + \alpha\, {\bf A}(t-\tau)]^2}{2}
  +
  I_p
  =
  0
  ,
\\
  \label{eq2:hhg_saddlepoint_t}
  \partial_{t} S({\bf p},t,\tau)
  &=
  \frac{[{\bf p} + \alpha\, {\bf A}(t)]^2}{2}
  +
  \frac{[{\bf p} + \alpha\, {\bf A}(t-\tau)]^2}{2}
  =
  2N + 1
  ,
\end{align}
\end{subequations}
where $S$ is the classical action, and $N$ is a positive integer number.
Each of these equations has a physical implication.
Equation~\eqref{eq2:hhg_saddlepoint_p} states that the returning electron recombines exactly at the same position where the electron has been initially ionized.
Equation~\eqref{eq2:hhg_saddlepoint_tau} defines the tunnel time $\tau$ [cf. Eq.~\eqref{eq2:sfa_simple}].
Only imaginary $\tau$ can fulfill Eq.~\eqref{eq2:hhg_saddlepoint_tau}. 
Equation~\eqref{eq2:hhg_saddlepoint_t} in combination with Equation~\eqref{eq2:hhg_saddlepoint_tau} relates the photon energy of the emitted photon to the $(2N+1)$-th harmonic of the driving frequency $\omega$.

HHG has also become very popular for molecular systems (for reviews see Refs.~\cite{LiLu-JPhysB-2010,HaCa-JPhysB-2011,KoPf-AAMO-2012}).
For molecular HHG spectra, the non-spherical Coulomb potential can no longer be ignored.
It has been shown that SFA calculations miss several important features in the HHG spectrum~\cite{AuMo-JMO-2011,AuFa-MPLB-2012}.
The quantitative rescattering theory (QRT)~\cite{LeLi-PRA-2009} has become a very prominent way to calculate HHG spectra for molecules.
It relates the recombination step to inverse photoionization.
Hence, in QRT the recombination cross section gets replaced with the well-studied photoionization cross section obtained from high-level calculations or experiments (see Sec.~\ref{p1c2:photo}).

For time-dependent propagation methods as presented in Sec.~\ref{p1c2:theory}, the Larmor formula has to be used to calculate HHG spectra.
The Larmor formula connects the acceleration of a charged particle, $\ev{\hat {\bf a_c}}(t)$, to the emitted radiation, $S(\omega)$~\cite{Jackson-book}.
Just as the photoionization cross section can be described with a range of equivalent expressions [see discussion of Eq.~\eqref{eq2:cs_vel}], so too are there several equivalent ways to express the HHG spectrum,
\begin{align}
  \label{eq2:hhg_spectrum}
  S(\omega)
  &\propto
  \left|
    \int_{-\infty}^\infty \hskip-2ex dt \
    e^{-i \omega t}
    \ev{ \hat {\bf a}_c}(t)
  \right|^2
  =
  \left|
    \int_{-\infty}^\infty \hskip-2ex dt \
    e^{-i \omega t}
    \left[\big. \partial_t \ev{\hat {\bf v}}(t) \right]
  \right|^2
  =
  \left|
    \int_{-\infty}^\infty \hskip-2ex dt \
    e^{-i \omega t}
    \left[\big. \partial^2_t \ev{\hat {\bf r}}(t) \right]
  \right|^2
  .
\end{align}
The acceleration [$\ev{\hat {\bf a}_c}(t)$], the velocity [$\ev{\hat {\bf v}}(t)$], and the length [$\ev{\hat {\bf r}}(t)$] forms are shown in Eq.~\eqref{eq2:hhg_spectrum}.
As in the case of the photoionization cross section, all three expressions are only equivalent when the exact wavefunction is known. 
For atomic hydrogen, the equalities of the expressions in Eq.~\eqref{eq2:hhg_spectrum} have been confirmed numerically~\cite{HaMa-PRA-2010}.
For multi-electron systems, approximations have to be made to the wavefunction and/or the Hamiltonian.
Hence, the equalities in Eq.~\eqref{eq2:hhg_spectrum} do not strictly hold anymore.
Results based on the TDCIS method (see Sec.~\ref{p1c2:tdcis}) have shown that for larger atoms the discrepancy in the HHG spectrum increases between the length and the acceleration forms.
This indicates that the approximations made in TDCIS (and even more for the SFA and SAE models) become less valid for heavier atoms.
For the two lightest noble gas atoms helium and neon, both ways of calculating the HHG yield identical results.

\subsubsection{Extending the Cut-Off Energy \& Phase Matching}

The attribute ``high'' in HHG is justified, since the harmonic order $n$ can easily be a triple-digit number ($n \geq 100$). 
In a very recent experiment, harmonic orders of $n>5000$ have been achieved~\cite{PoCh-Science-2012}.
Interestingly, Eq.~\eqref{eq2:hhg-cutoff} states that higher cut-off energies can be achieved with lower driving frequencies (longer wavelengths).
The maximum harmonic order achievable goes with $\omega^{-3}$.
Long driving wavelengths have been proven to be a successful way to increase the cut-off energy into the water window ($\approx 300--500$~eV)~\cite{ChAr-PRL-2010} and even up into hard x-ray regime above 1~keV~\cite{PoCh-Science-2012}.
Equation~\eqref{eq2:hhg-cutoff} shows that increasing the wavelength is not the only way to increase the cut-off energy.
Higher electric field strengths (or intensities) als move the cut-off energy into the keV regime~\cite{SeSe-APL-2006}.

\begin{figure}[b!]
  \centering
  \includegraphics[width=.95\linewidth]{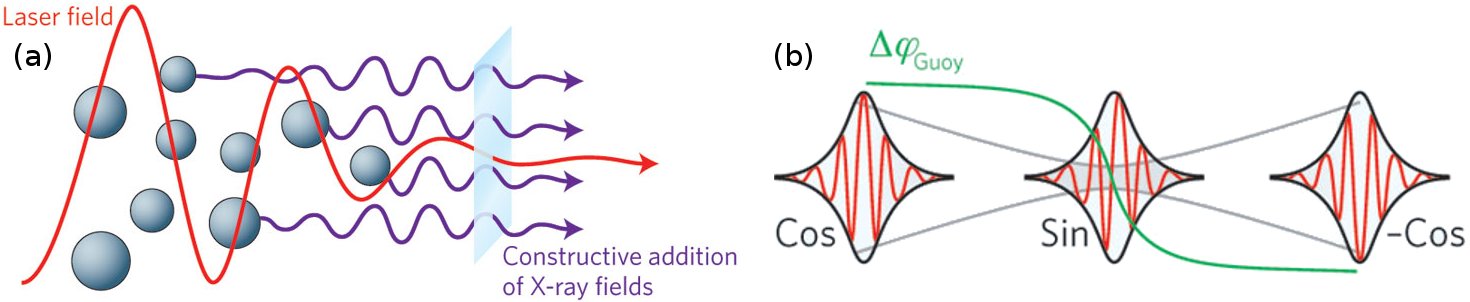}
  \caption{
    (a) The concept of phase matching and the coherent build up of the macroscopic HHG signal are illustrated.
    (b) The geometrical Guoy phase is shown. 
    It changes particularly around the focal point of the NIR driving field.
    The Guoy phase has to be considered to ensure phase matching.
    The figure is taken from Ref.~\cite{PoCh-NatPhoton-2010}.
    Copyright~\copyright~2010 Nature Publishing Group (NPG).
  }
  \label{fig2:hhg_macro}
\end{figure}

Both approaches for extending the cut-off energy (using longer wavelengths or higher intensities) have drawbacks.
High intensities automatically lead to higher ionization probabilities. 
Highly ionized media, however, are not wanted.
They result in strong dispersion effects.
As a consequence, the HHG spectrum significantly changes while traveling through the medium.
Dispersion effects need to be minimized to ensure the HHG light is always in phase with the driving field such that the maximum gain in the HHG yield can be achieved (see Fig.~\ref{fig2:hhg_macro}).
If the HHG light is out of phase with the driving field, the HHG light produced from different atoms no longer adds up constructively~\cite{LoVa-PRL-2005,StPf-PRA-2006}.

The ionization potential is another knob that can be turned (by targeting the appropriate atom) to increase the cut-off energy.
Large ionization potentials are favorable not just because of Eq.~\eqref{eq2:hhg-cutoff}.
They reduce the ionization probability and make it, therefore, easier to ensure phase-matching.
Hence, the most ideal atoms are noble gas atoms, particularly He~\cite{PoCh-PNAS-2009}.

Locking of the phase velocities of the HHG field to that of the driving field is known as phase-matching (for a review see~\cite{PaGi-QuanElec-2006}).
In Fig.~\ref{fig2:hhg_macro}(a) the idea of phase matching is illustrated.
Figure~\ref{fig2:hhg_macro}(b) shows the Guoy phase~\cite{Gouy-phase-1890}, which changes mostly around the focus point of the laser field and arises due to geometrical reasons~\cite{FeWi-OptLett-2001}.
The correct Guoy phase is important for the HHG spectrum.
This has been shown for Xe~\cite{VoNe-NJP-2011} where the overall HHG spectrum changes dramatically depending on the geometrical Guoy phase.
The Guoy phase is controlled by the position of the laser focus relative to the gas medium (see Fig.~\ref{fig2:hhg_macro}).

Phase-matching is the main bottleneck for generating broad HHG spectra ranging into the x-ray regime~\cite{RuDu-Science-1998,KoPf-AAMO-2012}, since it is difficult to ensure phase matching over such a large energy range.
Long wavelengths avoid high ionization rates but suffer low conversion efficiencies, which scale microscopically with $\lambda^{-6.5}$~\cite{TaAu-PRL-2007,ShIs-PRL-2007,FoMa-PRL-2008,ShTr-PRL-2009}.
%

Several novel ideas~\cite{KoWi-PRL-2007,HoLi-JOSAB-2008,Is-PRL-2003,Fl-PRA-2008,ZhWu-PRA-2009,BuKo-OptLett-2011,BuHe-arixv-2012}, which go beyond a single active electron~\cite{KoWi-PRL-2007} and a single driving frequency~\cite{HoLi-JOSAB-2008,Is-PRL-2003,Fl-PRA-2008,ZhWu-PRA-2009,BuKo-OptLett-2011,BuHe-arixv-2012}, have been proposed to extend the HHG energy cut-off.
Unfortunately, the conversion efficiency is quite reduced compared to conventional HHG.
Instead of ionizing just one electron, an idea was put forward~\cite{KoWi-PRL-2007} to ionize two electrons and let them recombine simultaneously.
This leads to a second plateau region in the HHG spectrum.
The efficiency is strongly reduced by up to 12 orders of magnitude.
Multi-color driving fields~\cite{HoLi-JOSAB-2008} have been explored as well as combinations of an intense NIR pulse with an assisting non-resonant~\cite{Is-PRL-2003,Fl-PRA-2008,ZhWu-PRA-2009} or resonant~\cite{BuKo-OptLett-2011,BuHe-arixv-2012} UV/x-ray pulse.

\subsubsection{Isolated Attosecond Pulses}
The discrete harmonic peaks in the HHG spectrum (as in Fig.~\ref{fig2:hhg}) lead to attosecond pulse trains rather than an isolated pulses.
To generate isolated attosecond pulses~\cite{BaUd-Nature-2003}, a frequency filter is used (often a thin metal foil) to filter out a continuous energy window of the HHG spectrum in the cut-off region~\cite{YaIv-OptExp-2007}. 
Phase stability over the entire spectrum is important for generating pulses with a finite duration.
Having control over the phase of each single frequency allows one to create pulses with arbitrary pulse shapes~\cite{WiGo-Science-2011}.
For Fourier-limited pulses, the pulse duration is inversely proportional to the bandwidth of the HHG spectrum, i.e., $\tau=4\ln(2)/(d\omega)$, where $\tau$ and $d\omega$ are the FWHM-duration and FWHM-bandwidth, respectively, of the intensity profile.
Hence, subfemtosecond pulses require bandwidths of a few eV whereas subattosecond pulses require bandwidths reaching into the keV regime~\cite{PoCh-CLEO-2011}.

\begin{figure}[ht!]
  \centering
  \includegraphics[width=.75\linewidth]{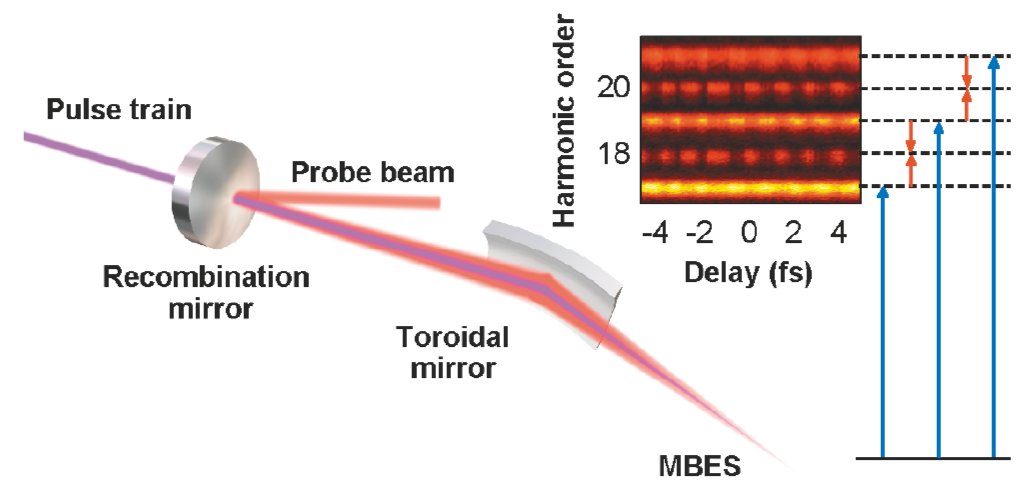}
  \caption{
    Illustration of a RABITT measurement.
    The overlap of HHG and driving fields leads to two ambiguous ionization pathways (right figure) 
    The resulting interferences in the electron energy as a function of the time delay (upper right figure) contains information about the relative phases between neighboring harmonics in the HHG spectrum. 
    The figure is taken from Ref.~\cite{LoVa-PRL-2005}.
    Copyright~\copyright~2005 American Physical Society (APS).
  }
  \label{fig2:rabitt}
\end{figure}

To experimentally prove that the generated HHG spectrum really leads to attosecond pulses, it is necessary to measure phase and amplitude of the frequency components of the HHG spectrum.
There exist two main techniques to characterize attosecond pulses: 
\begin{enumerate}
 \item Reconstruction of attosecond beating by interference of two-photon
transition \\ (RABITT)~\cite{PaTo-Science-2001,MaBo-Science-2003,LoVa-PRL-2005}, 
  \item Frequency-resolved optical gating (FROG)~\cite{KaTr-OptLett-1993-FROG,MaQu-PRA-2005,SaBe-Science-2006}. 
\end{enumerate}
The pulse durations obtained by both methods agree within 10\%~\cite{KiKa-arxiv-2007}.
RABITT can only be used to determine the pulse duration of attosecond pulses in pulse trains (not isolated attosecond pulses).
RABITT (see Fig.~\ref{fig2:rabitt}) is based on interferences of two-photon-processes, which lead to photoelectron energies corresponding to even harmonics of the driving frequency $\omega$.
This is done by overlapping the attosecond pulse train with the driving NIR field.
The interference appears due to two ionization pathways: (1) absorbing an UV photon and an NIR photon [$(2n-1)\omega + \omega$], and (2) absorbing an UV photon and emission of an NIR photon [$(2n+1)\omega - \omega$].
The relative phase between the harmonics $2n\pm1$ can be measured by changing the time delay between the attosecond pulse train and the NIR driving field.

FROG is an auto-correlation technique. 
It is based on an iterative method to reconstruct the amplitude and the phase of the electric field~\cite{KaTr-OptLett-1993-FROG}. 
FROG has been widely used for optical fields.
A modified version of FROG called FROG-CRAB (Frequency-resolved optical gating for complete reconstruction of attosecond bursts)~\cite{MaQu-PRA-2005,QuMa-JMO-2005} can characterize isolated attosecond pulses, and it is nowadays widely used experimentally~\cite{ThBa-OptExp2009,FeCa-NatPhoton-2010}.
FROG-CRAB measures the FROG trace between a low-frequency field and the isolated attosecond pulse.
To do so, the attosecond UV pulse ionizes the system with a one-photon step and the low-frequency field dresses the ionized electron in the continuum (like attosecond streaking discussed in Sec.~\ref{p1c2:streaking}).
The FROG trace is, then, given by the spectrum of the ionized electron $|a({\bf v},\tau)|^2$, where $a({\bf v},\tau)$ is the transition amplitude to a final continuum state with velocity ${\bf v}$ [cf. Eq.~\eqref{eq2:trans_amp}].
The term $\tau$ is the time delay between the two pulses. 
The low-frequency, dressing field is used like a temporal phase gate $e^{i\phi(\tau)}	$, where $\phi(\tau)$ is the accumulated phase of the ionized electron in the dressing field.

\subsubsection{Probing Electronic Structure and Dynamics}
\begin{figure}[b]
  \centering
  \includegraphics[width=.57\linewidth]{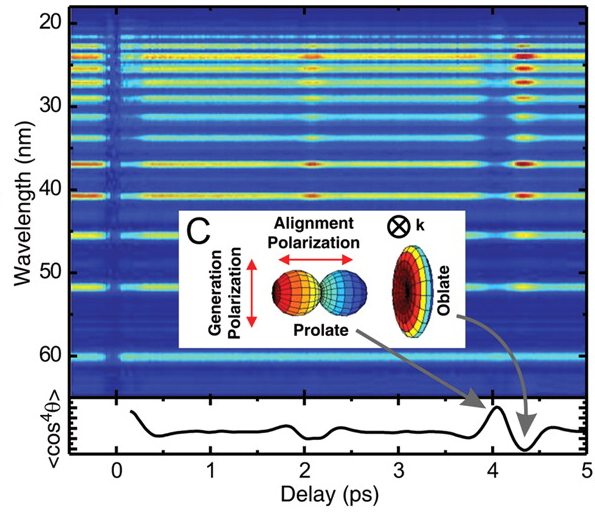}
  \caption{
    HHG spectrum of SO$_2$ is shown as function of pump-probe delay.
    With the help of impulsive alignment, HHG spectra from differently aligned molecules can be obtained.
    Particularly at pump-probe delays where the alignment is non-isotropic, new feature appear in the HHG spectrum.
    The figure is taken from Ref.~\cite{McGu-Science-2008}.
    Copyright~\copyright~2008 American Association for the Advancement of Science (AAAS).
  }
  \label{fig2:hhg_multiorbital}
\end{figure}

Up to now, HHG has been discussed as a tool to generate attosecond UV/x-ray pulses used for studying ultrafast processes.
The popularity of HHG, however, does not stop here.
HHG has become more and more a scientific tool in itself to study electronic structure and dynamics of atoms and molecules~\cite{HaCa-JPhysB-2011}.
Particularly popular is tomographic imaging of the outer-most molecular orbitals (HOMOs)~\cite{ItLe-Nature432-2004}.
The idea behind this tomographic approach is that the HHG spectrum contains structural information of the orbitals.
The HHG spectrum depends on the recombination matrix elements $\bra{\varphi_i}\hat{\bf r}\ket{\varphi_e}$, where $\varphi_i$ is the hole wavefunction, $\varphi_e$ is the wavefunction of the returning electron.
Assuming that the returning electron can be approximated as a plane wave ($\varphi_e({\bf r}) \approx e^{i{\bf k}\cdot{\bf r}}$) the HHG spectrum contains spatial Fourier components of $\varphi_i$, which can be used to reconstruct the hole orbital~\cite{ItLe-Nature432-2004,VoNe-NatPhys-2011}.

It is even possible to apply this technique to HHG spectra with multiple orbital contributions~\cite{McGu-Science-2008}.
Here the ability to laser-align molecules (see Sec.~\ref{p1c1}) is crucial, since tunnel ionization and recombination are highly angle-dependent processes.
In Fig.~\ref{fig2:hhg_multiorbital}, the HHG spectrum for laser-aligned SO$_2$ is shown.
At pump-probe delays where the molecular alignment is non-isotropic (alignment or anti-alignment), new features appear in the HHG spectrum.
Depending on whether molecules are aligned or anti-aligned, certain harmonics are suppressed or enhanced, indicating multi-orbital contributions. 

Normally it is assumed that the HHG electron only comes from the most weakly bound (outer-most) orbital.
This is not true anymore when the ionization potentials of neighboring states are quite similar.
Consequently, the hole wavefunction of the ion is not stationary anymore, and can be written as a superposition of several ionic states.
Correlation effects within the ion can also lead to additional hole motions~\cite{CdDo_AdvChemPhys65,KuLu-JPCA-2010}.
In SO$_2$ up to 3 orbitals (i.e., HOMO, HOMO-1, and HOMO-2) can contribute~\cite{SmIv-Nature-2009} to the HHG signal depending on the alignment of the molecule with respect to the strong field polarization direction.
Ionizing along the molecular axis can lead to a coherent superposition of HOMO and HOMO-2.
It has been even demonstrated that it is possible to retrieve the relative phase between the two orbitals at the time of tunnel ionization~\cite{SmIv-Nature-2009}.
However, other experimental~\cite{VoNe-NatPhys-2011} and theoretical~\cite{LiLu-JPhysB-2010} studies seem to suggest that there is no strong HOMO-2 contribution in the HHG spectrum and that the HOMO orbital is sufficient to explain the HHG spectrum.

\begin{figure}[b]
  \centering
  \begin{subfigure}[b]{0.47\textwidth}
    \centering
    \includegraphics[width=\linewidth,height=45mm]{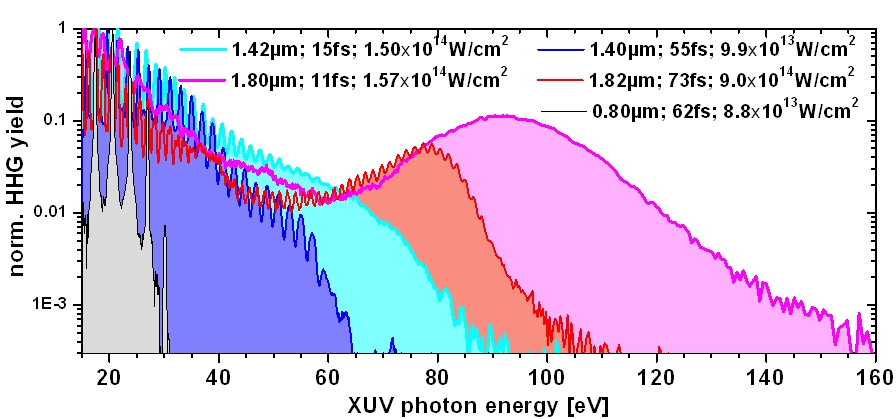}
    \caption{Experimental spectrum}
     \label{fig2:hhg_xenon_exp}
  \end{subfigure}
  \hspace{1mm}
  \begin{subfigure}[b]{0.47\textwidth}
    \centering
    \includegraphics[width=\linewidth]{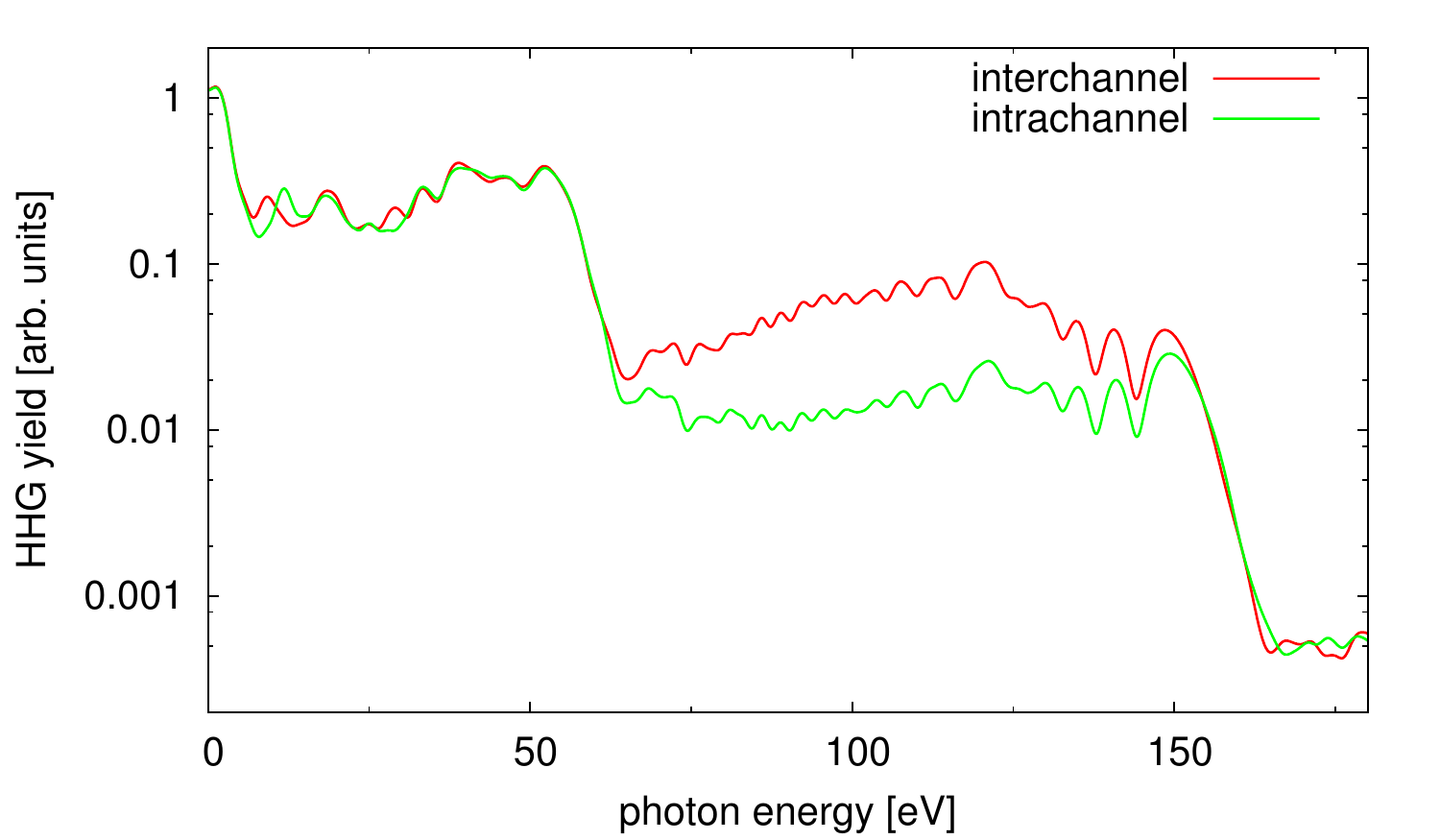}
    \caption{Theoretical spectrum}
    \label{fig2:hhg_xenon_theo}
  \end{subfigure}
  \caption{
    (a) The experimentally obtained HHG spectrum of Xe is shown for different driving pulses leading to different energy cut-offs.
    The figure is taken from Ref.~\cite{ScSh-JPB-2012}.
    Copyright~\copyright~2012 IOP Publishing.
    (b) The theoretical spectrum obtained with a TDCIS approach is shown with (red line) and without (green line) interchannel coupling.
  }
  \label{fig2:hhg_xenon}
\end{figure}

Multi-orbital contributions can also play a role in atoms~\cite{WoVi-PRL-2009,ShVi-NatPhys-2011,ZhGu-PRL-2013}, but they are less prominent due to higher symmetries and larger energy differences between the orbitals.
For larger atoms, multi-electron effects start to emerge.
A very prominent example is the giant dipole resonance in Xe as discussed in Sec.~\ref{p1c2:photo} in terms of the photoionization cross section.
The giant dipole resonance (originating from the $4d$ subshell) also affects the photoionization cross sections of the $5s$ and $5p$ subshells due to interchannel coupling.
As a result, signatures of the giant dipole resonance appear in the recombination step ($\varepsilon l \rightarrow 5p$) and, therefore, also in the HHG spectrum, provided the cut-off energy is large enough to reveal this feature ($E_\text{cutoff}>100$~eV).
This has been confirmed experimentally for the first time in Ref.~\cite{ShVi-NatPhys-2011}.
Figure~\ref{fig2:hhg_xenon_exp} shows the experimental HHG spectrum for different cut-off energies. 
The giant dipole resonance starts to appear with increasing $E_\text{cutoff}$.
Results from theoretical {\it ab initio} calculations based on TDCIS (see Sec.~\ref{p1c2:tdcis}) are shown in Fig.~\ref{fig2:hhg_xenon_theo}.
No giant dipole resonance can be seen in the HHG spectrum when interchannel interactions are ignored.

Also the Cooper minimum in Argon, which has been discussed in Sec.~\ref{p1c2:photo} in terms of the photoionization cross section, can be seen in the HHG spectrum~\cite{WoVi-PRL-2009,FaSc-PRA-2011,HiFa-PRA-2011,PaGr-PRA-2012}.
In Sec.~\ref{p1c2:apps_hhg} the influence of multiple orbitals on the location and shape of the Cooper minimum is discussed. 
Analogous to atoms, the Cooper minimum does also exist in molecules~\cite{CaKr-JCP-1982}, and, therefore, also in the HHG spectrum of molecules.
This has been recently experimentally demonstrated on N$_2$\cite{BeWo-PRL-2012} and CS$_2$~\cite{WoLe-PRL-2013}.

\subsection{Attosecond Streaking}
\label{p1c2:streaking}
After the discussion of photoionization and tunnel ionization---the two most important processes in attosecond physics---and how they are involved in generating attosecond pulses, the focus now shifts to applications and prominent attosecond techniques involving both types of ionization processes.
Pump-probe experiments are an ideal tool to study fundamental physical mechanisms in a time-resolved fashion~\cite{KrIv-RMP-2009}.
One popular pump-probe experiment is streaking~\cite{ItQu-PRL-2002} and has been used for atomic~\cite{DrKr-Nature-2002}, molecular~\cite{BaMa-PRL-2010}, and solid state~\cite{CaMu-Nature449-2007} systems.
As discussed in Sec.~\ref{p1c2:tunnel}, the exact eigenstates of a free electron in the presence of an electric field are the Volkov states~\cite{Wo-ZPhys-1935}.
The instantaneous velocity ${\bf v}(t)$ at time $t$ can be directly connected to the instantaneous velocity ${\bf v}(t')$  at any other time $t'$ [see Eq.~\eqref{eq2:sfa_velocity}].
Due to the importance of Eq.~\eqref{eq2:sfa_velocity} for streaking, it is stated again: ${\bf v}(t)={\bf v}(t') + \alpha\,{\bf A}(t') - \alpha\,{\bf A}(t)$.

\begin{figure}[t!]
  \centering
  \includegraphics[width=.53\linewidth]{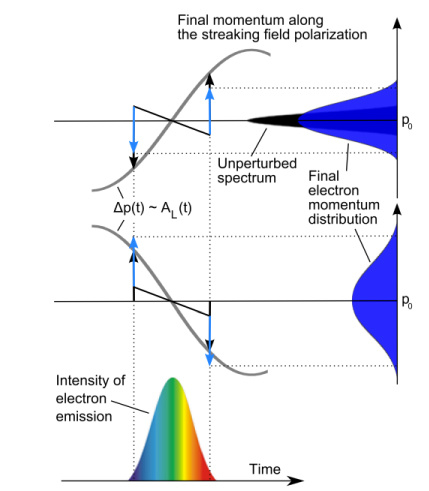}
  \caption{
    Illustration of a streaking measurement with a chirped electron wavepacket.
    Depending on the vector potential, the final electron momentum distribution is broadened or narrowed.
    The figure is taken from Ref.~\cite{HoSc-OptExp-2011}.
    Copyright~\copyright~2011 Optical Society of America.
  }
  \label{fig2:streaking}
\end{figure}

When the electron hits the detector and its energy gets measured, no electric field is present such that ${\bf A}(t')=0$.
Additionally, when the initial velocity ${\bf v}(t)$ is known, it is possible to extract uniquely the time $t$ from Eq.~\eqref{eq2:sfa_velocity} for an appropriate shape of the streaking pulse.
``Appropriate'' means in this context that the vector potential ${\bf A}(t)$ changes over the period of interest such that ${\bf A}(t)$ can be mapped one-to-one to the time $t$.
In Fig.~\ref{fig2:streaking}, the principle idea of a streaking experiment is illustrated.
Due to the time-dependent vector potential, the temporal structure of the electron wavepacket can be uniquely mapped onto the kinetic energy of the electron, which can be measured quite precisely.
This makes it possible to measure the chirp of an electronic wavepacket as illustrated in Fig.~\ref{fig2:streaking}.
Depending on the energy resolution of the detector and the gradient of the vector potential $\partial_t {\bf A}(t)$, a time resolution on the attosecond scale can be easily achieved~\cite{UiKr-Nature-2007,FrWi-NatPho-2009,ScYa-Science-2010} and even zeptosecond resolution is feasible~\cite{KoWo-OptExp-2011}.

Streaking can be also used to determine the shape of the streaking pulse itself~\cite{GaGo-AppPhysB-2008} or the shape of the photoionizing pulse~\cite{ItQu-PRL-2002,HoSc-OptExp-2011} (for a review see Ref.~\cite{Fr-JPB-2011}).
When the streaking pulse is well-known, photoionization directly connects the substructure of the electron momentum distribution to information about the chirp and pulse duration of the ionizing pulse.
The shift of the center of the electron momentum distribution ($p_0$ in Fig.~\ref{fig2:streaking}) is a direct measure of the vector potential.
The entire vector potential ${\bf A}(t)$ can be measured by varying the pump-probe delay.
The streaking pulse can vary from terahetz (THz)~\cite{FrWi-NatPho-2009} to IR frequencies~\cite{GaGo-AppPhysB-2008}.
It is important that the jitter (i.e, fluctuation in the pump-probe delay) is small compared to the wavelength of the streaking field.

\subsubsection{Studying Ultrafast Electron Motion}
One of the first demonstrations of streaking has been done on atomic krypton~\cite{DrKr-Nature-2002}.
After photoionizing the inner-shell $3d$ electron, the $3d^{-1}$ krypton ion decays further to Kr$^{2+}$ via the prominent MNN-Auger decay~\cite{AkAk-PRA-1984,JuKi-RPA-2001}.
The time at which each single Auger decay event happened was measured by streaking the Auger electron.
From the resulting statistics, which show an exponential decay behavior ($\propto e^{-t/\tau_h}$), the lifetime of the $3d$ hole was found to be $\tau_h=7.9\pm1.0$~fs. 
The corresponding linewidth of $\Gamma=84\pm10$~meV agrees well with the $88\pm4$~meV energy-domain measurements~\cite{JuKi-RPA-2001}.
The pump pulse, which defines the reference $t=0$, must be short in comparison to the Auger decay; hence, it needs to be on the subfemtosecond scale.

Streaking measurements make it possible to study very fundamental questions about the dynamics of electrons that could not have been addressed before.
For instance, when several subshells of an system (e.g., atom) can be directly ionized: Does the ionization happen in all subshells at the same time?
Streaking measurements on neon showed that the $2p$ electron is ionized 21~as later than the more tightly bound $2s$ electron~\cite{ScYa-Science-2010}.
This finding sparked a wave of theoretical investigations~\cite{KhIv-PRL-2010,MoLy-PRA-2011,KlDa-PRL-2011,NaPa-PRA-2012,IvKh-PRA-2012,IvSm-PRL-2011} not just in neon but also in other noble gas atoms~\cite{KlDa-PRL-2011}.
Time delays between different ionization channels have been measured in other atoms as well~\cite{NaPa-JPB-2011,GuKl-PRA-2012}.
Most theoretical studies~\cite{KhIv-PRL-2010,MoLy-PRA-2011,NaPa-PRA-2012} predict a time delay of no more than $\approx10$~as---half the experimental value.
Some studies point towards an additional measurement-induced time delay originating from the IR streaking pulse~\cite{IvSm-PRL-2011}.
For a review about attosecond delays in photoionization see Ref.~\cite{DaHu-JPB-2012}.

With the high time resolution of streaking experiments it is possible to address whether tunneling takes a finite amount of time (i.e., tunneling time) or happens instantaneously~\cite{HaSt-RMP-1989,LaMa-RMP-1994}.
In the classically forbidden tunneling limit, the tunneling time is purely imaginary indicating that tunneling is an instantaneous process in real time.
A recent experiment~\cite{EcPf-Science-2008} on helium strengthens this statement by setting the upper limit for the tunneling time at 12~as due to experimental uncertainties.
Experimentally it is quite difficult to define an initial time just before the electron tunnels  and a final time when the electron appears outside the barrier.

The authors of Ref.~\cite{EcPf-Science-2008} were able to do so by using elliptically polarized light.
The maximum of the electric field defines the initial time of tunneling.
The angular kick the electron experienced due to the elliptically polarized light determines the time the electron appears in the continuum~\cite{PfCi-NatPhys-2012}.
The difference between these two times yields the tunneling time.
In contrast to previous examples, the angular rather than the radial vector potential component is exploited.
Another modification to common streaking experiments is that only one pulse has been used, which operates as pump and as probe.
A separate pump pulse is not needed due to the highly non-linear tunnel ionization rate, which naturally selects time ``zero'' to be the time when the electric field reaches its maximum.

\subsection{Attosecond Transient Absorption Spectroscopy}
\label{p1c2:tas}
In the previous section, the strong-field pulse has been used to probe the ionized electron.
In the following, another pump--probe setup is explored, where the longer pulse is used as pump and the shorter (UV attosecond) pulse is used as probe.
The longer pulse could be an NIR pulse, which tunnel-ionizes the system.
The shorter UV attosecond pulse probes the system via 1-photon absorption.
However, instead of probing the outgoing electron, as done in streaking, the attosecond pulse is used to probe the ionized system that is left behind.
The discrete electronic excitation structure of an ion makes it possible to use resonances, which do not exist in a continuum-continuum transition.
Resonant transitions have the advantage of possessing a strongly enhanced transition probability and being energetically localized.
Both aspects make resonant transition features useful markers.

Since the electron is not probed, one does not look at electronic momentum distributions but rather at the transmitted/absorbed probe signal.
Therefore, this technique is called (attosecond) transient absorption spectroscopy~\cite{GoKr-Nature-2010}.
Transient absorption spectroscopy was not invented by the attosecond community, it has been already used on the femtosecond scale for probing chemical~\cite{DaRo-JCP-1987,AsKl-AnChem-2003} and solid state systems~\cite{KlMc-PRL-1998}.
However, it has been recently extended into the attosecond regime~\cite{GoKr-Nature-2010,WiGo-Science-2011,SaYa-PRA-2011}.
The time resolution arises by systematically changing the pump-probe delay $\tau$.
The quality of the time resolution is determined on the one hand by the duration of the probe pulse and on the other hand by the jitter in the pump-probe delay.

The rapid progress in attosecond technology~\cite{BrKr-RMP-2000,KrIv-RMP-2009}, particularly, in controlling the phases of single frequency components in ultrashort pulses~\cite{WiGo-Science-2011} decreased the jitter uncertainty to a few attoseconds.
Beside the high attosecond resolution, there is a nice side effect that comes along with attosecond pulses.
Due to their short durations, they provide a broad spectrum.
With a broad spectrum, several ionic excitations can be accessed with a single pulse.
This has the advantages that phase relations between different states can be probed.
At first glance, the high temporal and the high spectral resolutions seem to contradict each other due to Heisenberg's uncertainty principle.
This is, however, not the case.
The high temporal confinement, due to the pulse duration, results in an excited electronic state which is not well-defined in energy.
This has, however, no consequences for the high spectral resolution of the transmitted attosecond pulse at the detector.
In a more technical language, the high temporal and the high spectral resolutions are not conjugated variables to each other and, therefore, can be independently measured from each other.
\begin{figure}[t!]
  \centering
  \includegraphics[width=.6\linewidth]{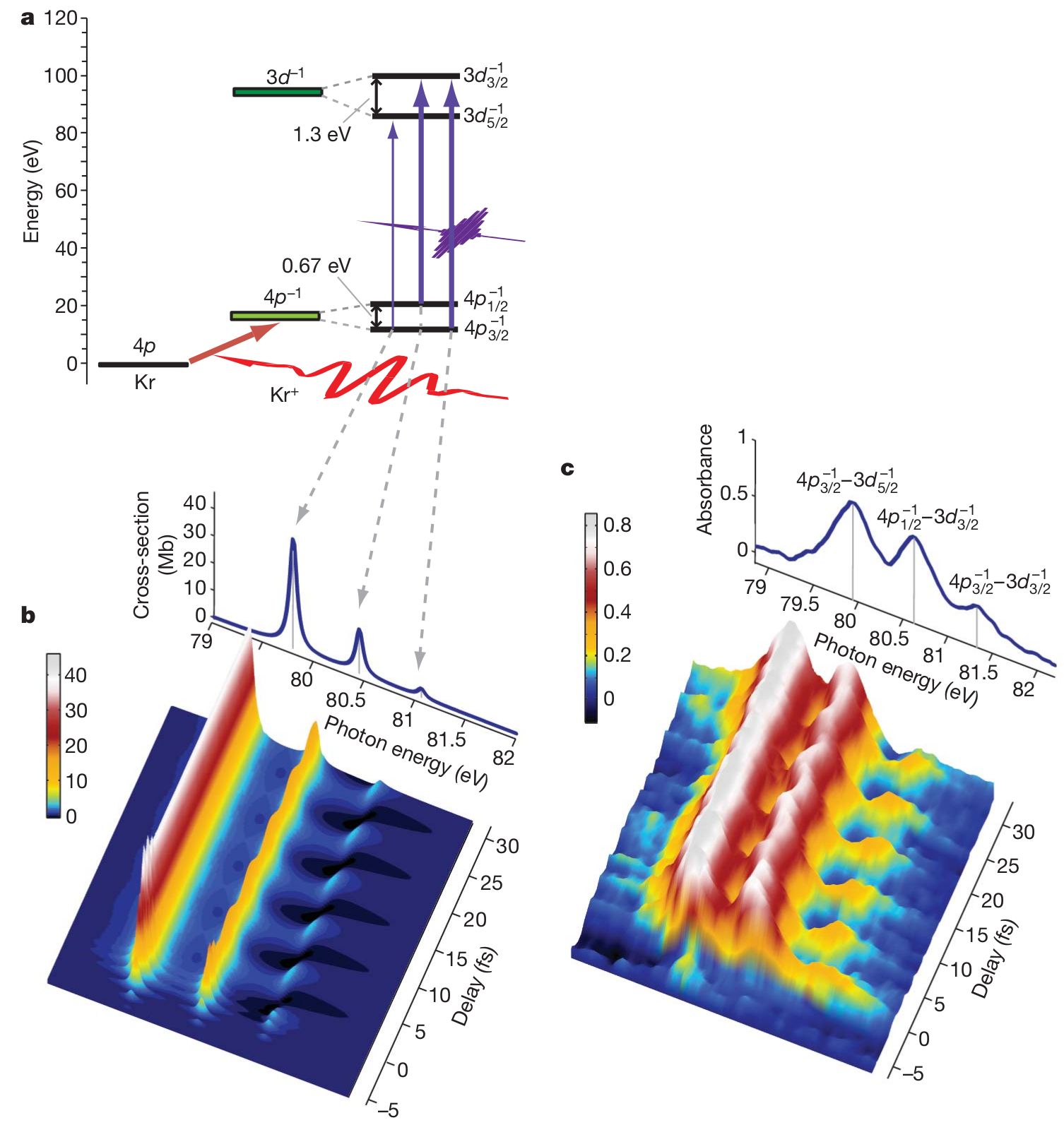}
  \caption{
    (a) Illustration of the attosecond transient absorption experiment with atomic krypton.
    The NIR pump pulse tunnel ionizes the atom and the EUV attosecond pulse probes the ion by resonantly exciting the ion further.
    The transition lines of the transient absorption spectrum contains population and coherence informations of ionic states.
    The theoretically predicted (b) and the experimentally measured (c) spectra are shown. 
    The figure is taken from Ref.~\cite{GoKr-Nature-2010}.
    Copyright~\copyright~2010 Nature Publishing Group (NPG).
  }
  \label{fig2:tas}
\end{figure}

\subsubsection{Studying Dynamical Processes in Ionic Systems}
\begin{figure}[t!]
  \centering
  \includegraphics[width=.52\linewidth]{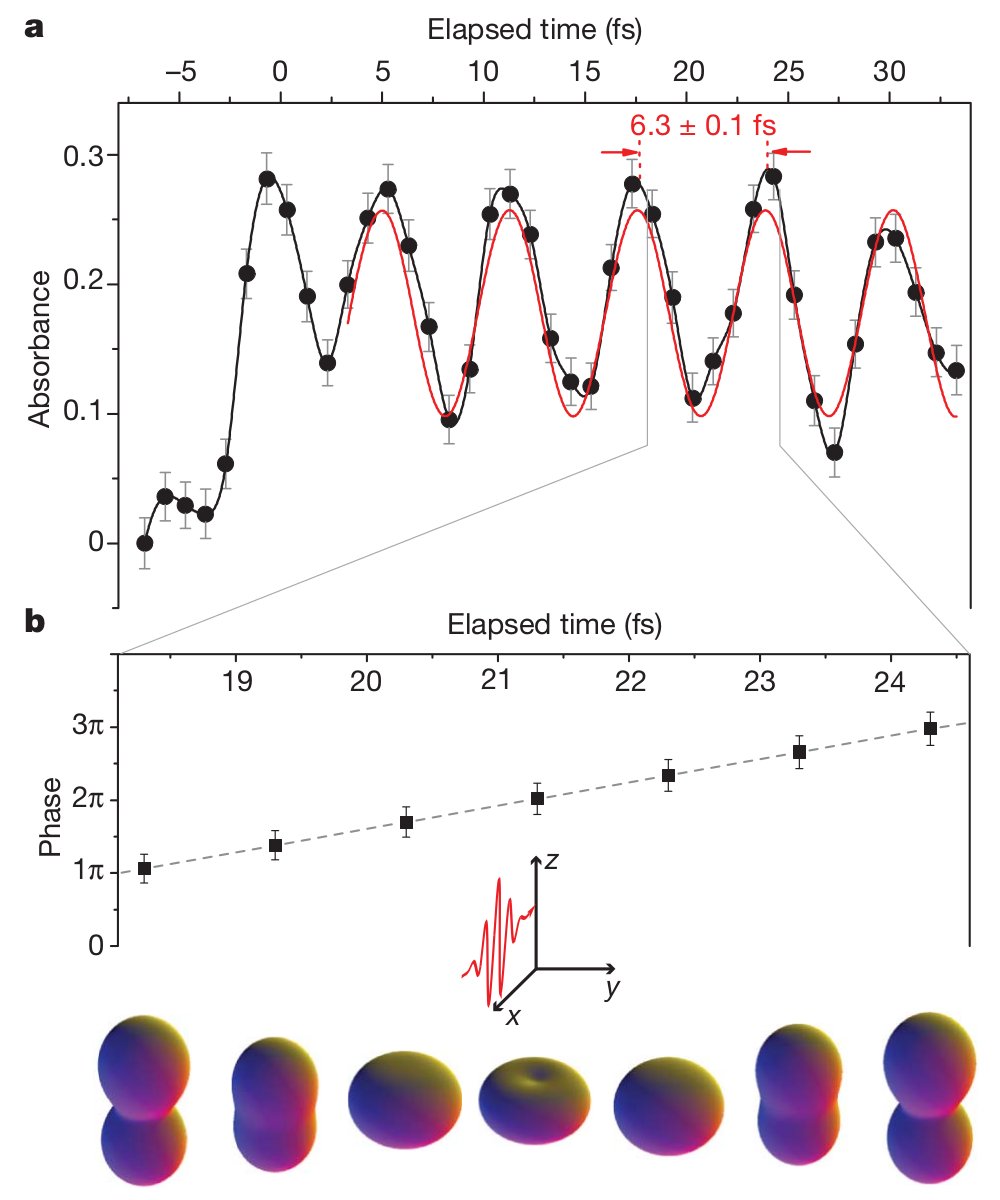}
  \caption{
    (a) The beating of the $[4p_{3/2}^{m}]^{-1} \rightarrow [3d_{3/2}^{m}]^{-1}$ transition line is shown.
    This beating gives a direct measure for the coherence between the $[4p_{3/2}^{\pm 1/2}]^{-1}$ and $[4p_{1/2}^{\pm 1/2}]^{-1}$ ionic states.
    (b) The reconstructed phase between the two ionic states is shown.
    After having determined the populations and phase between ionic state, the full hole dynamics, which is shown at the bottom, can be reconstructed.
    The figure is taken from Ref.~\cite{GoKr-Nature-2010}.
    Copyright~\copyright~2010 Nature Publishing Group (NPG).
  }
  \label{fig2:tas_holemotion}
\end{figure}

The first demonstration of attosecond transient absorption spectroscopy has been done with atomic krypton~\cite{GoKr-Nature-2010}.
The NIR pump pulse tunnel-ionizes the atom by creating a hole in the $4p^{-1}$ manifold.
The attosecond pulse probes the ionic states by exciting the ion into states with a hole in the $3d^{-1}$ manifold.
In Fig.~\ref{fig2:tas} a sketch the pump and probe steps (a) as well as the theoretically predicted (b) and experimentally measured (c) transient absorption spectrum are shown.
Due to spin-orbit interaction, there are three transition lines instead of just one visible between the $4p^{-1}$ and $3d^{-1}$ manifolds.
The corresponding transitions are $[4p_{3/2}^{m}]^{-1} \rightarrow [3d_{5/2}^{m}]^{-1}$,$[4p_{1/2}^{m}]^{-1} \rightarrow [3d_{3/2}^{m}]^{-1}$, and $[4p_{3/2}^{m}]^{-1} \rightarrow [3d_{3/2}^{m}]^{-1}$.
The population probabilities are directly proportional to the corresponding transition strengths~\cite{SaYa-PRA-2011}.
Even coherence properties can be extracted from the spectrum via the shape of the transition lines.
In Fig.~\ref{fig2:tas_holemotion}(a), the influence of the line shape on the effective line strength is shown.
It reveals the hole dynamics due to the coherence superposition between the ionic states $[4p_{3/2}^{\pm 1/2}]^{-1}$ and $[4p_{1/2}^{\pm 1/2}]^{-1}$.
Consequently, the full state of the ionic system, which is given by the ion density matrix (IDM), can be measured. 

An analytic expression has been derived in Ref.~\cite{SaYa-PRA-2011} which treats the probe pulse in $1^\text{st}$-order perturbation theory.
By fitting the analytic expression to the experimental data, all entries of the IDM (within the $4p^{-1}$ manifold) can be extracted.
After doing this for many pump-probe delays $\tau$, the entire hole dynamics of Kr$^+$ is reconstructed.
The relative phase between the ionic states $[4p_{3/2}^{\pm1/2}]^{-1}$ and  $[4p_{1/2}^{\pm 1/2}]^{-1}$ and the resulting spatial motion of the hole state are shown in Fig.~\ref{fig2:tas_holemotion}(b).
Furthermore, even traces of doubly ionized krypton states have been seen.
This makes it possible to directly test many-body physics of multielectron ionization dynamics with transient absorption spectroscopy~\cite{WiSa-ChemPhys-2012}.

An important assumption in the above discussion is that the ion is an isolated system meaning the ionized electron is far away and cannot influence the ion anymore.
When pump and probe pulses do overlap, this assumption does not hold anymore.
It is, therefore, not clear to which extent the transient absorption signal can be still related to the instantaneous IDM. 
However, it turns out that transient absorption spectroscopy can even offer insight into laser-induced electron-ion interactions.
This becomes particularly important when sub-cycle ionization dynamics are studied.

\subsubsection{Studying Excited Electronic States}
Attosecond transient absorption spectroscopy refers only to how the system gets probed.
Therefore, there are many ways the system can be prepared (by the pump pulse).
Instead of ionizing krypton with an intense NIR pulse, another attosecond pulse can be used to initially ionize krypton~\cite{BaLi-PRA-2012}.
Helium is another popular system that has been studied by transient absorption spectroscopy~\cite{GaBu-PRA-2011,ChZh-PRL-2012,TaGr-PRA-2012,OtKa-arxiv-2012,ChBe-PRA-2012}.
The pump step here is an excitation to Rydberg states rather than tunnel or photoionization.
Many phenomena can be observed with such pump schemes.
If the excited state is a metastable state, the autoionization dynamics can be studied~\cite{WaCh-PRL-2010,TaGr-PRA-2012}.
The Autler-Townes splitting of the excited states can also be probed by applying an additional NIR pulse that induces a coupling to nearby Rydberg states~\cite{GaBu-PRA-2011,TaGr-PRA-2012,OtKa-arxiv-2012,ChBe-PRA-2012}.
A sub-cycle Stark-shift driven by an NIR pulse has also been observed in an excited helium atom~\cite{ChZh-PRL-2012}.

It is also possible to probe the recombination dynamics of HHG mechanism by transient absorption~\cite{HoSc-PRL-2011}. 
Here, an attosecond pulse train rather than an isolated attosecond pulse is used as probe.
First, one attosecond pulse excites the electron into a Rydberg state. 
The NIR pulse drives this electron away and back to the ion (as in HHG).
When the electron returns, the subsequent attosecond pulse excites another electron.
This leads to inferences in the HHG spectrum, which can be used to study the underlying electronic dynamics.

\subsection{Numerical Methods In Strong-Field Physics}
\label{p1c2:theory}
There exist two major theoretical approaches for describing the underlying ultrafast processes.
One approach is based more on analytical expressions (like the SFA).
Many simplifications have to be made in order to arrive at these analytical expressions, limiting applicability.
More general approaches are no longer analytically solvable and require computational help.
This means numerically solving the time-dependent Schr\"odinger equation (TDSE) [cf. Eq.~\eqref{eq2:tdse}].
However, even the TDSE cannot be solved numerically for many-electron systems without making some approximations~\cite{LaMa-PR-1998}.

For systems that are not highly correlated such as noble gas atoms, it is useful to exploit an independent particle picture, where the $N$ electrons only interact indirectly with each other via a mean-field potential created commonly by all $N$ electrons.
The most common mean-field theory is Hartree-Fock (HF)~\cite{GrRu-book,SzOs-book}, which approximates the full $N$-electron wavefunction $\ket{\Psi}$ by a single Slater determinant $\ket{\Phi_0}$ as described in Sec.~\ref{p1c2:photo}.
Most many-body theories~\cite{BarSt-inbook} (e.g., configuration-interaction or couple-cluster) use the HF state as a reference state to build up a multi-configuration wavefunction consisting of several Slater determinants, which is able to capture correlation effects going beyond the independent particle picture.
However, these many-body theories have the disadvantage that they quickly become numerically costly.
Hence, they are mainly used for ground state properties of molecular systems.

In strong-field calculations, it is necessary to describe a wide range of continuum states, since an electron can be freed from the system and may even return to it at a later time.
The highly delocalized continuum states required in the calculations do not favor an orbital description based on Gaussian-type functions as done in most quantum chemistry approaches~\cite{AbMa-PRA-2010}.
The non-adiabatic character of the electronic motion makes it quite challenging to exploit many-body theories including a high degree of correlations.

For multi-cycle pulses, it is advantageous to exploit the periodicity of the pulse by using Floquet theory~\cite{Sh-PR-1965,SoCh-PRA-2008}.
The time-dependent Hamiltonian is transformed into a time-independent, block-trigonal Hamiltonian, where the diagonal blocks correspond to the number of photons that have been absorbed or emitted.

Many ultrafast processes in closed-shell systems are dominated by a single particle.
Adapting the theoretical approach by incorporating a single-active-electron (SAE) picture is quite advantageous.
A single-active electron (SAE) picture means only one electron is involved in the process (active electron) whereas all other electrons are just spectators (passive electrons) and are not affected by the dynamics of the active electron or the external field.
On the one side, it dramatically reduces the numerical challenges (see Sec.~\ref{p1c2:sae}).
On the other side, it also favors a more intuitive physical picture.

However, many-body effects going beyond the SAE picture do exist and indeed become even very important for larger atoms and molecular systems as discussed in Sec.~\ref{p1c2:hhg}.
A time-dependent configuration-interaction singles (TDCIS) approach makes use of a single-active electron picture.
However, it does not restrict from which orbital the electron gets ionized.
It is also possible for the active electron to modify the ionic state after it has been ionized (i.e., interchannel interactions).
In Sec.~\ref{p1c2:tdcis}, the basic aspects of TDCIS are discussed.
In Sec.~\ref{p1c2:apps}, ultrafast phenomena are reviewed that require many-body theories like TDCIS and cannot be explained by a SAE picture.
First, however, the SAE approach with its advantages and disadvantages is discussed.

\subsubsection{Single-Active-Electron Models}
\label{p1c2:sae}
The most common numerical approach to tackle strong-field problems is solving the TDSE [cf. Eq.~\eqref{eq2:tdse}] with the SAE approximation.
The SAE reduces the Hamiltonian to a one-particle Hamiltonian
\begin{align}
  \label{eq2:sae_ham}
  \hat H 
  &=
  \frac{\hat p^2}{2}
  + 
  \hat V
  ,
\end{align}
where $\hat V$ is a one-particle potential resulting from the interaction with the remaining electrons.
In the SAE, $\hat V$ is commonly described by a local model potential~\cite{HiFa-PRA-2011}.
For atomic systems, the model potential is also assumed to be spherically symmetric.
In the limit of very short and very large distances, the behavior of $\hat V$ is known: $\lim_{r\rightarrow 0} V(r) = -Z/r$ and $\lim_{r\rightarrow\infty} V(r) = -1/r$.
Very close to the atomic core, the electronic potential is dominated by the nuclear potential (i.e., $-Z/r$).
Far away from the atom, the nuclear potential is screened by the remaining $N-1$ electrons and the ionized electron feels an attractive potential of a singly-charged point-particle (i.e., $-1/r$).
Between these two limits, the model potential is constructed such that the ionization potential and excitation energies are well reproduced~\cite{Mu-PRA-1999}.
A common example of a model potential looks like~\cite{HiFa-PRA-2011}
\begin{align}
  \label{eq2:modelpotential}
  V(r)
  &=
  -\frac{1 + A e^{-Br} + (Z-1-A) e^{-Cr}}{r}
  ,
\end{align}
where $A,B$ and $C$ are the coefficients adjusting the potential.
Despite the simplicity of such model potentials, they can explain and reproduce quite well experimental observations like ATI and HHG spectra~\cite{Mu-PRA-1999}.
The success of SAE models confirms that a wide range of strong-field and attosecond processes are mainly one-electron processes.
This is particularly true for noble gas atoms, specifically the lighter ones.
For heavier noble gas atoms~\cite{ShVi-NatPhys-2011} or molecules~\cite{WaSm-JPhysB-2010} many-electron effects and the interaction of several orbitals start to emerge.
In this case, it is not possible to capture the dynamics with the SAE approximation, since SAE assumes that all other electrons are frozen.

The SAE approximation does also imply that the remaining ion can only be in one specific state, normally the ionic ground state.
SAE is a single-channel theory, where the term ``channel'' refers to the state of the $N-1$ electrons.
It is necessary to use multi-channel theories when studying ionic excitations and superpositions of several ionic states (as discussed in Sec.~\ref{p1c2:tas})

The spherically symmetric character of the ionic potential is strictly speaking not true, since mostly the outer-most p$_0$ orbital gets ionized by linearly polarized light.
Hence, the electron distribution in the ion is not spherical and, therefore, also the resulting potential is not spherical.
For atomic systems,  that may not be so critical.
For molecular systems, however, the potential is obviously non-spherical and also electron-electron interactions become more important.

One way to construct a molecular SAE potential is by following the spirit of density functional theory (DFT)~\cite{AbMa-PRA-2010,ZhJi-PRA81-2010}.
Here, a one-particle potential is reconstructed from the electron density of the neutral ground state, which can be calculated with standard quantum chemistry codes (e.g., {\sc dalton}~\cite{dalton}).
The overall SAE potential can be written as~\cite{AbMa-PRA-2010}
\begin{align}
  V({\bf r})
  &=
  V^x_{ee}({\bf r})
  +
  V_{en}({\bf r})
  +
  V^d_{ee}({\bf r})
  +
  G_c({\bf r})
  ,
\end{align}
where $V^{d/x}_{ee}({\bf r})$ is the direct and exchange contribution of the electron electron interaction, respectively, $V_{en}({\bf r})$ is the attractive electron-nuclei interaction, and $G_c({\bf r})$ is a long-range correction in order to obtain the right $-1/r$ long-range behavior.
The exchange term $V^{x}_{ee}$ is generally non-local.
The local density approximation (LDA)~\cite{PaYa-DFT-book} is, however, a common way to make the potential local by expressing $V^{x}_{ee}$ in terms of the local electron density $\rho({\bf r})$.
Other DFT approximations have also been considered to improve the SAE potential for molecules~\cite{ZhJi-PRA82-2010}.
The quantity that determines the success of these potentials is the angle-dependent ionization rate~\cite{AbMa-PRA-2010,MuSp-PRL-2011} (see Fig.~\ref{fig2:sfa_angular} and discussion in Sec.~\ref{p1c2:tunnel}).
The correct angle dependence is particularly important for imaging molecular orbitals~\cite{MuSp-PRL-2011}.

\subsubsection{Time-dependent Configuration-Interaction Singles (TDCIS)}
\label{p1c2:tdcis}
A common post-Hartree-Fock method is configuration-interaction (CI), which adds systematically higher excitation classes to the wavefunction i.e.,
\begin{align}
  \label{eq2:ci}
  \ket{\Psi}
  &=
  \alpha_0\, \ket{\Phi_0}
  +
  \sum_{n=1}^N \sum_{ a_1,\ldots,a_n \atop i_1,\ldots i_n }
    \alpha^{a_1,\ldots,a_n}_{i_1,\ldots,i_n}\, 
    \underbrace{\prod_{h=1}^n \cre_{a_h}\ann_{i_h} \ket{\Phi_0}}_{
      \ket{\Phi^{a_1,\ldots,a_n}_{i_1,\ldots,i_n}} }
\\\nonumber
  &=
  \alpha_0\, \ket{\Phi_0}
  +
  \sum_{a_1,i_1}
    \alpha^{a_1}_{i_1}\, \ket{\Phi^{a_1}_{i_1}}
  +
  \sum_{a_1,a_2,i_1,i_2} 
    \alpha^{a_1,a_2}_{i_1,i_2}\,\ket{\Phi^{a_1,a_2}_{i_1,i_2}}
  +
  \ldots
  ,
\end{align}
where $\Phi_0$ is the HF ground state and $\ket{\Phi^{a_1,\ldots,a_n}_{i_1,\ldots,i_n}}$ are  $n$-particle-$n$-hole ($n$p-$n$h) configurations with $\cre_a,\ann_i$ being creation and annihilation operators of the corresponding orbitals.
The HF ground state is used as reference state from which all $n$p-$n$h configurations are defined.
The indicies $a$ and $i$ refer to unoccupied (virtual) and occupied orbitals in $\Phi_0$, respectively.
The indicies $p,q,r,s$ are used to refer to all orbitals (occupied + unoccupied).
If $n$ goes up to the total number of electrons, one speaks of full CI (FCI).
Including only singly excited configurations is called CI-Singles (CIS).
Similarly, including only doubly excited configurations is called CI-Doubles (CID), and including singly and doubly excited configurations is called CI-Singles-Doubles (CISD).

The size of the $n$p-$n$h-configuration space is $(N_aN_i)^n$, where $N_a$ and $N_i$ are the numbers of unoccupied and occupied orbitals, respectively.
Since $N_a$ is quite large (in principle infinitely large) it is computationally not feasible to go to very high excitation classes.
To calculate exact ground state properties (e.g., response functions), it not necessary to include states that are very delocalized and extend far from the atom/molecule.
Hence, $N_a$ is relatively small and higher $n$p-$n$h-excitations classes and, therefore, higher-order correlations can be included in the calculations.
In strong-field physics, where it is important to describe a wide range of the continuum states, $N_a$ can easily be a 5-digit number.
Including, therefore, doubly or triply excited configurations quickly becomes infeasible.

Time-dependent configuration-interaction singles (TDCIS), with time-dependent coefficients $\alpha^a_i(t)$, seems to be an ideal extension to the SAE model for investigating strong-field and attosecond phenomena.
The TDCIS wavefunction reads
\begin{align}
  \label{eq2:tdcis_wfct}
  \ket{\Psi(t)}
  &=
  \alpha_0(t)\, \ket{\Phi_0}
  +
  \sum_{a,i}
    \alpha^a_i(t)\, \ket{\Phi^a_i}  
  .
\end{align}
This wavefunction ansatz still makes use of the fact that most strong-field processes in atoms are one-electron processes but it also acknowledges that the active electron may not just come from the least weakly bound (outer-most) occupied orbital.
TDCIS also takes into account that the active electron can influence the state of the parent ion.

The TDCIS method discussed here was first presented by Rohringer {\it et al.}~\cite{RoSa-PRA-2006} for atomic systems and has been later extended to include spin-orbit splitting for the occupied orbitals~\cite{RoSa-PRA-2009} and the inclusion of the exact residual Coulomb interaction $\hat H_1$~\cite{GrSa-PRA-2010}.
Later both extensions have been combined in order to study electron-ion correlation effects on orbitals that are split due to spin-orbit interaction~\cite{PaSy-PRA-2012}.
The Hamiltonian\footnote{In Ref.~\cite{GrSa-PRA-2010,PaSa-PRL-2011,PaGr-PRA-2012,SyPa-PRA-2012,PaSy-PRA-2012} the charge of the electron is $q_e=1$ such that $-E(t)$ rather than $E(t)$ appears.} 
\begin{align}
  \label{eq2:tdcis_ham}
  \hat H(t)
  &=
  \hat H_0 + \hat H_1 + E(t)\,\hat z 
\end{align}
is preferably partitioned into three parts:
\begin{itemize}
 \item 
 $\hat H_0$ is the Fock operator defining the HF ground state $\Phi_0$ and the one-particle orbitals $\varphi_p$ with their orbital energies $\varepsilon_p$ ($\hat H_0 \ket{\varphi_p}=\varepsilon_p \ket{\varphi_p}$).
 
 \item
 $\hat H_1$ is the residual electron-electron interaction that cannot be captured by the mean-field potential that is included in $\hat H_0$.
 
 \item
 The light-matter interaction $E(t)\,\hat z$, expressed in the length form after the dipole approximation is made (no space dependence of $E$). 
\end{itemize}
The residual Coulomb interaction $\hat H_1$ is the only two-body operator in Eq.~\eqref{eq2:tdcis_ham} and captures all effects beyond an independent particle picture.
The detailed expression reads $\hat H_1=\hat V_{ee} - \hat V_\text{HF} - E_\text{HF}$, where $\hat V_{ee}=\frac{1}{2}\sum_{i \neq j} \frac{1}{|\hat{\bf r}_i - \hat{\bf r}_j|}$ is the exact electron-electron interaction, $\hat V_\text{HF}$ is the HF mean-field potential, and $E_\text{HF}$ is the HF ground state energy.
Subtracting $E_\text{HF}$ is only been done out of convenience, shifting all energies such that the HF ground state has zero energy.
Inserting Eq.~\eqref{eq2:tdcis_wfct} in Eq.~\eqref{eq2:tdse}, the equations of motion for the CI-coefficients emerge, which read
\begin{subequations}
\label{eq2:tdcis_eom}
\begin{align}
  \label{eq2:tdcis_eom_gs}
  i\alpha_0(t)
  =&
  E(t) \sum_{a,i} \bra{\Phi_0}\hat z \ket{\Phi^a_i} \alpha^a_i(t)
  ,
\\
  \label{eq2:tdcis_eom_ai}
  i\alpha^a_i(t)
  =& \ 
  (\varepsilon_a - \varepsilon_i)\, \alpha^a_i(t)
  +E(t) \left[
     \bra{\Phi^a_i} \hat z \ket{\Phi_0}\, \alpha_0(t)
     +
     \sum_{b,j} \bra{\Phi^a_i} \hat z \ket{\Phi^b_j}\, \alpha^b_j(t)
   \right]
\\\nonumber 
  & +
  \sum_{b,j}
    \bra{\Phi^a_i} \hat H_1 \ket{\Phi^b_j}\, \alpha^b_j(t)
  .
\end{align}
\end{subequations}
The terms $\bra{\Phi_0}\hat z \ket{\Phi^a_i}$ and $\bra{\Phi^a_i} \hat z \ket{\Phi_0}$ describe the light-matter interaction that couples the neutral ground state to the singly-excited states. 
Note that $\hat H_1$ does not lead to couplings between the ground state and the singly-excited states.
Hence, without any external field the atom remains in the ground state.

By setting the matrix elements $\bra{\Phi^a_i} \hat H_1 \ket{\Phi^b_j}$ with $i \neq j$ to zero, the interchannel interaction can be switched off in the TDCIS model.
When $\hat H_1$ is switched off the intrachannel interactions are also ignored.
This makes it possible to systematically study dynamical effects that go beyond an independent particle picture.
The transition matrix elements $\bra{\Phi^a_i} \hat z \ket{\Phi^b_j}$ separate into two independent single-particle transitions [cf. Eq.~(4) in Ref.~\cite{PaSy-PRA-2012}], one between occupied orbitals (ionic transition) and one between virtual orbitals (electronic transition).

\begin{figure}[ht!]
  \centering
  \includegraphics[width=.85\linewidth]{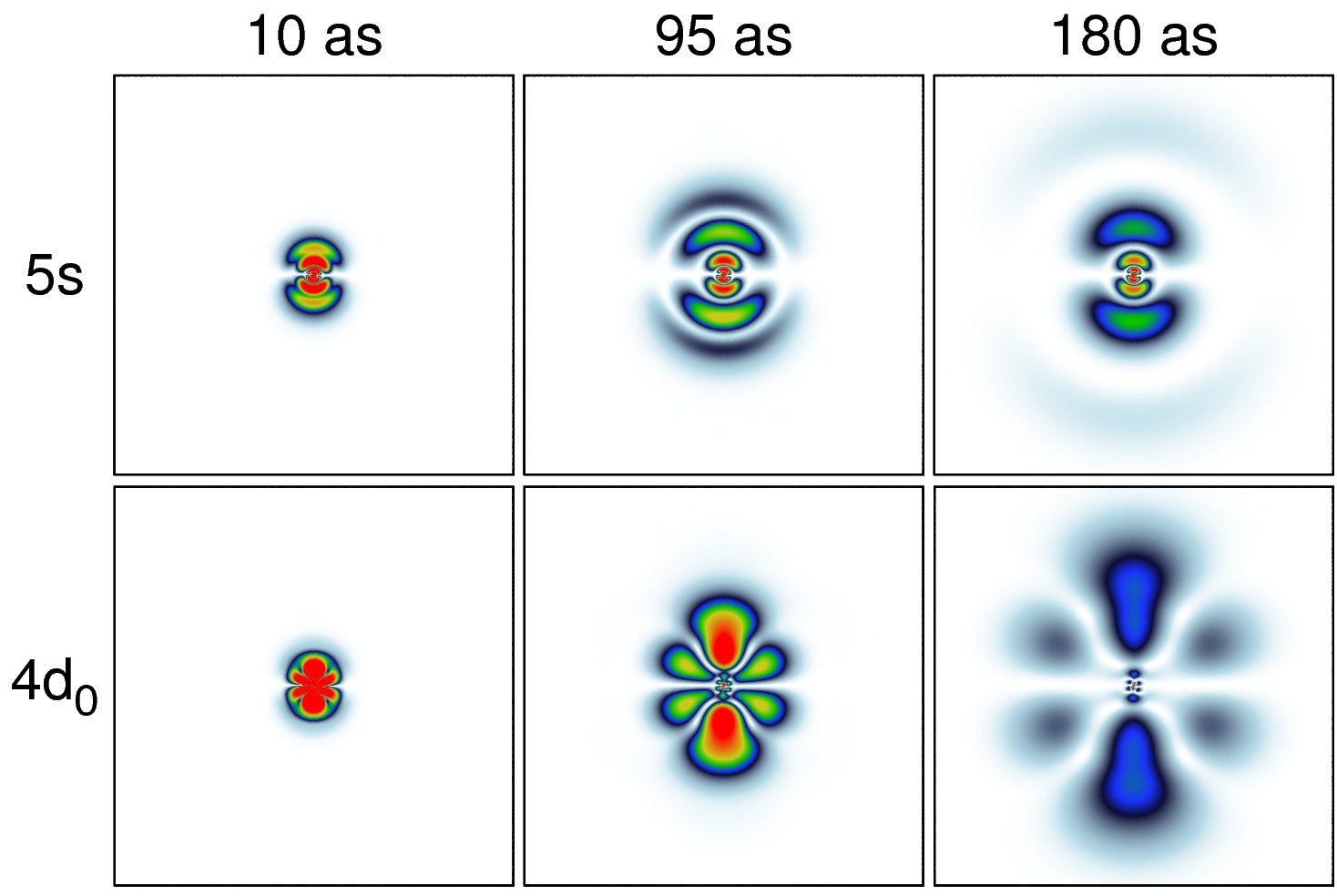}
  \caption{
    Photoelectron wavepackets originating from the $4d_0$ and $5s$ orbitals are shown for several time delays after the ionizing UV attosecond pulse has hit the atom.
    The mean photon energy is 136~eV and the pulse duration is 10~as.
  }
  \label{fig2:tdcis_ionize}
\end{figure}

The calculated photoelectron wavepacket after ultrafast UV photoionization (10~as pulse) is shown for the $5s$ and the $4d_0$ ionization channels in Fig.~\ref{fig2:tdcis_ionize}.
The $l=1$ and the $l=3$ characteristics of the electron wavepackets are clearly visible.
As noted in Sec.~\ref{p1c2:photo} the photoelectron prefers to increase its angular momentum rather than to decrease it when absorbing the photon.
Therefore, the photoelectron of the $4d_0$ channel possesses mainly $l=3$ character.

The residual Coulomb interaction within each channel ($\bra{\Phi^a_i}\hat H_1\ket{\Phi^b_i}$) is generally not local and not spherically symmetric---approximations that are normally made in SAE models.
Interchannel interactions ($\bra{\Phi^a_i}\hat H_1\ket{\Phi^b_j}$ with $i \neq j$) do fulfill this assumption even less.
Only asymptotically do they hold.
In Sec.~\ref{p1c2:apps_hhg}, the effect of these approximations are discussed in terms of the HHG spectrum of argon.

\paragraph{Ion Density Matrix}
TDCIS describes two subsystems, the active electron and the parent ion.
In Secs.~\ref{p1c2:apps_decoh} and \ref{p1c2:apps_tas}, situations are reviewed, where the interaction between these two subsystems leads to measurable effects in the ionic state.
To extract information about the ionic subsystem from the full $N$-body system, the trace of the full $N$-electron density matrix $\hat\rho(t)=\ket{\Psi(t)}\bra{\Psi(t)}$~\cite{Fa-RMP-1957} over the other subsystems (i.e., the electron) has to be performed~\cite{Blum-book,BrPe-book}. 
As a result, one obtains the reduced density matrix of the ionic subsystem
\begin{subequations}
\label{eq2:idm}
\begin{align}
  \label{eq2:idm_operator}
  \hat \rho^\IDM(t)
  &=
  \text{Tr}_a \left[ \hat\rho(t) \right],
\\[1.5ex]
  \label{eq2:idm_element}
  \rho^\IDM_{i,j}(t)
  &=
  \sum_a \braket{\Phi^a_i}{\Psi(t)\big.} \braket{\Psi(t)}{\Phi^a_j}
  .
\end{align}
\end{subequations}
also called ion density matrix (IDM).
This IDM uniquely characterizes the state of the ion.
Since more than one occupied orbital (channel) can be ionized, it is possible to create a superposition of ionic eigenstates.
Figure~\ref{fig2:tas_holemotion} shows the hole motion in singly-ionized krypton.
The IDM is also an ideal quantity to study coherences in the ionic subsystem.
After the atom has been ionized, the parent ion does not need to be in a coherent state as the discussion of Sec.~\ref{p1c2:apps_decoh} shows.

\paragraph{Orbital Representation}
For atomic systems, it is convenient to use spherical coordinates, where the field-free one-particle orbitals can be  factorized into a radial and an angular part, i.e., 
\begin{align}
  \label{eq2:orbital_spherical}
  \braket{\bf r}{\varphi_p}
  &=
  \varphi_p({\bf r}) 
  = 
  \frac{u_{n_p,l_p}(r)}{r}\, Y_{l_p,m_p}(\Omega)
  ,
\end{align}
where $n_p, l_p$ and $m_p$ are the radial, the angular momentum, and the magnetic quantum numbers of the orbital $\varphi_p$, respectively.
$Y_{l,m}(\Omega)$ are spherical harmonics~\cite{Zare-book}, where $\Omega=(\vartheta,\varphi)$ stands for the two angular coordinates.
The factorization between radial and angular parts has several advantages. 
First, the 3D-integrals needed for evaluating the matrix elements in Eq.~\eqref{eq2:tdcis_eom_ai} factorize into three 1D-integrals, where the two angular integrals can be solved analytically (in the form of Clebsch-Gordan coefficients).
Furthermore, the magnetic quantum number of the full $N$-electron state, $\Psi(t)$, is conserved ($M=0$) for linearly polarized light\footnote{
Circularly or elliptically polarized light breaks this symmetry.}
.
For TDCIS, this has the consequence that the magnetic quantum number of the active electron cannot be changed ($m_a=m_i$).
This restriction in $m$ is quite advantageous for numerical purposes. 

The underlying radial grid is chosen to be nonlinear such that more grid points are closer to the origin (i.e., the atomic nucleus).
In particular, the M\"obius transformation $ x\longmapsto r(x) = r_\text{max}\zeta/2\, (1+x)(1-x+\zeta)^{-1}$ is exploited~\cite{WaCh-PRA-1994,GrSa-PRA-2010} with $x\in[-1,1]$.
The nonlinearity of the mapping and the usage of pseudo-spectral techniques~\cite{WaCh-PRA-1994} significantly reduce the number of grid points needed to a few hundreds for system sizes of around 100~$a_0$ ($a_0=\nolinebreak 5.29\cdot 10^{-11}$~m).

\paragraph{Complex Absorbing Potential}
For ionization scenarios, it is common that the ionized electron separates from the ion with great speed. 
Before the pulse is over, the electron may have traveled several hundreds or thousands of $a_0$.
Such large grids are computationally not feasible.
To avoid artificial reflections from the grid boundary, a complex absorbing potential (CAP) is introduced~\cite{RiMe-JPB-1993}, which absorbs the outgoing electron just before the grid ends.
By putting the absorbing potential at the end of the grid, the absorbed electron is far enough away such that it does not influence the ion anymore.
Hence, the absorbing potential does not influence the physics near the atom/ion.

Unfortunately, introducing a CAP results in a non-hermitian Hamiltonian ($\hat H \rightarrow \hat H - i\eta \hat W$).
Hence, the norm of the wavefunction is not conserved anymore.
The specific shape of the absorbing potential used in all examples presented in Sec.~\ref{p1c2:apps} is given by $W(r)=(r-r_c)^2\,\Theta(r-r_c)$~\cite{GrSa-PRA-2010}, where $\Theta(x)$ is the Heaviside function.
Having a non-hermitian $\hat H_0$ also means that the orbital energies $\varepsilon_p$ become complex.

If the CAP is chosen correctly, the imaginary part of $\varepsilon_p$ can be directly related to the lifetime of the orbital~\cite{RiMe-JPB-1993}.
This is particularly interesting for autoionizing states~\cite{SaCe-PhysRep-2002}.
An alternative method is exterior complex scaling (ECS), where the radius $r$ is replaced by a complex radius $R(r)$~\cite{RiMe-JPB-1993}.
ECS is formally an exact method, which make an analytical continuation of the Hamiltonian into the complex plane~\cite{RiMe-JPB-1993}.
However, the scaled Hamiltonian can also be separated into an unscaled and a scaled part.
The scaled part can then be treated as a CAP, which is now non-local.

Absorbing the electron results in a reduced norm in the ionic subsystem.
To restore the norm, at least for the ionic subsystem, which should not be affected by the CAP, a correction term for $\hat \rho^\IDM$ has been derived.
An explicit derivation is given in Sec.~II.D of Ref.~\cite{GrSa-PRA-2010}.
The CAP correction can be also used to determine the ionization rate or the ionization cross section~\cite{SyPa-PRA-2012}, since each ionized electron has to hit the CAP eventually.
One has to be careful when the electron is only excited and still bound.
In this case the CAP correction cannot be used as a measure of ionization.

A more efficient way to determine the photoionization cross section is by using the autocorrelation function $g(t_1-t_0)=\braket{\Psi(t_1)}{\Psi(t_0)}$ with the initial state $\ket{\Psi(t_0)}=\hat{p} \ket{\Phi_0}$~\cite{FaCo-RMP-1968,ToTo-PRA-2010}.
Here, one assumes a delta-like kick at time $t_0$ such that all frequency components are equally included. 
Consequently, all energy eigenstates are involved in the system response.
The photoionization cross section in terms of the correlation function reads
\begin{align}
  \label{eq2:crosssection_corrfct}
  \sigma(\omega)
  =
  \frac{2\pi\,\alpha}{\omega}
  \int_{-\infty}^\infty \!\! dt\
    g(t)\, e^{i\omega\,t}
  =
  \frac{4\pi\,\alpha}{\omega} \
  \text{Re}\left[
    \int_{0}^\infty \!\! dt\
      g(t)\, e^{i\omega\,t}
  \right]
  ,
\end{align}
where in the second relation the property $g(t)=g^*(-t)$ has been used.

\paragraph{Limitations}
TDCIS is a multi-channel theory, where only one electron can be ionized.
Hence, multiple ionization processes cannot be described by TDCIS.
Due to the residual Coulomb interaction, this outgoing electron can alter the ionic state.
The degree of freedom of the ionic state is, however, quite limited. 
The only allowed ionic states are one-hole configurations $\ket{\Phi_i}=\ann_i\ket{\Phi_0}$, where an electron is removed from the HF ground state.
In these configurations, all other electrons are frozen.
Hence, the only ionic motion allowed is the hopping of the hole between occupied orbitals.
More complicated dynamics require at least one more electron to be non-frozen meaning 2p-2h configurations $\ket{\Phi^{a_1,a_2}_{i_1,i_2}}$ are needed.

The consequences of the CIS restriction for the ionic states have been discussed in Ref.~\cite{PaSy-PRA-2012})in terms of the polarizability of the ion.
The comparison with higher-order approaches like CASSCF\footnote{CASSCF - Complete Active Space Self-Consistent Field Theory} has shown that polarizability is underestimated by CIS.
Particularly the rearrangement of the remaining electrons is crucial to obtain a more accurate picture of the ionic subsystem.

\subsection{Applications}
\label{p1c2:apps}

All strong-field applications presented here focus on many-body and correlation effects that go beyond an independent particle picture.
Already in a fundamental process like photoionization these effects are present.
In Sec.~\ref{p1c2:apps_decoh}, correlation effects in ultrafast attosecond photoionization is presented.
There, a surprising phenomenon has been found in the parent ion resulting from interchannel coupling effects.
The single-active-electron model is tested in Sec.~\ref{p1c2:apps_hhg}, where the influence of multi-orbitals and multipole effects in terms of the HHG spectrum of argon is investigated.
The state of the multi-level ion is uniquely defined by the ion density matrix (IDM).
The IDM is experimentally accessible with transient absorption spectroscopy as discussed in Sec.~\ref{p1c2:tas}.
Even field-induced dynamics on a sub-cycle time scale can be studied with overlapping pump and probe pulses.
In Sec.~\ref{p1c2:apps_tas}, studies are discussed that demonstrated this possibility on atomic krypton exposed to an intense NIR pulse.

\subsubsection{Decoherence in Attosecond Photoionization}
\label{p1c2:apps_decoh}
Correlation effects can appear in the most fundamental processes like photoionization.
Closely connected is the question to which degree a coherent hole wavepacket can be initiated in the ion and how correlations between the photoelectron and the ion may affect the ionic state. 
As we known, ionization is not an instantaneous process and the photoelectron needs a finite amount of time to separate from the ion.
During this finite time window, the electron is still close to the ion and via electron-electron interactions, the photoelectron can change the ionic state. 
Such an interaction entangles the electronic wavefunction with the ionic wavefunction.
This ionization process can be also viewed in a system+bath picture, where the parent ion is the system and the ionized electron is the bath~\cite{BrPe-book}.
Note that the bath is in this case the smaller subsystem (contrary to common system-bath scenarios in chemistry where the bath is the chemical environment).
It is well-known from system-bath models that the interaction between the two subsystems influences, or more precisely reduces, the coherent properties within each subsystem.
Exactly the same is true for photoionization.

From a theoretical point of view, two common approximations have to be abandoned to be able to describe an entanglement between ion and photoelectron.
First, the sudden approximation cannot be made, which assumes that the ionized electron is instantaneously removed from the ion and no interaction between the ionized electron and the ion is possible.
Second, the remaining $N-1$ electrons in the ion cannot be assumed to be spectators which are frozen during the ionization process. 
The ion has to be described as a multi-level system such that the state of the ion is able to change due to the residual Coulomb interaction with the photoelectron.
The TDCIS approach is, here, ideal to describe these correlation dynamics.
In photoionization, this change in the ionic state is known as interchannel coupling (see Sec.~\ref{p1c2:photo} for a detailed discussion).
If the interaction with the photoelectron does not change the ionic state, the entanglement between the electron and the ion is strongly reduced and the coherence in the ionic subsystem is preserved.

Atomic xenon is chosen to demonstrate the importance of interchannel coupling effects in ultrafast photoionization.
Here, the principal idea is to create a dynamical hole wavepacket in singly ionized xenon.
The energetically most favorable option (ignoring spin-orbit splitting) for creating a coherent hole state via one-photon ionization with linearly polarized light is through a superposition of the ionic states $4d_0^{-1}$ and $5s^{-1}$, where $4d_0$ and $5s$ refer to the occupied orbitals the ionized electron originates from.
The finale state of the photoelectron must be the same to be able to create a coherent hole wavepacket in the ion.
This is necessary, since it must be impossible to tell from which orbital the electron came.
Since linearly polarized light is used, the angular momentum projections $m_l$ of both occupied orbitals must be the same.
The angular momentum $l$ of both orbitals must be the same or differ by 2, since photoionization always changes $l$ of the electron by 1.

\begin{figure}[t!]
  \centering
  \includegraphics[width=.8\linewidth]{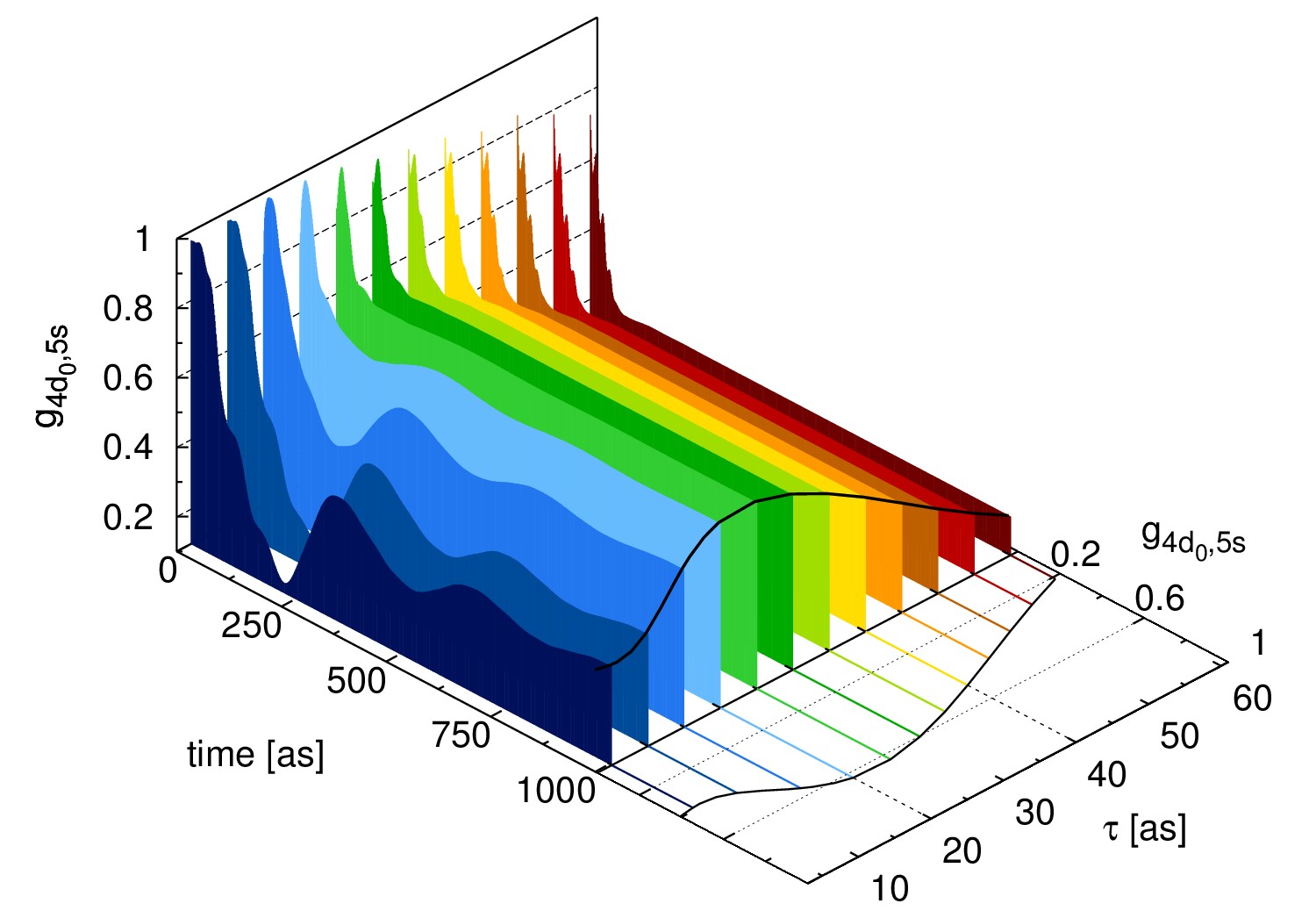}
  \caption{
    The dynamical behavior of the degree of coherence $g_{I,J}(t)$ between the ionic states $I=4d_0^{-1}$ and $J=5s^{-1}$ is shown for different pulse durations.
    The mean photon energy is kept fixed at 136~eV.
    The final degree of coherence at 1~fs after the pulse is projected onto a 2D graph shown on the right.
    The figure is taken from Ref.~\cite{PaSa-PRL-2011}.
    Copyright~\copyright~2011 American Physical Society (APS).
  }
  \label{fig2:coh}
\end{figure}

The final energy of the photoelectron must also be indistinguishable for the two channels.
This is only possible when the spectral bandwidth of the pulse is larger than the energy difference between both ionic states.
In xenon, the energy difference between the ionic states $4d_0^{-1}$ and $5s^{-1}$ is $\approx 50$~eV.
For a Fourier-limited Gaussian pulse that means a full-width-half-maximum (FWHM) duration of $\approx 36$~as.
To classify the entanglement between the parent ion and the photoelectron, the lack of coherence (i.e., decoherence) in the ionic subsystem is investigated.
The degree of coherence $g_{I,J}(t)$ between the two ionic states $I$ and $J$ is given by
\begin{align}
  \label{eq2:doc}
  g_{I,J}(t)
  &=
  \frac{\left|\rho^\IDM_{I,J}(t)\right|}
  {\sqrt{\rho^\IDM_{I,I}(t) \rho^\IDM_{J,J}(t)}}
  ,
\end{align}
where $g_{I,J}=1$ stands for perfect coherence and $g_{I,J}=0$ stands for a perfectly incoherent system. 
In Fig.~\ref{fig2:coh}, the degree of coherence between the ionic states $4d_0^{-1}$ and $5s^{-1}$ is shown as a function of time for different pulse durations.
The mean photon energy is kept constant at 136~eV, well above the ionization potential of the $4d_0$ and $5s$ orbital (75~eV and 25~eV, respectively).

There are several interesting aspects encoded in Fig.~\ref{fig2:coh}.
First, the initial ($t=0$) and the final ($t=1~fs$) degrees of coherence decrease for long pulse durations.
This drop is well understood and is directly related to how well the spectral bandwidth of the pulse can cover the energy difference between $4d_0^{-1}$ and $5s^{-1}$.
For longer pulses, the spectral bandwidth decreases such that it becomes possible to determine energetically from which orbital the photoelectron came.

Second, the initial degree of coherence at $t=0$ increases monotonically with shorter pulses.
When the spectral bandwidth becomes larger than the energy difference between the ionic states, $g_{I,J}(0)$ approaches perfect coherence.
Third, interchannel coupling destroys this initial coherence by entangling the photoelectron with the ion.
When interchannel interactions are switched off, the high initial degree of coherence is preserved for later times (see Fig.~4 in Ref.~\cite{PaSa-PRL-2011}).
For very short pulses in particular, the ionization dynamics of the electron becomes strongly entangled with the ion. 
Consequently, the ionic coherence starts to drop directly after the pulse.
The oscillations in $g_{I,J}(t)$ also indicate that there exists a strong, dynamic interaction between the ion and the photoelectron.

The reason why interchannel coupling increases in importance for ultrashort pulses lies in the nature of the broad spectrum.
Even though the mean photon energy is sufficiently large, there exist many energetically lower photons in the broad spectrum, which result in a photoelectron with almost no kinetic energy.
Thes low energy part of the photoelectron does strongly interact with the parent ion and leads to an enhanced entanglement.
Our studies on the degree of coherence as a function of the mean photon energy with a fixed spectral bandwidth (see Fig.~3 in Ref.~\cite{PaSa-PRL-2011}) confirm that the enhanced entanglement is due to the low rather than the high energy part of the photoelectron.

The lesson learned from this study is that the required spectral bandwidth needed to create a coherent superposition is not sufficient to ensure that the hole wavepacket in the ion is coherent.
Besides the interest in understanding fundamental processes, the question whether or not it is possible to create a coherent hole wavepacket in an ion via photoionization is of wider interest.

In molecular systems, where the valence-shells are highly delocalized, the hole dynamics can be quite complex~\cite{KuBr-JCP-2005,NeRe_NJoP10-2008}.
Creating a coherent hole wavepacket is important for the subsequent dynamics. 
Without coherences (between the exact ionic eigenstates) no hole motion can exists.
It has been shown that higher order correlation effects within the ion can also be a driver of hole dynamics~\cite{BrCe-JCP-2003}.
Transient absorption spectroscopy is an ideal tool to study these hole motions.
In combination with UV and even more so with x-ray light, it is possible to probe local, site-specific consequences of the delocalized electronic motion in molecules~\cite{Ch-ARPC-2005,HeZh-JPCL-2012}.

\subsubsection{Multiorbital and Multipole Effects in the HHG Spectrum of Argon}
\label{p1c2:apps_hhg}
Multiorbital effects and particularly interchannel effects are well known in photoionization (cf. Sec.~\ref{p1c2:photo}).
They lead to strong modification in the partial cross sections of valence and inner-shell orbitals.
The coherent properties of the ion are also influenced by interchannel interactions as discussed in Sec.~\ref{p1c2:apps_decoh}.
In the tunnel ionization regime, it has been shown that the electron can come from multiple orbitals~\cite{GoKr-Nature-2010,WiGo-Science-2011}.
Tunnel ionization and recombination are the first and the last step of HHG, respectively.
Therefore, it is quite natural to ask to which extent multiorbital effects have to be considered in HHG, where they are normally neglected (cf. Sec.~\ref{p1c2:hhg}).

In molecular systems, multi-orbital contributions to the HHG spectrum have been observed as a function of alignment angle~\cite{McGu-Science-2008}. 
In atomic systems, multiorbital effects are generally less important due to the higher orbital symmetries and the larger energy splittings between them.
However, this does not mean they are absolutely absent, and the larger the atom the more prominent are multiorbital effects.
For example, the HHG spectrum of xenon can only be understood when interchannel interactions between the $4d$ and $5p$ orbitals are taken into consideration (see Fig.~\ref{fig2:hhg_xenon}). 

\begin{figure}[t!]
  \centering
  \begin{subfigure}[b]{0.46\textwidth}
    \centering
    \includegraphics[width=\linewidth]{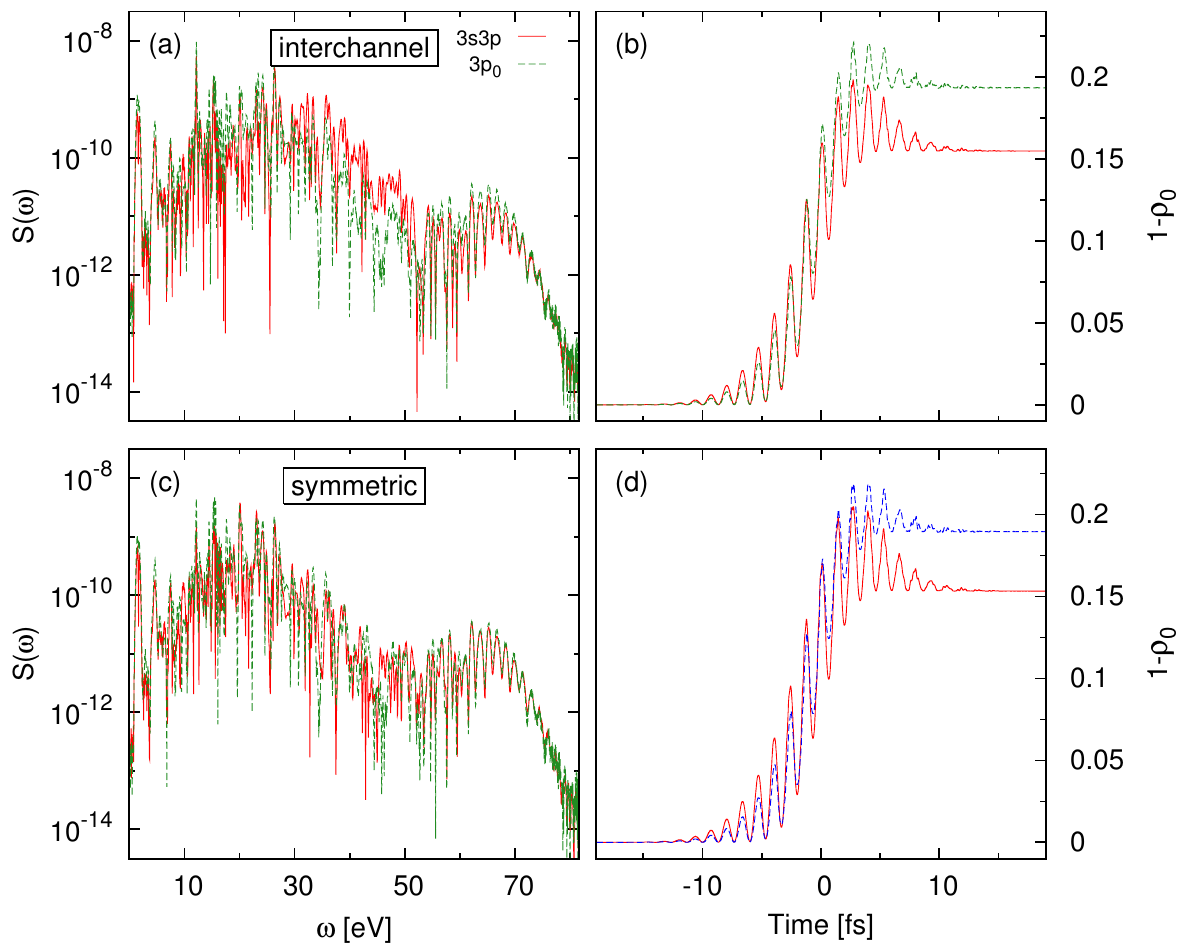}    
    \caption{HHG spectrum and depopulation}
    \label{fig2:hhg_argon_1}
  \end{subfigure}
  \begin{subfigure}[b]{0.5\textwidth}
    \centering
    \includegraphics[width=\linewidth]{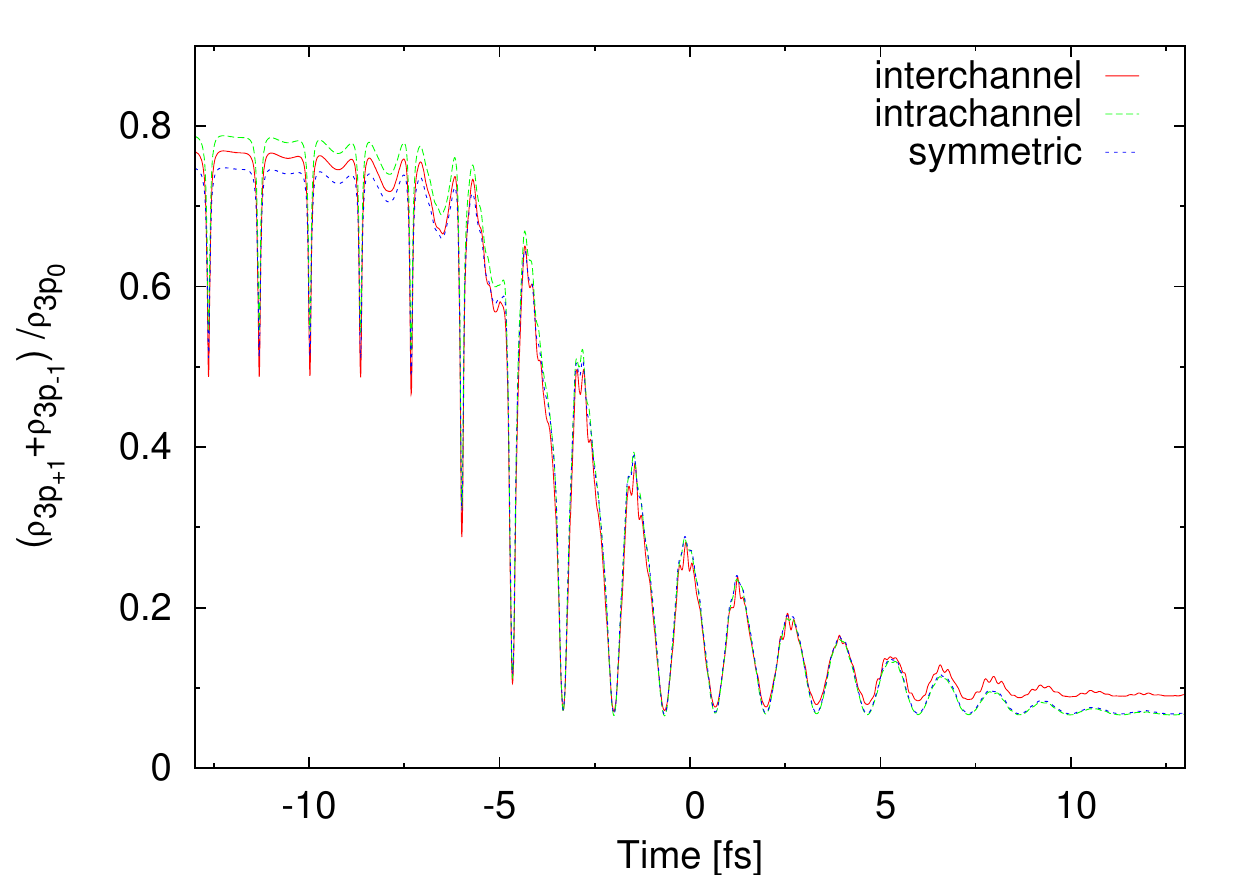}
    \caption{Population ratio}
    \label{fig2:hhg_argon_2}
  \end{subfigure}
  \caption{
    (a) The HHG spectrum $S(\omega)$ and the ground state depopulation $1-\rho_0$ of Ar are shown for different combinations of active occupied orbitals and approximations of the residual Coulomb interaction.
    Different sets of active occupied orbitals are compared for the full and the symmetrized version of $\hat H_1$.
    (b) The ratio of the hole populations $3p_0^{-1}$ and $3p_{-1}^{-1}+3p_1^{-1}$ is shown as a function of time.
    These figures are taken from Ref.~\cite{PaGr-PRA-2012}.
    Copyright~\copyright~2012 American Physical Society (APS).
  }
  \label{fig2:hhg_argon}
\end{figure}

Multiorbital as well as multipole effects in the HHG spectrum as been investigated on atomic argon~\cite{PaGr-PRA-2012}.
Argon is an ideal atom, since, with 18 electrons, it is not one of the lightest noble gas atoms~\footnote{ 
The calculations for heavier atoms, like krypton and xenon, are more costly, since for large HHG cut-off energies the $3d$ and $4d$ orbitals have to be considered, respectively.
}.
Additionally, in the photoionization cross section and also in the HHG spectrum of argon there exists a special feature called the Cooper minimum, which can be used as a marker to help quantify differences between models.
Note that the existence of the Cooper minimum is not a result of multiorbital effects but its position and shape are.
Multipole effects are normally ignored in atoms, where the effective potential of the active electron is modeled as spherically symmetric [see Eq.~\eqref{eq2:modelpotential}].
Since the electron gets mainly ionized out of the $m=0$ orbital, the remaining ion is predominantly in the $P_{m=0}$ state.
Consequently, the resulting Coulomb potential of the ion has, at short range, a quadrupolar rather than a spherically symmetric character.
To simulate a spherically symmetric potential, the residual Coulomb interaction $H_1$ is averaged over all orbital magnetic quantum numbers (for details see Ref.~\cite{PaGr-PRA-2012}).
The spherical averaging has the additional consequence that interchannel interactions drop out of the spherically symmetrized $\hat H_1$.
Hence, the symmetrized $\hat H_1$ includes only intrachannel interactions.

Both multiorbital and multipole effects are investigated separately.
First, multi-orbital effects are studied by allowing ionization out of $3s$ and all $3p$ orbitals, out of all $3p$ only, or out of $3p_0$ only.
Second, multipole effects are studied by using interchannel and intrachannel interactions, only intrachannel interactions, and the spherically symmetric intrachannel version of $\hat H_1$.
In Fig.~\ref{fig2:hhg_argon_1}, the HHG spectrum and the ground state depopulation are compared for a single-orbital ($3p_0$) and a multiorbital ($3s$ and all $3p$) model of argon.
When the symmetrized residual Coulomb interaction is used, the differences in the HHG are very small indicating that the $3p_0$ orbital is the dominant contributer to the HHG spectrum.
When interchannel interactions are included in the calculations, the HHG spectrum becomes sensitive to whether or not the orbitals $3p_{\pm 1}$ are considered.
The influence of $3s$ is minimal and can be ignored.
The direct contributions of $3p_{\pm 1}$ are negligible but their indirect influence on $3p_0$ is not.
In the energy region of 30-50~eV, the HHG spectrum is enhanced by interchannel interactions by up to one order of magnitude.

In Fig.~\ref{fig2:hhg_argon_2}, the population ratio between the orbitals $3p_{\pm 1}$ and $3p_0$ is shown as a function of time during NIR driving field, which is centered around $t=0$ and has a FWHM-width of 10~fs.
Interestingly, the ratio decreases monotonically (up to the oscillations synchronized with the electric field).
The low final ratio illustrates that mainly $3p_0$ gets ionized.
Small final hole populations in $3p_{\pm1}$, which are often used as an argument to ignore these channels, are not as small during the pulse (relative to the $3p_0$ population) as shown in Fig.~\ref{fig2:hhg_argon_2}.
Taking additionally interchannel coupling effects into account shows that the population-based argumentation of excluding the $np_{\pm1}$ orbitals is not universally valid and has to be used with caution.
Also in atomic xenon, it is due to the interchannel effects that the $4d$ orbitals cannot be neglected even though they are deeply bound and the direct contributions of $4d$ are negligible.

\subsubsection{Attosecond Transient Absorption Spectroscopy for Overlapping Pump and Probe Pulses}
\label{p1c2:apps_tas}
In Sec.~\ref{p1c2:tas}, the basic ideas of attosecond transient absorption spectroscopy (ATAS) have been discussed.
ATAS is an ideal probe for studying the electronic dynamics in the parent ion.
The high time resolution accompanying ATAS makes it possible to study ultrafast electronic motion on a subfemtosecond time scale.
Since an optical cycle of 800~nm (NIR) is 2.6~fs, subcycle ionization dynamics can be studied as demonstrated in Ref.~\cite{WiGo-Science-2011} with atomic krypton.

The final degree of coherence of $g=0.85 \pm 0.05$ that has been experimentally achieved agrees with the TDCIS predictions of $g=0.82$ after taking into account propagation effects, which increase the degree of coherence measured at the detector by up to 14\%  ($g=0.72$ before propagation effects).
In order to theoretically study the coherences between the ionic states $[4p^{\pm1/2}_{1/2}]^{-1}$ and $[4p^{\pm 1/2}_{3/2}]^{-1}$, the {\sc xcid} package was extended to include spin-orbit effects in the occupied orbitals\footnote{
The notation of the ionic states is explained in Sec.~\ref{p1c2:tas}.
}.
The exact residual Coulomb interaction, $\hat H_1$, has also been included to account for coherence losses due to interchannel effects.

\begin{figure}[t!]
  \centering
  \includegraphics[width=.82\linewidth]{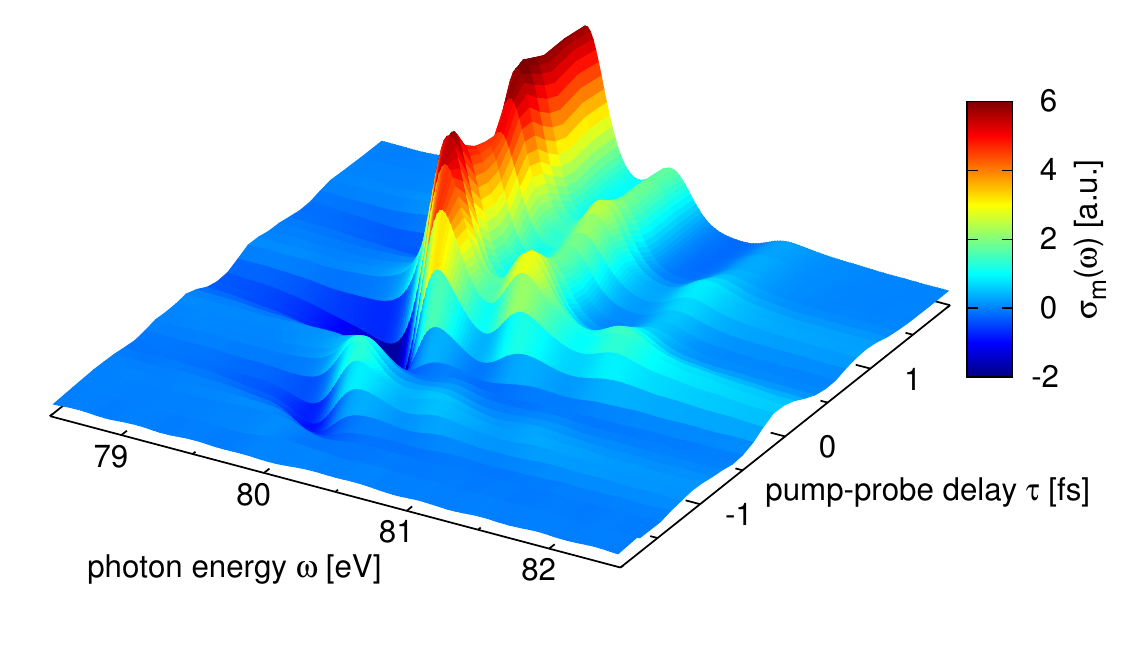}
  \caption{
    Attosecond transient absorption spectrum of krypton during tunnel ionization is shown as a function of photon energy $\omega$ and pump-probe delay $\tau$.
    The figure is taken from Ref.~\cite{PaSy-PRA-2012}.
    Copyright~\copyright~2012 American Physical Society (APS).
  }
  \label{fig2:tas-3d}
\end{figure}

For well separated and non-overlapping pump and probe pulses, the three transient absorption lines in Kr$^+$ corresponding to the transitions $[4p_{3/2}^{m}]^{-1} \rightarrow [3d_{5/2}^{m}]^{-1}, [4p_{1/2}^{m}]^{-1} \rightarrow [3d_{3/2}^{m}]^{-1}$, and $[4p_{3/2}^{m}]^{-1} \rightarrow [3d_{3/2}^{m}]^{-1}$ (see Fig.~\ref{fig2:tas-3d}) contain all information about the tunnel-ionized Kr$^+$ ion with a hole in the $4p_j$ orbital manifold.
As theoretically~\cite{SaYa-PRA-2011} and experimentally~\cite{GoKr-Nature-2010} shown, this information can be used to reconstruct the full ion density matrix (IDM).
Establishing a direct mapping between the transient absorption spectrum and the instantaneous IDM requires that the ionized electron does not influence the ion during and after the probe step. 
This condition does not hold anymore when the pump step (i.e., tunnel ionization) is probed, since the ionized electron as well as the pump field itself influence the ion state.
Therefore, it is not clear to which extent the instantaneous IDM can be probed during the ionization process.

Additionally, strong modification of the transient absorption spectrum has been observed during tunnel ionization (see Fig.~\ref{fig2:tas-3d}) indicating that the krypton ion cannot be treated as isolated and field-free.
Both aspects arising with probing the tunnel ionization process are have been addressed.~\cite{PaSy-PRA-2012} 

The comparison between the hole populations obtained from a calculated transient absorption spectrum and the corresponding instantaneous hole populations revealed that the obtained hole populations match quite well with the instantaneous ones.
A delay in the extracted hole motion of up to 200~as was found (see Fig.~5 in Ref.~\cite{PaSy-PRA-2012}).

Beside the question of population dynamics, a new phenomenon has been identified during the ionization process.
The transition lines do not just rise in strength as the hole populations do.
They also show strong deformations in their shape as shown in Fig.~\ref{fig2:tas-3d}.
Before trying to understand where these deformations come from it is important to understand the basic mechanism behind the transient absorption spectrum.

The reduction in transmitted photons can also be described in a semi-classical picture, where the electric field is not quantized and treated as a classical field. 
In classical electrodynamics, the Larmor formula describes the generation of radiation due to the acceleration of charged particles~\cite{Jackson-book}.
Quantum mechanically, the electron motion is captured in the dipole moment $\ev{\hat z}(t)$.
The generated and the absorbed radiation of an ion is, therefore, determined by the dynamics of the ionic dipole moment $\ev{\hat z}_\text{ion}(t)$.
A field-independent photoionization cross section $\sigma(\omega)$ (see Sec.~\ref{p1c2:photo}) reads in terms of the ionic dipole moment
\begin{align}
  \label{eq2:cs}
  \sigma(\omega)
  &=
  4\pi\,\alpha\ \omega
  \text{Im} \left[ \frac{\ev{\hat z}_\text{ion}(\omega)}{E(\omega)} \right]
  ,
\end{align}
where $\omega$ is the photon energy and $E(\omega)$ is the spectrum of the incident electric field.

\begin{figure}
  \centering
  \begin{subfigure}[b]{0.5\textwidth}
    \centering
    \includegraphics[width=\linewidth]{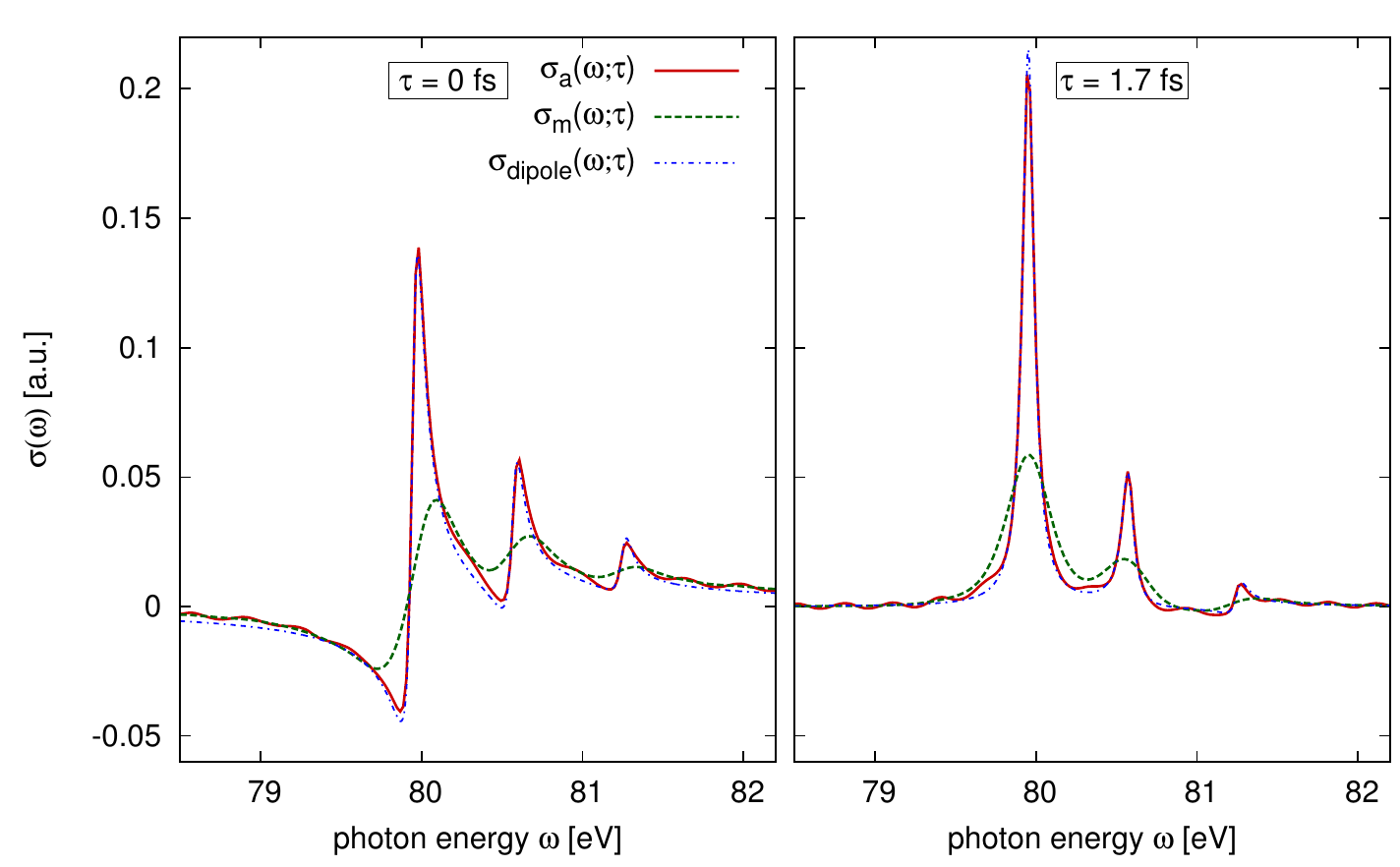}
    \caption{Transient absorption spectrum}
    \label{fig2:tas-fit_1}
  \end{subfigure}
  \begin{subfigure}[b]{0.45\textwidth}
    \centering
    \includegraphics[width=\linewidth]{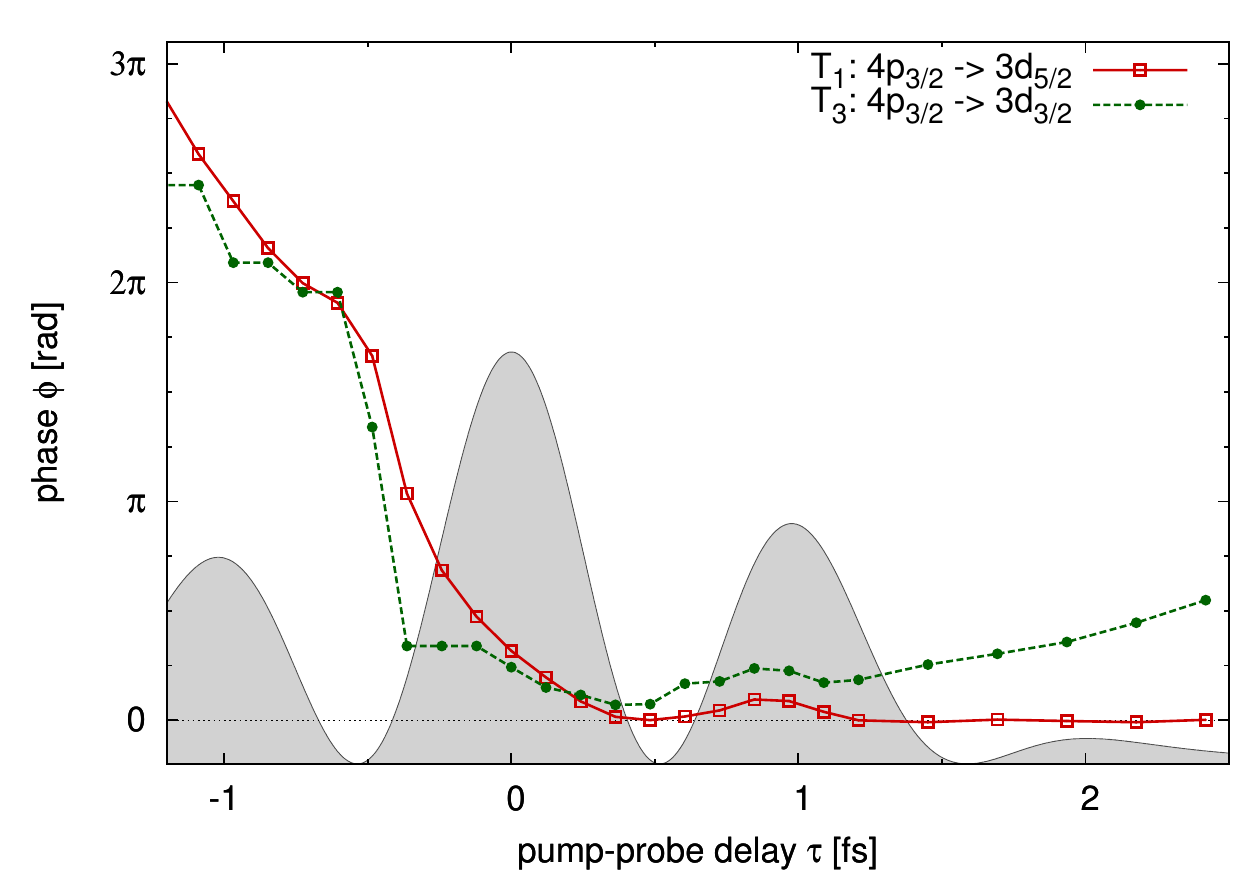}      
    \caption{Ionic dipole phase $\phi(\tau)$}
    \label{fig2:tas-fit_2}    
  \end{subfigure}
  \caption{
    (a) The calculated atomic cross sections (red solid line), the measured cross sections (green dashed line), and the fit obtained from a simplified oscillating dipole model (yellow dotted line) are shown for the pump-probe delays $\tau=0$~fs and $\tau=2.4$~fs, respectively.
    (b) Ionic dipole phase $\phi(\tau)$ obtained from the fits for the transition lines $4p_{3/2}^{-1} \rightarrow 3d_{5/2}^{-1}$ (red solid line) and $4p_{3/2}^{-1} \rightarrow 3d_{3/2}^{-1}$ (green dashed line) are shown.
    The figures are taken from Ref.~\cite{PaSy-PRA-2012}.
    Copyright~\copyright~2012 American Physical Society (APS).    
  }
  \label{fig2:tas-fit}
\end{figure}

In the case of absorption, the ionic dipole, $\ev{\hat z}_\text{ion}(t)\propto \sin(\omega_0\,t)$, oscillates $180^\circ$ out of phase with respect to $E(t)$ such that the spectral strength $E(\omega_0)$ gets reduced for the specific oscillation frequency $\omega_0$.
If the dipole oscillates in phase, the electric field gets enhanced leading to an emitting behavior.
The energetically lowest absorption line $[4p_{3/2}^{m}]^{-1} \rightarrow [3d_{5/2}^{m}]^{-1}$  in Fig.~\ref{fig2:tas-3d} shows a purely absorbing (Lorentzian) behavior when pump and probe pulses do not overlap\footnote{
The other two absorption lines do not show purely absorbing behavior due to coherences between the ionic states $[4p^{\pm1/2}_{1/2}]^{-1}$ and $[4p^{\pm 1/2}_{3/2}]^{-1}$ (see Sec.~\ref{p1c2:tas}).
}.
The widths of the transition lines are determined by the lifetime of the $3d^{-1}$ ionic states and the detector resolution.

When the phase shift $\phi$ in the oscillating ionic dipole [$\ev{\hat z}(t)\propto \sin(\omega_0\,t +\phi)$] is not a multiple of $\pi$, the transition line shapes are not Lorentzian anymore (see Fig.~\ref{fig2:tas-fit_1} for $\tau=0$).
Figure~\ref{fig2:tas-fit_1} shows cuts of the transient absorption spectrum of Fig.~\ref{fig2:tas-3d} for the pump-probe delays $\tau=0,2.4$~fs.
At $\tau=0$~fs, the NIR field peaks and all three transition lines are strongly deformed.
The ionic phase shifts $\phi(\tau)$ are shown in Fig.~\ref{fig2:tas-fit_2} for the energetically smallest and largest transition lines.
They are obtained from the TDCIS results by fitting Eq.~\eqref{eq2:cs} with $\ev{\hat z}_\text{ion}(t)=z_0\,\sin(\omega\,t + \phi)$ for each pump-probe delay.
The increasing phase shift in the $4p_{3/2}^{-1} \rightarrow 3d_{3/2}^{-1}$ transition for large $\tau$ is due to the coherent superposition of the ionic states $4p_{3/2}^{-1}$ and $4p_{1/2}^{-1}$ (see Sec.~\ref{p1c2:tas}).
There are three mechanisms that could contribute to $\phi$:

\begin{enumerate}
 \item 
    the quadratic Stark shift of the ionic energy levels due to the high electric field strength of the
    ionizing pulse,
  \item
    the residual Coulomb interaction between the ion and the electron, 
  \item
    field-driven coupling between the freed electron and the parent ion via the neutral ground state $\Phi_0$.
\end{enumerate}

It was found that the influence of the residual Coulomb interaction on $\phi$ is negligibly small.
The ionic Stark shift is quite large for each ionic state.
However, what matters are the differences in the energy shifts between the ionic states such that the transition energy gets modified.
Even though the absolute energy shifts could be as large as 1~eV, the relative energy differences do not change by more than 100~meV, which leads to a maximum phase shift of $\approx \frac{\pi}{10}$.
The phase shift shown in Fig.~\ref{fig2:tas-fit_2} is, however, much larger such that the only possible mechanism is the field-driven dressing of the neutral ground state.
The large phase shifts disappear when the field-driven coupling to the neutral ground state is immediately switched off after the ion has been probed (see Fig.~9 in Ref.~\cite{PaSy-PRA-2012}).
Interestingly, the electric field cannot directly couple the ion with the ionized electron.
Only via the neutral ground state (i.e., $\bra{\Phi_0}\hat z\ket{\Phi^a_i}$) is it possible to create a field-driven correlation interaction the ionic and electronic subsystems.

This study has shown that attosecond transient absorption spectroscopy is a versatile technique to investigate ultrafast electronic motion.
It can be used to probe the electronic state of the ionic subsystem, and at the same time it is sensitive to interactions of the ionic subsystem with the environment.


\section{Summary}
\label{p3c1:conclusion}
Ultrafast sciences are young and rapidly developing.
The characteristic time scales of electronic and nuclear dynamics range from picoseconds ($10^{-12}$~s) to attoseconds ($10^{-18}$~s). 
No mechanical and electronic devices operate on such time scales.
Therefore, the only way to investigate this nanoworld is by literately shining light on it.
Within the last decade tremendous technological progress has been made in generating short and shorter pulses, and at the same time possessing better control of their shape and timing with respect to other pulses.
These are essential prerequisites for watching and controlling ultrafast motions of atoms and molecules.
Along with the reduction in the pulse duration comes a steady increase in the photon energy of these pulses.
This wide tunability makes it possible to study a vast range of motions spanning from molecular rotations and vibrations to electronic dynamics of outer-valence, inner-shell, or even core electrons.

On the theoretical side, much progress has also been made in tackling the challenges that emerge in the ultrafast world.
Many dynamics triggered by light pulses lead to complex motions requiring a large configuration space in order to fully capture the motion.
Laser alignment of molecules or the high-harmonic generation (HHG) process are two examples where high rotational states with angular momenta $J>50$ are needed to accurately describe the underlying dynamics. 
The large variety of pulse durations, intensities, and photon energies make it difficult to develop a theoretical model that can cope with all the possible kinds of dynamics that can be triggered by such a large range of pulse parameters. 
Particularly challenging are laser-matter interactions that cannot be treated perturbatively, such as strong-field ionization or non-adiabatic alignment of molecules in three dimensions.

In this review, two prominent classes of ultrafast dynamics have been discussed:
\begin{itemize}
 \item laser-alignment dynamics of molecules,
 \item ultrafast ionization dynamics of noble gas atoms,
\end{itemize}
where the characteristic time scale of molecular rotations is in the picosecond range, and ionization dynamics happens on the scale of attoseconds to a few femtoseconds.
In order to describe a large variety of pulse interactions with the molecule or atom, the corresponding Schr\"odinger equation has to be solved by an explicit time propagation.
Solving numerically the time-dependent Schr\"odinger equation is computationally demanding but, therefore, also more general.
Exploiting the underlying symmetries has proven to be essential in making these calculations more feasible and pushing them even further.
Often speed-ups of a factor 2 or more can be easily achieved.

\subsection{Laser Alignment of Molecules}
The rotational motion of molecules is triggered by non-resonant laser pulses.
The induced dipole moment in the molecule interacts with the electric field of the pulse and creates an angle-dependent potential, which forces the molecules to be aligned in the polarization direction(s) of the pulse.
Using pulses that are short or long with respect to the rotational dynamics allows the generation of complex rotational motions.
Combinations of several short pulses, long and short pulses, or static and optical fields are also exploited.
This is often done to maximize the degree of alignment without ionizing the molecule.
Multi-pulse methods have been predominantly used to study impulsive alignment dynamics of linear or symmetric-top molecules due to their unique alignment dynamics.
Impulsive alignment offers the possibility to study these aligned molecules under field-free conditions.
This is particularly interesting for strong-field or attosecond experiments on molecular systems where the alignment is needed but the subsequent processes should not be disturbed by the presence of an aligning laser pulse.

In the case of asymmetric-top molecules, standard multi-pulse schemes, which are used to align symmetric-top molecules in 1D, cannot be easily applied to 3D alignment.
Just replacing elliptically polarized pulses with linearly polarized pulses is not enough to impulsively align asymmetric-top molecules in 3D.
Particularly, the rotational revival feature of linear and symmetric-top molecules does not exist in asymmetric-top molecules. 
Even though asymmetric-top molecules can show some kind of revival behavior, the dephasing between rotational states is too rapid to be recoverable by a subsequent pulse.
The only promising multi-pulse approach that can overcome the rapid dephasing is a fast sequence of pulses very closely spaced to each other.
The number of pulses is unfortunately limited because the spacing between them has to decrease with every pulse since the rotational dynamics becomes faster and faster.
Studies on SO$_2$ have shown that the alignment of the most and the second most polarizable axes increases with the number of pulses.
The degree of alignment of the third axis is, however, almost unchanged from its initial isotropic value.

Aligning molecules is not done out of self-interest. 
Aligning molecules is essential for many experiments.
Hence, the understanding of alignment motion is crucial to find new ways of improving the alignment even further.
All kinds of laser-matter interactions are highly dependent on the relative alignment/orientation of the molecule.
This can be seen in the alignment dependence of the HHG spectrum, and also in x-ray diffraction studies of laser-aligned molecules.
Investigations of x-ray diffraction from laser-aligned naphthalene molecules have shown that the degree of 3D alignment is essential for getting useful, high-$Q$ diffraction information needed to reconstruct the molecular structure with atomic resolution.
For naphthalene, degrees of alignment of $\ev{\cos^2\theta}>0.9$ are needed to resolve atomic positions.
This can only be achieved with adiabatic alignment at the moment, since impulsive alignment schemes for asymmetric molecules do not reach this high degree of alignment.

Finite rotational temperatures of the gas phase molecules are the main cause of a reduced alignment quality.
Temperatures below 1~K are required to achieve the necessary alignment, since the maximum intensity of the adiabatic alignment pulse cannot be arbitrarily high.
If the pulse intensity exceeds the tunnel ionization threshold, naphthalene molecules start to get ionized. 
As long as the x-ray pulse duration is comparable or shorter than the alignment dynamics, the effective resolution in the reconstruction is not affected by the finite pulse duration.
Studies have shown that at rotational temperatures of 1~K it is possible to image molecular structure with a resolution such that the two-ring structure of naphthalene is visible.
At temperatures of $T=10$~mK, the position of the individual carbon atoms starts to emerge.

\subsection{Ultrafast Ionization Processes in Noble Gas Atoms}
When the pulse intensity reaches values of $10^{13}--10^{14}$~W/cm$^2$, the electric field is so strong that an electron can be ripped out from an atom.
This ionized electron can strongly interact with the ion and changes its state as long as it is in the vicinity of the ion.
Studying these types of effects requires a multichannel theory such as time-dependent configuration-interaction singles (TDCIS).
TDCIS combines the one-electron nature of many strong-field processes with electronic structure theory, where correlation effects between electrons are included that go beyond the independent particle model.

HHG is a very prominent mechanism connecting strong-field physics with attosecond physics.
Nowadays, HHG is not just used as a tool to produce UV attosecond pulses; the HHG spectrum itself has moved into the scientific focus.
It can provide information about the electronic structure and electronic motion.
To successfully investigate the electronic structure, it is essential that correlation effects are included in the theoretical analyses.
Our findings on the HHG spectrum of argon demonstrate that multi-orbital and multipole effects have to be included in the theory.
Not doing so can lead to HHG yields that are up to one order of magnitude off.
The HHG cut-off region seems to be quite robust to these effects but at lower energies the changes can be significant.
In argon, the benchmark was the position and the shape around the characteristic Cooper minimum.
Similar to the giant dipole resonance in xenon, the existence of the Cooper minimum is not a multielectron effect.
But its strength, shape, position, and indirect side effects in neighboring subshells depend strongly on multielectron and correlation effects.

Ultrafast correlation dynamics have been investigated in several studies.
In the case of photoionization correlation effects introduce new decoherence phenomena in the ionic system.
The photoelectron (particularly the slower ones) interacts with the parent ion during its detachment---a process neglected in the sudden approximation.
The energy gained by the photoelectron due to absorption of a photon can be redistributed between the electron and the ion via the residual Coulomb interaction.
Such an energy exchange (known as interchannel effects) leads subsequently to an entanglement between the photoelectron and the ion.
If at some later stage only the ionic subsystem is probed (e.g., with transient absorption spectroscopy), one finds that the ionic state is not in a pure state---meaning it cannot be described by a coherent ionic wavefunction and is only fully characterized by a density matrix description.
The entanglement between the photoelectron and the ion naturally results in a reduction in the coherences of the subsystems.
Particularly slow photoelectrons do strongly interact with the ion.
Ultrashort pulses with broad spectral bandwidths create both slow and fast photoelectrons.
The increase in slow photoelectrons with broader spectral bandwidths reduces the decoherence in the ionic subsystem despite the fact that the broad spectrum is favorable for the creation of a coherent hole wavepacket in the ion.
If no interchannel effects occur during ionization, the electron-ion entanglement is strongly reduced and the ultrashort pulse triggers an almost perfectly coherent hole motion.

The possible degree of coherence that can be generated via photoionization is highly interesting.
Photoionization is the first step in launching a hole wavepacket.
In molecular systems, the hole dynamics can be quite complex and are not restricted to the atomic site where it was initiated.
After some time, the ionic hole may have traveled to a different atomic site in the molecule where it can influence site-specific chemical reactions; a situation which has already been observed in recent experiments.

The preparatory ionization process as well as the subsequent hole motion in the ion can be investigated by transient absorption spectroscopy.
Not only can the ionic population be studied time-dependently, but also the phase relations between ionic states can be measured.
Studies on atomic krypton have shown that access to the relative phase makes it possible to directly probe field-driven interactions.
During the tunnel ionization of krypton, the field-driven interaction between the ion and the freed electron via the neutral ground state results in strong deformations in the shapes of the absorption lines in the transient absorption spectrum.

Transient absorption spectroscopy, together with complementary techniques such as streaking, opens the door to the ultrafast world of both partners---the photoelectron and the ion---and the interaction between them.
Combined with the control of molecular alignment, not just atomic but also molecular dynamics can be studied.

\section{Outlook}
\label{p3c1:outlook}

\subsection{Ultrafast Ionization Processes in Noble Gas Atoms}
The rapid progress in the last years of strong-field and attosecond physics can be expected to continue in the next years to come.
The steady push towards shorter pulses and higher photon energies from the HHG side as well as from big facilities, namely the development of free-electron lasers (FELs) in the UV (e.g., \href{http://hasylab.desy.de/facilities/flash/}{FLASH}) and in the x-ray regime (e.g., \href{http://lcls.slac.stanford.edu}{LCLS}, \href{http://xfel.riken.jp}{SACLA}, and the \href{http://www.xfel.eu}{European XFEL}), pave the way for a wide range of coherent pump-probe experiments on femtosecond and attosecond time scales.

Transient absorption spectroscopy with femtosecond x-ray pulses would allow the probing of delocalized valence hole dynamics in molecules at specific atom sites by tuning the x-ray energy near element-specific absorption edges.
Also nuclear motion and its interplay with electronic excitations, leading to non-Born-Oppenheimer dynamics, can be studied on a few femtosecond time scale.
On a picosecond time scale, structural deformations have already been measured in metal-ligand complexes.
Here, additional dynamics in the spin degrees of freedom occur that are less common in closed-shell systems.

The theoretical, or more precisely the numerical, demand in solving a molecular system is much more challenging than for an atomic closed-shell system.
The great advantage that comes with a spherically symmetric object is then lost and the angular part of the wavefunction no longer factorizes from the radial part.
For elliptically polarized pulses, this radial-angular factorization is lost as well, regardless of the system symmetry.
Elliptically or circularly polarized light introduces new dynamics in the angular projection $M$.
Particularly for magnetic (open-shell) systems, circularly polarized light is ideal for studying magnetic and spin dynamics.
Therefore, the extension of the {\sc xcid} package to circularly and elliptically polarized pulses opens the possibility of studying new classes of motion that are not possible in the current version of the program.

Molecular systems, however, offer a much richer electronic motion, potentially coupled to vibrational or rotational degrees of freedom.
The single-active electron picture, which works quite well for closed-shell atoms, becomes less and less appropriate for larger molecular systems.
A theoretical description of a dynamical process with several active electrons (e.g., TDCISD) is on the other hand very challenging.
Therefore, most studies that included rigorously multi-electron dynamics focus on the smallest two-electron system---helium.

In strong-field processes, it is necessary on the one side to describe delocalized continuum states.
On the other side, it is not always necessary to include high order correlations, as done in higher order configuration-interaction (CI) approaches.
The idea of reducing the correlation dynamics but leaving the degrees of freedom for the spatial and momentum dynamics untouched can be realized in a multi-configuration time-dependent Hartree-Fock (MCTDHF) ansatz.
In MCTDHF, configuration coefficients as well as one-particle orbitals are time-dependent.
The advantage of time-dependent orbitals is that the configuration space is dynamically optimized, which keeps the number of needed configurations small.
Within this smaller dynamical configuration space, a full CI (FCI) is performed.

Instead of performing a FCI within this dynamics subspace, which might be quite challenging for ultrafast scenarios, it seems more attractive to do a lower order CI like CIS.
This idea of merging the MCTDHF logic with a TDCIS ansatz seems to be more promising and especially a much more feasible approach for extending the current {\sc xcid} package.
In addition to the common set of equations of motion for the configuration coefficients, a new set of equations emerges for the time-dependent orbitals.
With such an approach, the remaining electrons in an ion are now allowed to adjust to the removal of the $N^\text{th}$ electron without the need to introduce higher-order correlation classes.
Already the polarizability studies of Kr$^+$ have shown that the spatial adjustments of the orbitals rather than higher correlation effects are crucial for obtaining more reliable results.

\subsection{Laser Alignment of Molecules}
For strong-field and attosecond experiments with molecular systems, the alignment or even the orientation of the molecule becomes an important aspect.
Unfortunately, despite the fact that the polarizability increases with system size, larger molecules become more and more difficult to align.
The growing negative influence of the finite rotational temperature outpaces the positive effect of the polarizability.
Larger molecular are also less stiff than smaller ones.
This means that larger molecules more easily get deformed by an external field making the rigid-rotor approximation questionable.

Furthermore, the polarizability tensor is generally not diagonal in the principal axes of inertia.
Only for small molecules with high internal symmetries (e.g., SO$_2$ or naphthalene) are the polarizability tensor and the moment of inertia tensor diagonal in the same body-fixed frame.
The larger the molecules, the less likely it is that these two sets of reference frames coincide.
This creates two kinds of challenges for larger molecules.
First, the alignment dynamics becomes more challenging, since the molecule rotates around the moment of inertia axes but the aligning field forces the molecule to align to the polarizability axes.
Therefore, it is not clear to which extent current techniques are still suitable for large molecules.
Second, the computational challenge grows as well. 
The increased complexity in the alignment dynamics gets reflected onto the more complex dynamics of the angular momentum projection $K$, which no longer changes in steps of 2 but rather of 1, doubling the number of accessible rotational states.

Another interesting extension for the {\sc xalmo} package is the possibility to treat longer wavelengths, where the alignment motion cannot be averaged over a cycle period of the aligning laser pulse.
Such pulses can orient polar molecules.
From the numerical point of view, this leads to another loss of symmetry in the angular projection $M$, which would not change in steps of 2 anymore. 
Together with the loss in symmetry of the $K$ quantum number for large molecules, the numerical demand is significantly increased in comparison to the alignment studies presented here.
Despite the challenges, this extension would offer new opportunities to discover novel alignment dynamics schemes that boost the efficiency of aligning and/or orienting molecules.

\subsection{Overall Perspective}
From rotational motion of molecules down to single electronic excitations, a wide range of nanoscale dynamics can nowadays be probed and controlled by ultrafast pulses, and the possibilities continue to expand to even more complicated systems.
With the advent of FELs, intense femtosecond UV and x-ray (short wavelength) pulses became available.
Pump-probe experiments in the UV and x-ray regimes can target atom-specific electronic dynamics.
Particularly for molecules, this raises the tantalizing prospect of controlling chemical reactions by triggering the break-up and formation of bonds. 
Optical and NIR (long wavelength) pulses, on the other hand, provide unmatched spatial and temporal coherence properties.
Due to the high degree of coherence, electron motion on an attosecond scale can be revealed with techniques like attosecond streaking.
Optical pulses can also be used to probe structural information via high harmonic generation.
With more moderate intensities, optical pulses control the rotational dynamics of molecules.
By continuing the rapid progress experimentally and theoretically, the dream of making a molecular movie and watching electrons and atoms move might become a reality sooner than we think.

\begin{acknowledgement}
I am very greatful for the fruitful scientific discussions with Robin Santra.
I want to thank Cassandra Hunt for comments on the manuscript.
The financial support from the Deutsche Forschungsgemeinschaft (DFG) under the program Sonderforschungsbereich (SFB) 925/A5 is acknowledged.
\end{acknowledgement}

\bibliographystyle{back/apsrev4-1-etal}
\bibliography{back/amo,back/books,back/solidstate}

\end{document}